%% file: LinePaper.tex
\documentclass[aps,prd,preprint,superscriptaddress]{revtex4-1}
\usepackage{lineno}

\usepackage{hyperref}
\usepackage{graphicx}
\usepackage{comment}
\usepackage{subfigure}
\usepackage{mathrsfs}
\usepackage{multirow}
\usepackage{amsmath}
\usepackage{url}

\begin{document}

\title{\emph{Fermi} LAT Search for Dark Matter in Gamma-ray Lines \\ and the Inclusive Photon Spectrum}

\include{PhotonLinePaperAuthorList_011812}

\collaboration{FERMI-LAT}

\date{\today}

\begin{abstract}
Dark matter particle annihilation or decay can produce monochromatic gamma-ray lines and 
contribute to the diffuse gamma-ray background. 
Flux upper limits are presented for gamma-ray spectral lines from 7 to 200 GeV and for the 
diffuse gamma-ray background from 4.8 GeV to 264 GeV obtained from two years of 
\emph{Fermi} Large Area Telescope data integrated over most of the sky.  
We give cross section upper limits and decay lifetime lower limits for dark matter models 
that produce gamma-ray lines or contribute to the diffuse spectrum, including models proposed 
as explanations of the PAMELA and \emph{Fermi} cosmic-ray data.
\end{abstract}

\maketitle


\section{Introduction}

The \emph{Fermi} Gamma-ray Space Telescope (\emph{Fermi}), with its main instrument, the 
Large Area Telescope (LAT) \citep{Atwood2009}, is exploring the gamma-ray sky in the 
energy range 20 MeV to above 300 GeV with unprecedented accuracy.  
If weakly interacting massive particles (WIMPs) constitute the dominant component of the dark matter
in the Universe, the LAT may be sensitive to gamma rays that are produced from their annihilation or decay 
in this energy range~\citep{Bergstrom1988,Jungman1996,Bergstrom:1997fj,Bertone2005,Bergstrom2000}.  

We search for monochromatic gamma rays (referred to as photons) from WIMP annihilation or decay.   
We select photons with energies 4.8 GeV -- 264 GeV over the region-of-interest (ROI) of $|b|>10^\circ$ 
plus a $20^\circ\times20^\circ$ square centered at the Galactic center (GC) with point sources removed.  
This inclusive spectrum is also used to set constraints on WIMP annihilation or decay into various final states 
that produce a \emph{continuous} photon spectrum up to the mass of the WIMP. 

If a WIMP ($\chi$) annihilates or decays directly into a photon ($\gamma$) and another 
particle ($X$), the photons are approximately monochromatic with energy 
\begin{equation}
E_{\gamma}=m_{\chi}\,\Big(1-\frac{m^2_X}{4m_{\chi}^2}\Big)
\end{equation}
for annihilations and replacing $m_{\chi}\rightarrow m_{\chi}/2$ for decays (we assume $v/c\sim 10^{-3}$).
Detection of one or more striking spectral lines would be convincing evidence for dark matter.  
We find no evidence for photon lines and now improve our published 
11-month analysis \citep{Bloom2010} by including two years of data, and by 
using tighter photon selection cuts.  In addition, the description of the analysis \cite{Edmonds2011} is 
much more complete.  
The published statistical limits improve by 15\%~\cite{Edmonds2011}. 
We extend the lower bound for the energy range of the line search from 30 GeV to 7 GeV.  
We keep the upper bound the same at 200 GeV. 
We show the inclusive spectrum used in 
obtaining the line limits, for the first time, over the energy range 4.8 GeV -- 264 GeV, 
and use the inclusive spectrum to derive constraints on WIMP annihilation or decay into $b$-quarks, 
gluons, $W$-bosons, electrons, muons, tau-leptons, and new force carriers that in turn decay to 
electrons or muons. This includes models proposed as an explanation of the PAMELA and \emph{Fermi} 
cosmic-ray (CR) data \cite{Adriani:2008zr,Adriani:2010rc,FermiCRE2009}.  

We use the Pass 6 photon selection for our analysis. 
In \S \ref{analysis}, we present the relevant details of the LAT detector, as well as the 
energy determination, the instrument response function to lines, and the exposure.  
\S \ref{spectrum_sec} presents the inclusive photon spectrum from 4.8 -- 264 GeV and discusses the 
spectral fitting in detail.  A careful discussion of instrument related systematics is included.  
This inclusive photon spectrum forms the basis for the line search, the subject of \S \ref{sec:linesearch}.  
There we also discuss in detail the statistical method used to search for lines, and present flux upper 
limits. 
The implications for the indirect detection of dark matter, including cross section and lifetime constraints for 
dark matter annihilation and decay, are presented in \S \ref{sec:DMimplications}.  
We summarize our results in \S \ref{sec:conclusions}. 
Many more details of the analysis can be found in \cite{Edmonds2011}.

\section{Methods}\label{analysis}

We use the Pass 6 photon selection and the Pass 6 DATACLEAN Instrument Response
Functions (IRFs) 
as a starting point \citep{Atwood2009,FSSC_P6V3_Caveats}, for which 
three methods are used to accurately reconstruct the energies of incoming photons 
over the LAT's large energy and angular phase space. 
However, the standard method of selecting the ``best'' reconstructed energy using a classification tree 
\citep{Atwood2009} can introduce artificial structure in the photon energy spectrum \cite{Edmonds2011,FSSC_P6V3_Caveats}.  
We therefore select only the shower profile (SP) reconstruction method 
\citep{Atwood2009,Edmonds2011}, 
which minimizes such systematic effects and is available for photons that deposit $> 1$ GeV.  
For our energy range, 4.8--264 GeV, the SP energy, $E_{r}$, is almost always available, and 
its use leads to a negligible loss in effective area.  Below 4.8~GeV, the SP energy is less available, becoming totally 
unavailable by 1~GeV.  Above 264~GeV, the IRFs were not available and the statistics for two years of data are low.  

Energy resolution and accurate energy scale calibration are particularly 
important for a line search.  
A photon in our energy range deposits a large fraction of its energy in the LAT calorimeter 
(this fraction decreases for higher energies), and deposits a relatively small 
amount in the tracker module \citep{Atwood2009}.  
Therefore the precision and accuracy of the photon energy reconstruction depend largely on the calorimeter design and quantities measured within the calorimeter.
Calibration of the calorimeter energy scale was performed on the ground and in orbit 
\citep{Eduardo2009,Collaboration:2011sy}.  
Beam tests were performed to validate the full energy reconstruction and to check the calibration of the absolute energy scale.  The uncertainty on the absolute calibration of the energy scale 
is [-10\%,+5\%] \citep{CRE2010,Edmonds2011}.

\subsection{Photon Selection}\label{event_selection}

\begin{figure*}[t]
\begin{center}
  \includegraphics[width=0.75\textwidth]{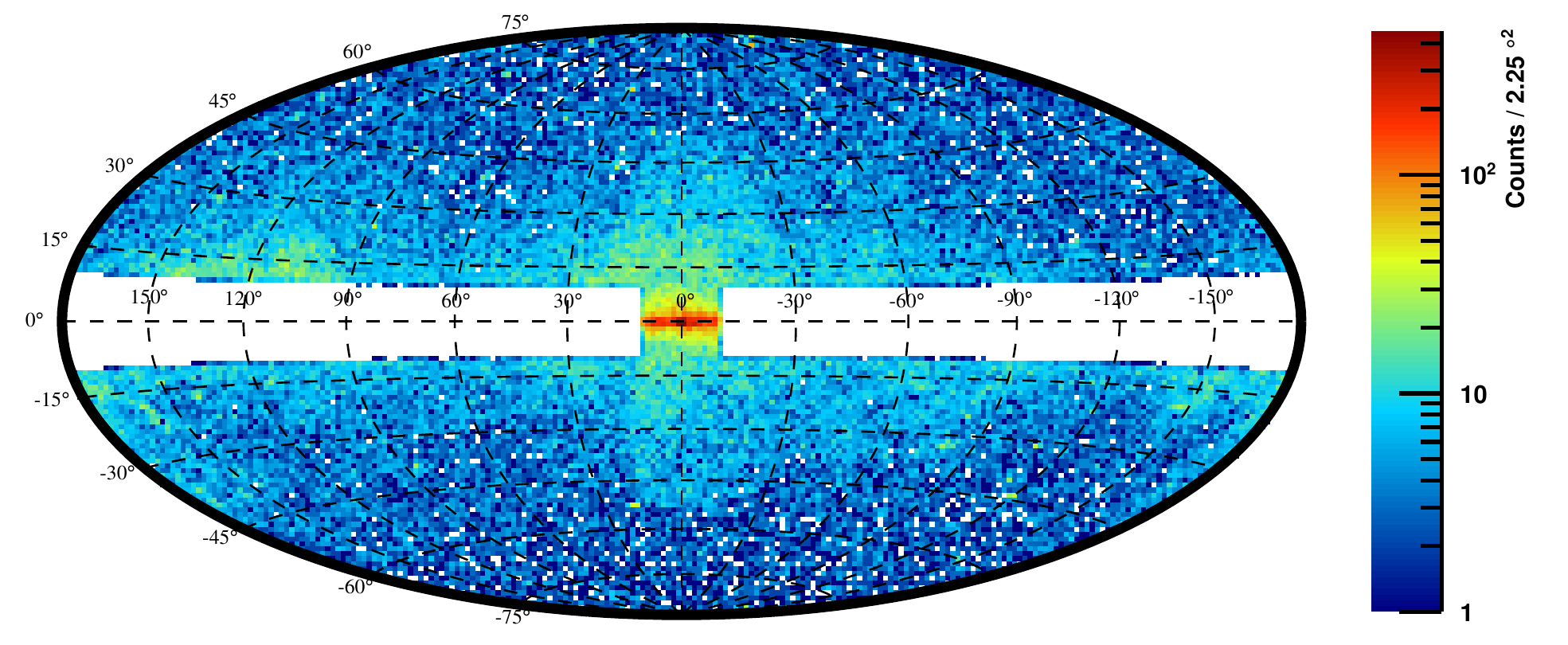}\\
  \caption{The line dataset binned in $1.5^\circ\times1.5^\circ$ spatial bins, plotted in galactic coordinates using a Hammer-Aitoff projection.  Photons with energy from 4.8 to 264 GeV are included.  The white areas away from the Galactic plane correspond to the locations of 1FGL point sources and have been masked here.   The area in white along the Galactic plane is excluded from this ROI. }\label{counts_map}
  \end{center}
\end{figure*}

\begin{figure*}[t]
\begin{center}
\includegraphics[width=0.7\textwidth]{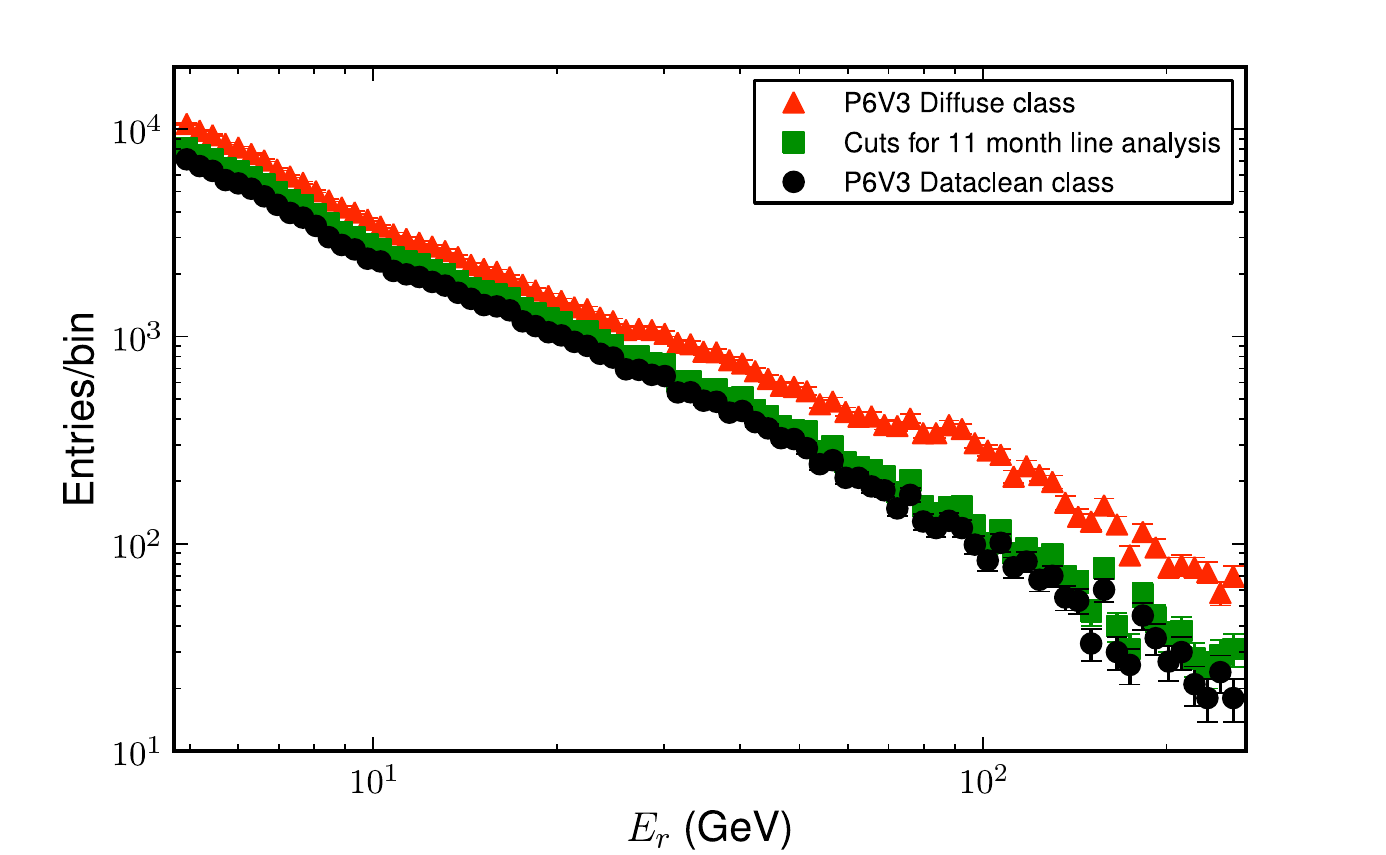}
\caption{Counts spectrum from 4.8 to 264 GeV in 5\% bins for photons in the  
Pass 6 DIFFUSE class (red triangles), the P6\_V3 11-month line analysis photon class (green square), and the Pass 6 DATACLEAN class (black circles) for the 
ROI $|b|>10^\circ$ plus a $20^\circ\times20^\circ$ square centered at the GC.
The dataclean class has the tightest cuts. 
Additional cuts on the inclination, zenith, LAT rocking angles, and point 
sources from the LAT 1-year catalog are included.}
\label{fig:fig_class_counts_spec}
\end{center}
\end{figure*}

In this section, we describe the selection criteria for photons included in our dataset.  
The dataset is comprised of Pass 6 DATACLEAN \cite{Ackermann2010,Edmonds2011} photons from 
the energy range 4.8 to 264 GeV and the 
ROI $|b|>10^\circ$ plus a $20^\circ\times20^\circ$ square centered at the GC.
We exclude the bulk of the Galactic plane, where the diffuse emission from interactions of cosmic rays with interstellar gas and the interstellar radiation field is strong, but include the GC where cuspy profiles should enhance the WIMP annihilation signal.  
Photons are removed that are near point sources (see below), 
that arrive when the rocking angle of the LAT is larger than $52^{\circ}$, 
or have zenith angles greater than $105^{\circ}$.  The selection cuts remove charged particles, atmospheric gamma rays from the Earth's limb (called ``albedo photons'' in this paper), and known astrophysical sources.  Our final sample consists of 
$\sim 105,000$ photons, with $\sim 10\%$ coming from the $20^\circ\times20^\circ$ square centered at the GC.  
Fig.~\ref{counts_map} shows the counts map for our ROI. 
Fig.~\ref{fig:fig_class_counts_spec} shows the counts spectrum for P6\_V3\_DIFFUSE class photons (red triangles), 
the P6V3 photon class used in the 11-month line analysis (green squares), and for Pass 6 DATACLEAN class 
photons (black circles) from 4.8 to 264 GeV, in 5\% energy bins.  
The Pass 6 DATACLEAN class has the smallest particle contamination 
and the same energy and direction reconstruction quality as the P6\_V3\_DIFFUSE class.  

In our ROI, there are 1087 sources in the LAT 11 month catalog (1FGL) \citep{Fermi1fgl}.  
We remove photons that are within an energy-dependent cut radius of each source.  
Based on Monte Carlo (MC) simulations we developed the following approximation for the energy 
dependence of the $68\%$ containment angle averaged over off-axis angles from $0^\circ$ to $66^\circ$~\cite{Atwood2009}:
\begin{equation}
\langle\theta_{68}(E)\rangle = \sqrt{\left[(0.8^\circ) \times (E_{GeV})^{-0.8}\right]^2 + \left[0.07^\circ\right]^2}\,.  
\end{equation}
However, the flight PSF for photons above 5 GeV may be larger than the P6\_V3 PSF derived from MC simulations.  
Above $\sim 30$ GeV, the MC PSF width may be underestimated by a factor of about 2 \citep{FSSC_P6V3_Caveats}, 
but most sources in any case have a negligible flux at high energy.  
We use the conservative cut radius $\max(0.2^\circ,2\times\langle\theta_{68}(E)\rangle)$, which 
equals $0.2^\circ$ at about 20 GeV.  
We keep the six sources within the $1^{\circ}\times1^{\circ}$ square centered on the GC.   
Over the entire energy range, a total of 10\% of the photons are removed, and $<1\%$ of the solid angle.

\subsection{Line Instrument Response Functions}\label{line_irfs_sec}

A line search requires accurate knowledge of the LAT energy resolution.
In the case that the line signal is extremely small in comparison to the background, an accurate 
probability distribution function (PDF) for spectral line photons will increase the confidence in and 
power of a statistical line search.  Accordingly, we simulate spectral lines reconstructed by the LAT, 
and parameterize their energy dispersion to construct line PDFs.

We use GLEAM, a GEANT4 based MC with the LAT geometry and material implemented, to model particle interactions with detector matter and perform full photon reconstruction \citep{Atwood2009}.  The GLEAM version corresponds to the P6\_V3 \emph{Fermi} data release instrument response functions.  The spectral lines were simulated at 5, 7, 10, 20, 50, 100, and 200 GeV.  Photons were selected using the applicable analysis cuts.  At each energy we simulate $\sim$40,000 photons for each spectral line.  For the MC photons, as in the case of the real data, we use $E_{r}$.  We have also checked that the MC reproduces well the detector response to photons as a function of incident angle (see \cite{Edmonds2011} for details).  

\begin{figure*}[t]
\begin{center}
\includegraphics[width=0.7\textwidth]{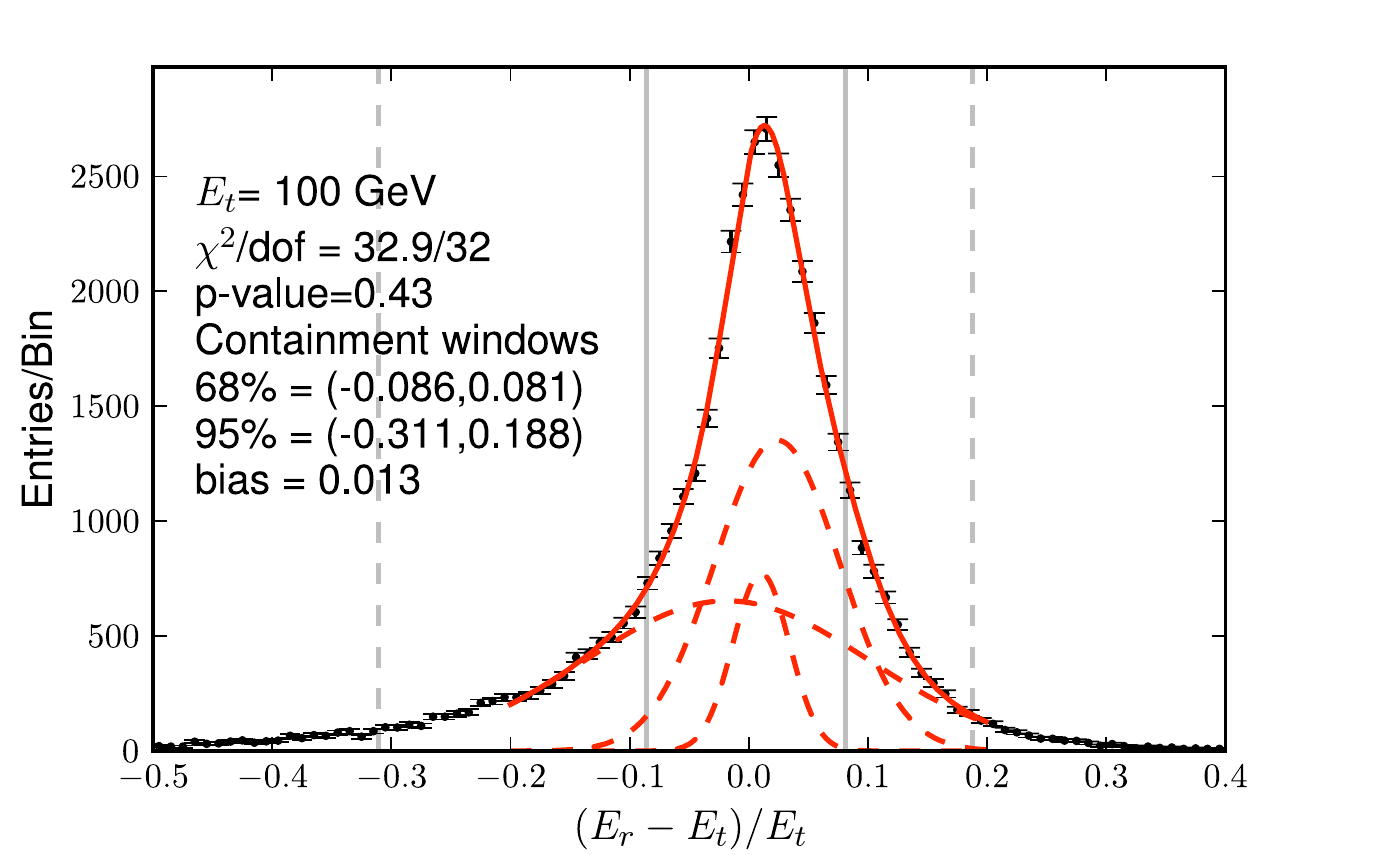}
 \caption{Energy dispersion for the 100 GeV MC spectral line.  The unbinned maximum likelihood best fit for $G(x)$ is shown with its component Gaussian functions.  The 68\% and 95\% containment windows are given by the solid and dashed vertical lines, respectively, and are based on $G(x)$.  The $\chi^2/\text{dof}$ and $p$-value for the binned data are included in the figure. 
 The fit of $G(x)$ to the MC data is cut off at $|x|>0.2$ (see text).}
 \label{mcfit_100GeVline}
\end{center}
\end{figure*}

Our choice of a function to parameterize the energy dispersion is empirically, not physically, motivated.  
We fit to a sum of three Gaussians, which provided a good fit to the line shape.  
Simpler models such as a single Gaussian or the sum of two Gaussians 
underfit the data in the peak region, degrading the line resolution.  
For each MC spectral line, we performed an unbinned maximum likelihood fit to
\begin{equation}\label{line_irf_eq}
G(x)={\sum_{k=1,2,3} a_{k}\cdot \frac{1}{\tilde{\sigma}_{k}\sqrt{2\pi}}e^{-(x-\tilde{\mu}_{k})^2/(2\tilde{\sigma}_{k}^2)}}\text{ and }a_{3}=1-a_{2}-a_{1},
\end{equation}
where $0<a_1,a_2,a_3<1$, $x=(E_{r}-E_{t})/E_{t}$, and $E_{t}$ is the true MC energy. 
The fits are made to functions of $\tilde{\mu} = (\mu-E)/E$ and $\tilde\sigma = \sigma/E$ that are normalized to the fit 
energy, as the IRFs normalized in this way only change slowly with energy.  To determine the signal PDF parameters 
as a function of energy when fitting for lines, the MC energy dispersion fit parameters are linearly interpolated, 
then transformed to reconstructed-energy space using $\sigma_{kj} = \tilde{\sigma}_{k,j}$ and 
$\mu_{k,j}=(\tilde{\mu}_{k,j}+1)E_j$~\cite{Edmonds2011}.
We average over the angular acceptance to produce a dispersion parameterization dependent on energy only.
Fig.~\ref{mcfit_100GeVline} shows a typical fit (100 GeV) to $G(x)$ for $|x| < 0.2$ to MC data, and the component Gaussian 
functions. 
$G(x)$ does not fit the Monte Carlo data well outside the range $|x| < 0.2$, owing to the non-Gaussian tails of the dispersion function.
Thus, we estimate a systematic error to the line limit analysis from the contribution of the $|x|>0.2$ tails of the energy dispersion distribution of 5\% for $E\le 130$ GeV and 20\% 
for $E > 130$ GeV (where the tails become much larger due to shower leakage). 

\begin{figure*}[t]
\begin{center}
    \includegraphics[width=0.7\textwidth]{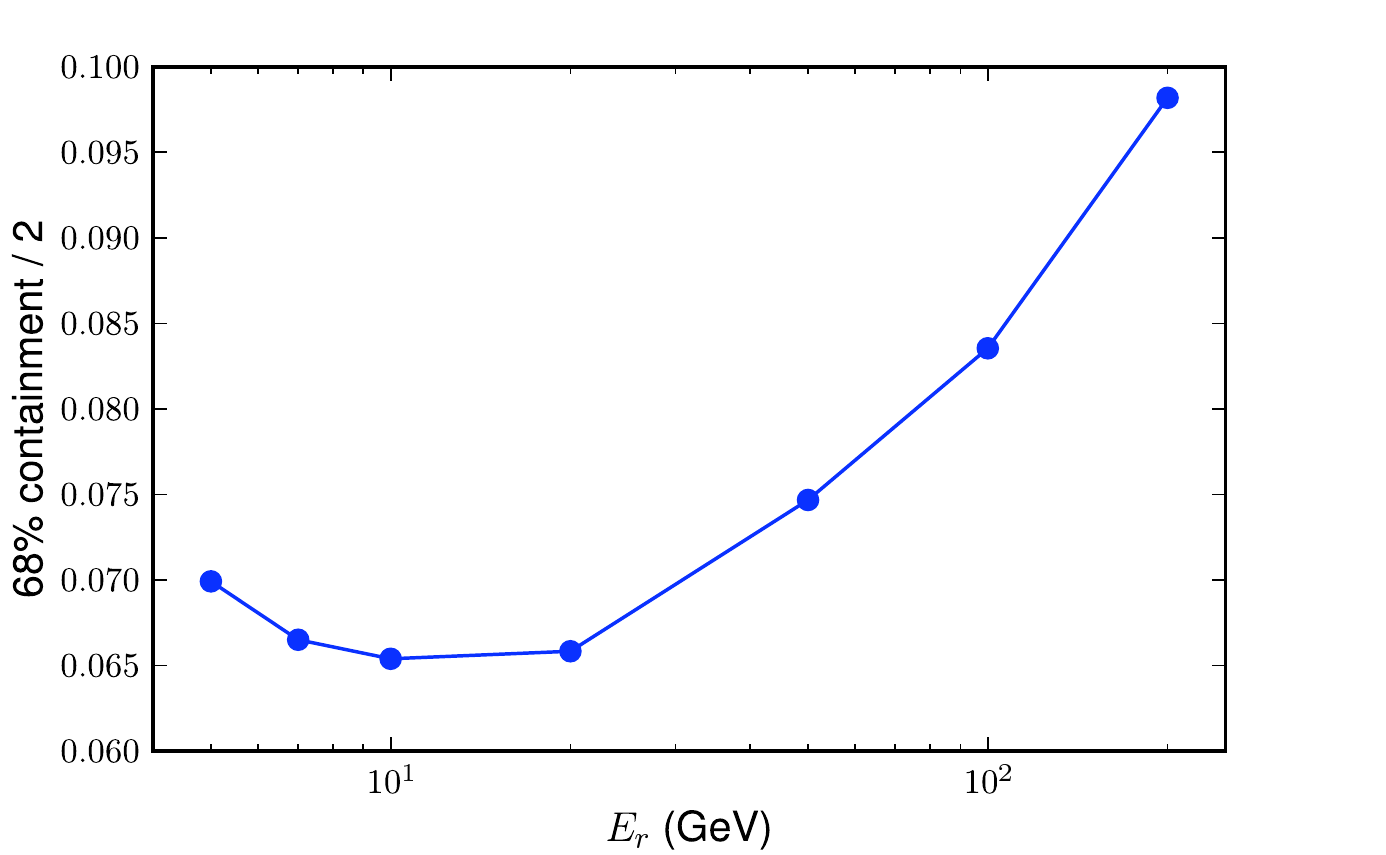}
  \caption{MC-derived energy resolution (68\% energy dispersion fractional containment divided by 2), integrated over the angular acceptance.}\label{line_resolution}
  \end{center}
\end{figure*}

We calculate the resolution for lines using asymmetric photon containment windows  \cite{CRE2010}.  Fig.~\ref{mcfit_100GeVline} shows the full 68\% (solid gray lines) and 95\% (dashed gray lines) containment windows and the bias and window limits in the figure text.  
The bias is the value of $x$ that maximizes $G(x)$.  
The positive (negative) 68\% containment window is the smallest window, beginning at $x=\text{bias}$, containing 68\% of the $x>\text{bias}$ ($x\leq\text{bias}$) photons.  The 95\% containment windows are calculated similarly.  In all cases, the absolute value of the bias is $\leq2.3\%$.  Fig.~\ref{line_resolution} shows the energy resolution (68\% energy dispersion containment divided by 2), integrated over the angular acceptance, for MC spectral lines as a function of line energy.  We estimate a systematic error of less than 10\% on the energy resolution.

\subsection{Exposure}\label{exposure_section}

\begin{figure*}[t]
\begin{center}
\includegraphics[width=0.86\textwidth]{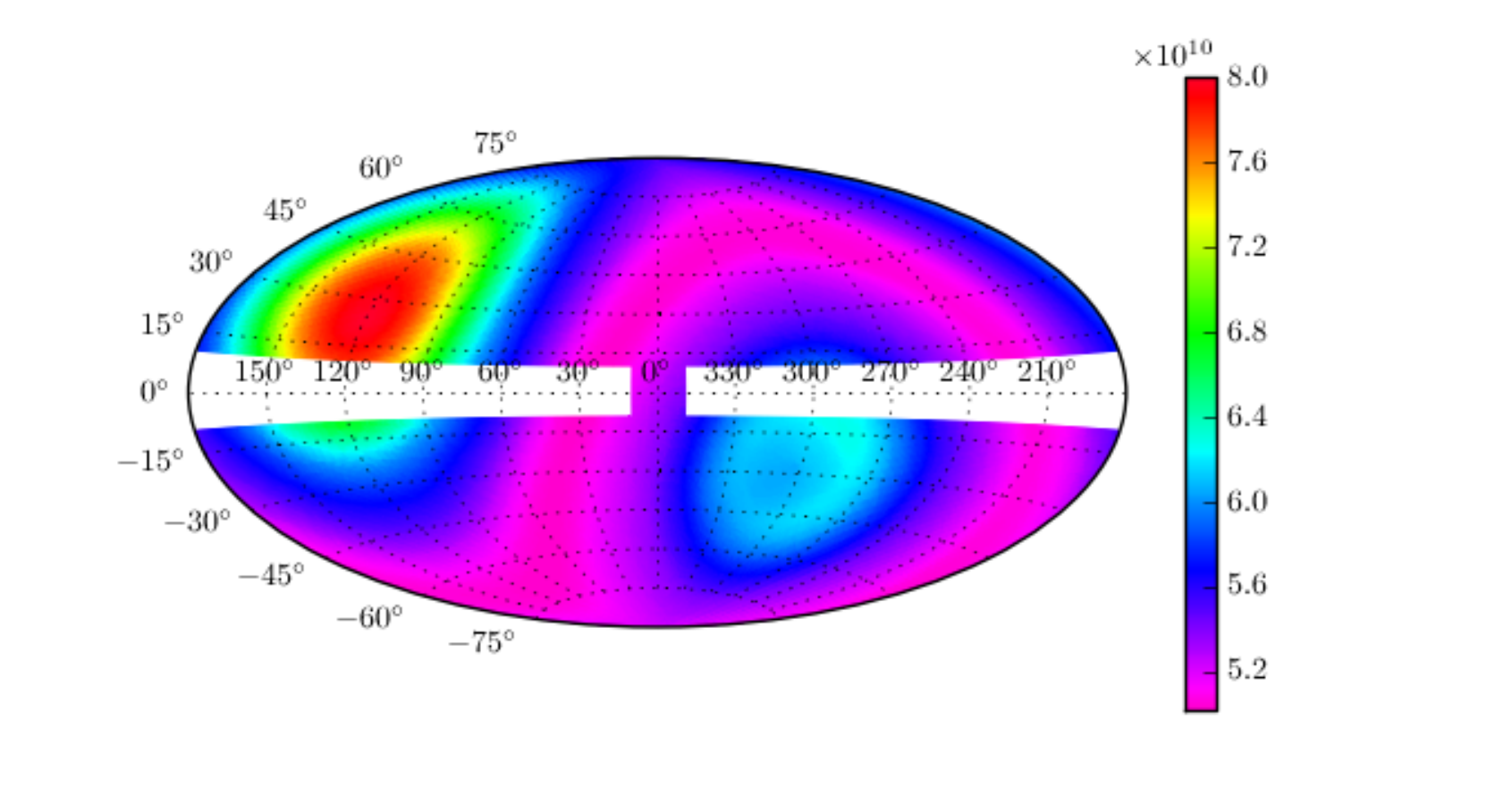}
\caption{Exposure map for 100 GeV in cm$^2$s in Hammer-Aitoff projection, cut for the ROI 
$|b|>10^\circ$ plus a $20^\circ\times20^\circ$ square centered at the GC.
The map average and square root of the variance is $(5.60\pm0.61)\times 10^{10}$ cm$^2$s.
The exposure is calculated over the line dataset energy range at 5\% intervals in log-space for use in 
the flux construction. The average exposure varies by only $\sim 25\%$ over our energy range.
}\label{fig:exposure100}
\end{center}
\end{figure*}

We use LAT Science Tools v9r18p3 to apply the class and angle selection cuts to our dataset, determine the good-time-intervals (GTIs), and calculate the exposure~\cite{Cicerone}.  The GTIs are the collection of time intervals during which the spacecraft rocking angle is less than $52^{\circ}$, the data quality is good, and the LAT is in normal science operations.  Our dataset from August 7, 2008 to June 30, 2010 covers 60 Ms of elapsed time.  
The GTIs sum to 49 Ms of observation time, 82\% of the elapsed time.
We calculate the integrated livetime $t$(RA, Dec, cos$\theta$) for the entire sky (represented as a HEALPix grid) as a 
function of inclination angle (cos$\theta$) using spacecraft pointing information and the GTIs \cite{Cicerone}. 
The exposure map for photons of energy $E$, $\varepsilon\left(E\right)$, is the product of the livetime 
and effective area $A_{\rm eff}$ 
($E$, cos$\theta$) integrated over solid angle, taking into account our maximum zenith and inclination angles. 
The effective area is from the P6V3\_DATACLEAN IRFs. 
The exposure is calculated at a specified energy in Galactic coordinates on a cartesian grid 
with $1^\circ \times 1^\circ$ grid spacing. The map is cut to select our ROI.  
Fig.~\ref{fig:exposure100} shows the exposure map for 100 GeV.  

\section{Spectrum}\label{spectrum_sec}

\begin{figure*}[t!]
\begin{center}
\subfigure{\includegraphics[width=0.49\textwidth]{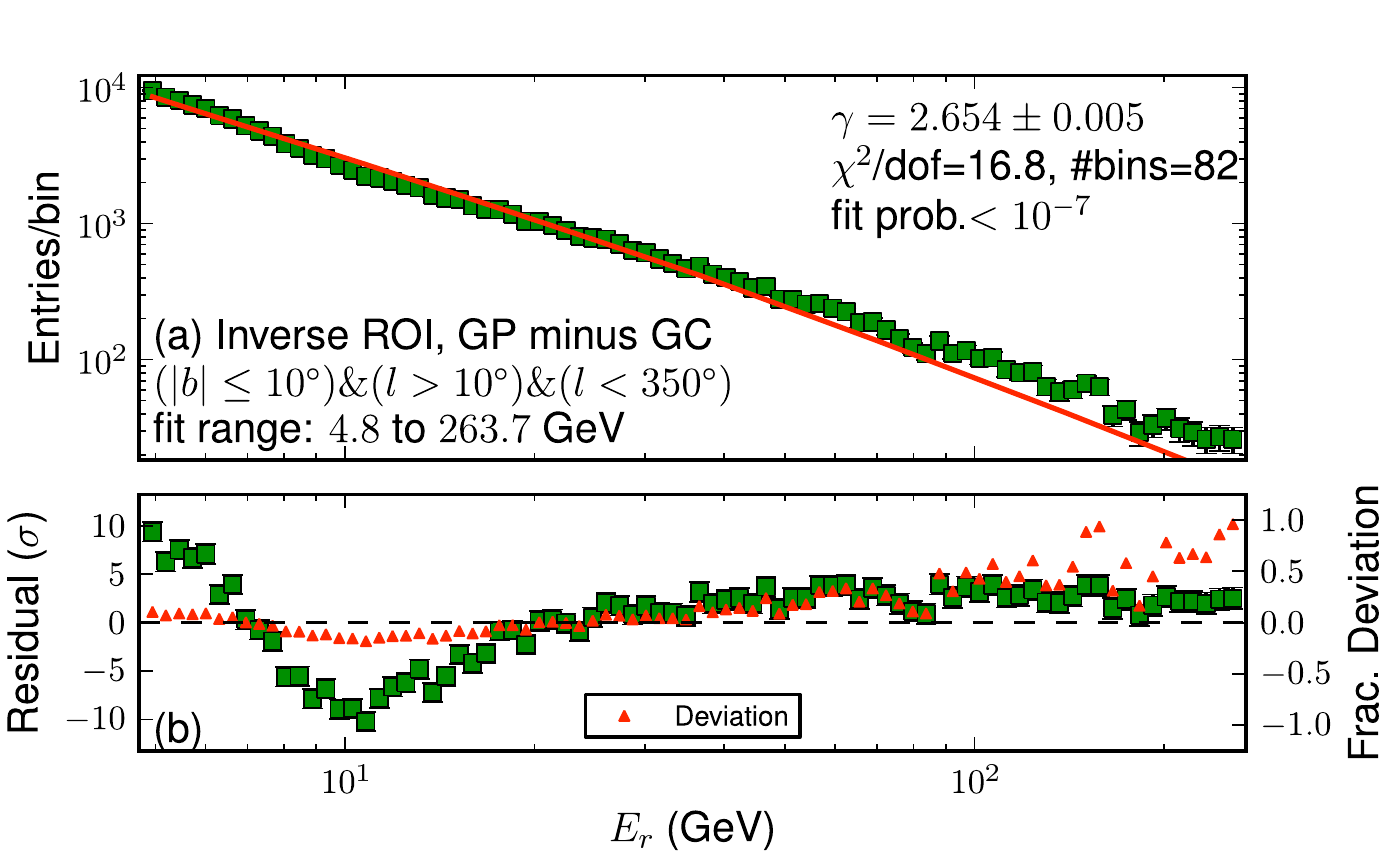}}
\hskip 2mm
\subfigure{\includegraphics[width=0.49\textwidth]{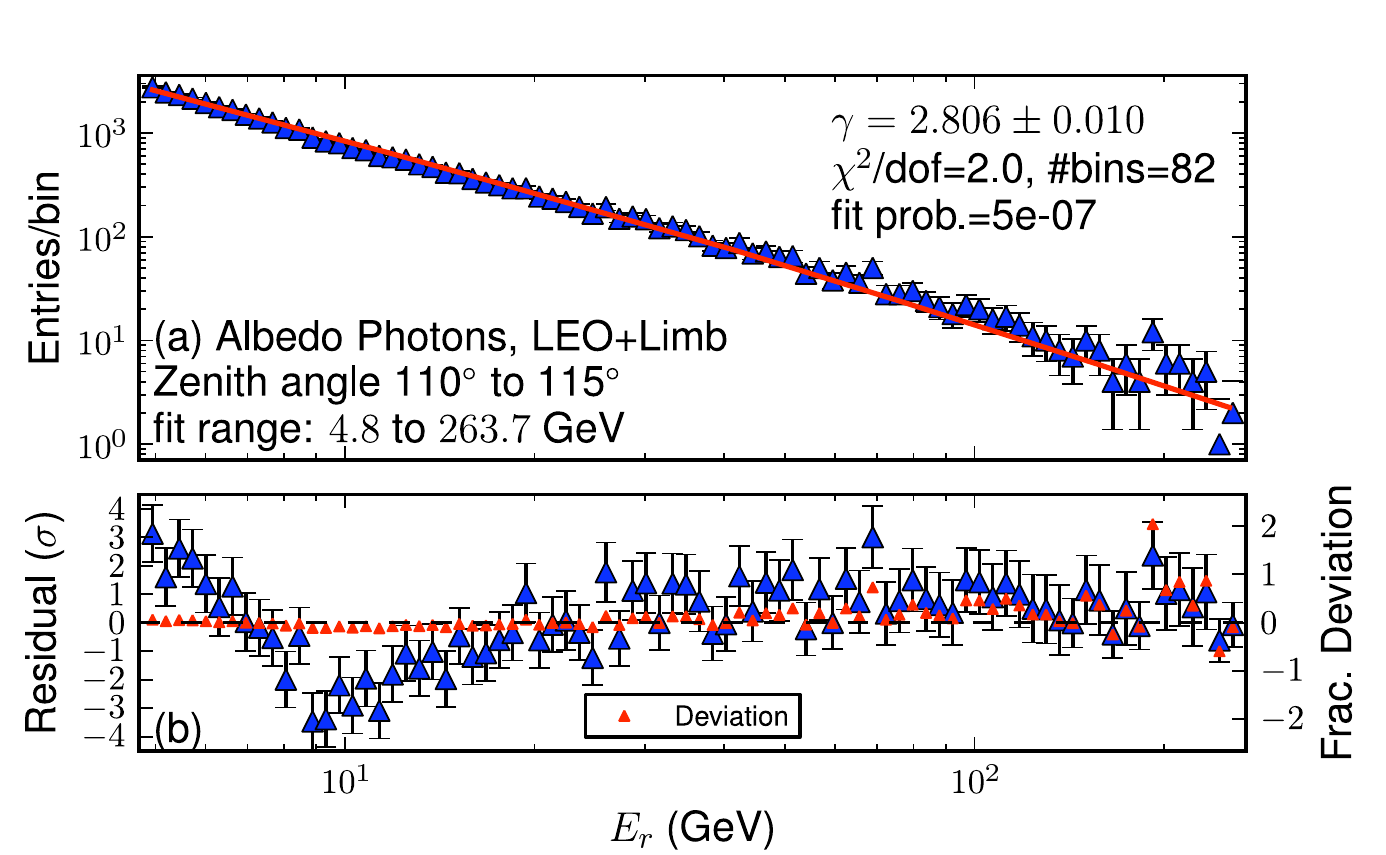}}\\
\subfigure{\includegraphics[width=0.49\textwidth]{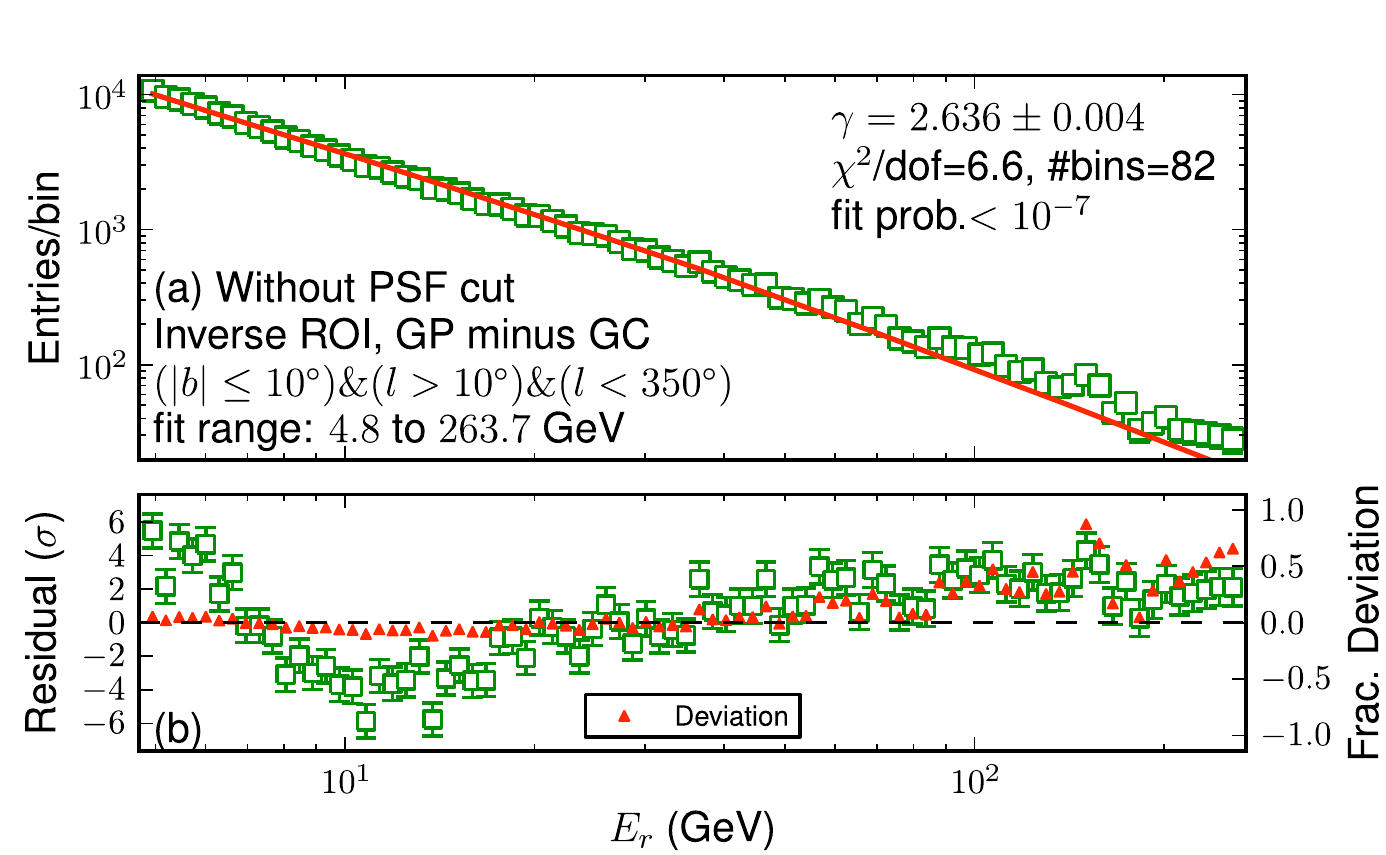}}
\hskip 2mm
\subfigure{\includegraphics[width=0.49\textwidth]{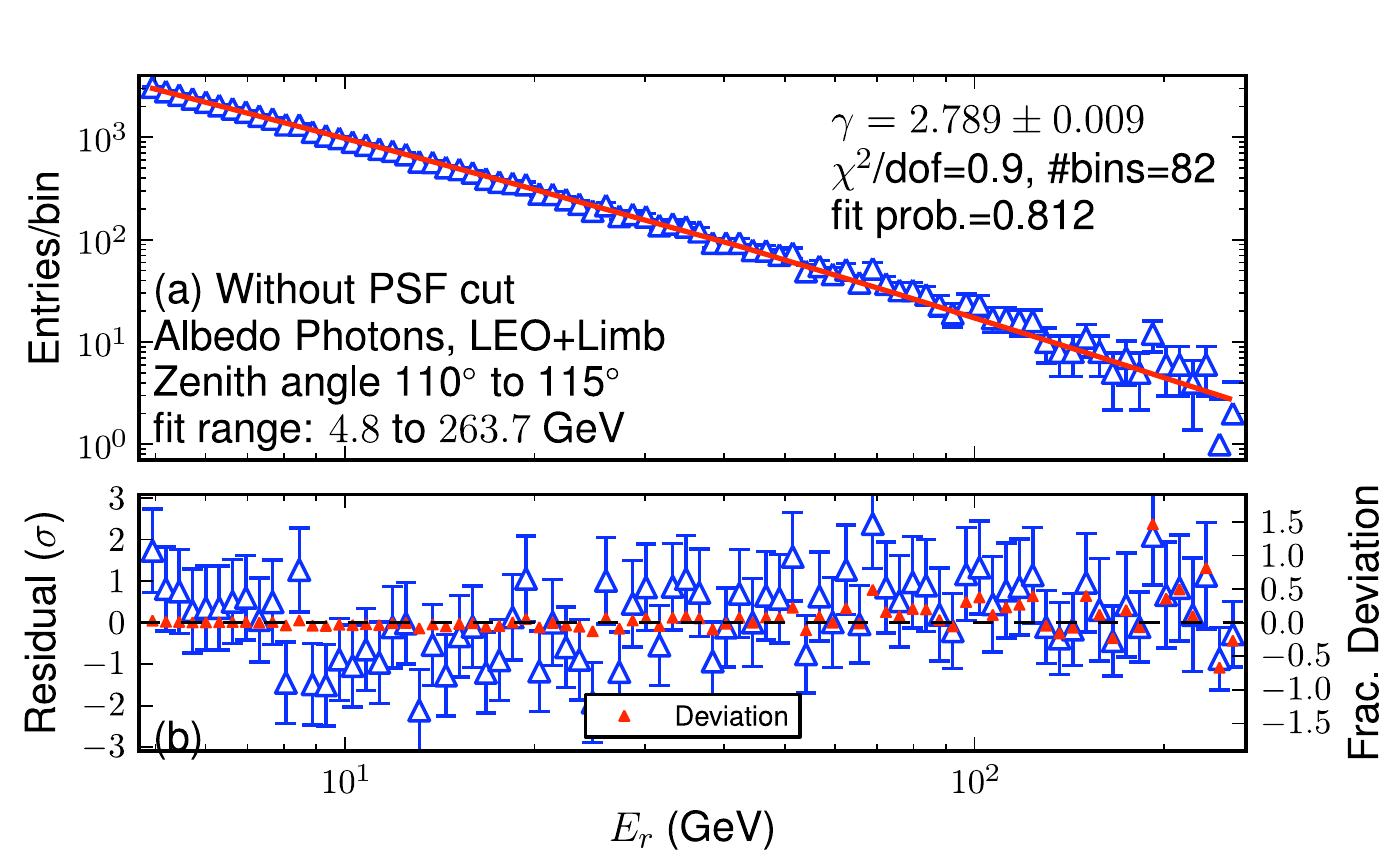}}
\end{center}
\caption{
The photon counts spectrum for the inverse ROI ($|b|<10^\circ$ excluding the 
$20^\circ\times20^\circ$ square at the GC) (solid squares for left top, open squares for left bottom) 
and for an albedo photon dataset (solid  triangles for right top, open triangles for right bottom).  
The top two plots show photons from the ``Pass 6 DATACLEAN'' class (see \S\ref{analysis}), while the bottom 
two plots in addition include photons that fail to pass the ``PSF cut'', which was 
designed to improve the LAT PSF.   
In each subfigure, the top plot (a) shows (with a solid line) a fit of the data to a power-law function 
$F\left(E,\gamma\right) = E^{-\gamma}$ times the appropriate exposure $\varepsilon\left(E\right)$ 
(see text); the bottom plot (b) shows the normalized fit residuals and fractional deviation 
(small triangles) of the data from the fit curve. 
The top panel shows a prominent bump and dip below about $\sim 13$ GeV, which is 
mitigated by removing the 
PSF cut in the bottom panel, indicating that these features are systematic effects.  
Note that the vertical scale of the residuals is different for the different panels.  
The errors bars on the entries/bin are $\sqrt{\rm{counts}}$, while the errors bars on 
the residuals are $\pm 1$ sigma to guide the eye. 
}
\label{fig:albedo_inverseROI}
\end{figure*}

In this section, we present the measured inclusive photon intensity $(\Phi)$ from 4.8 to 264 GeV for the line 
dataset.  In this work, intensity is related to flux by the solid angle of the ROI of Fig.~\ref{fig:albedo_inverseROI}, 
$\Delta\Omega = 10.5$ sr. 
We find that a power law describes the spectrum over a large portion of the energy range, 12.8 to 264 GeV.  Below 12 GeV, there are large deviations from power-law behavior due to a broad structure that also creates structures on the scale of the instrument resolution, a bump at $\sim$5 GeV and a dip at $\sim$10 GeV.  
The appearance of similar structures in two controls, a dataset of photons from the Galactic Plane 
excluding the Galactic Center (referred to as the ``inverse ROI''), and a dataset of albedo photons, 
which should show no sign of dark matter, indicates that these spectral structures are 
likely 
systematic effects from imperfect representation of the effective area in the Pass 6 DATACLEAN IRFs.  
Further study of one of the Pass 6 DATACLEAN selection criteria which is designed to improve the LAT PSF, ``PSF cut'' 
\footnote{Using Monte Carlo simulations, we estimate the probability that the reconstructed photon direction falls within the nominal 68\% containment angle of the PSF. The ``PSF cut'' requires this probability to be greater than 0.1. 
This cut significantly reduces the tails of the PSF, and thus the large angle leakage by point sources.  
The exact reduction depends on the energy, but the decrease in the 95\% containment of the PSF at high energies 
is about a factor of 2.  For the line analysis, we find that this cut introduces systematic errors in the acceptance of the 
instrument, leading to structured residuals that may be interpreted as a line signal at about 6.5~GeV.  
}, gives us confidence that these structures are a 
systematic \cite{FSSC_P6V3_Caveats,Edmonds2011}.  

\begin{figure*}[t]
\begin{center}
\subfigure{\includegraphics[width=0.49\textwidth]{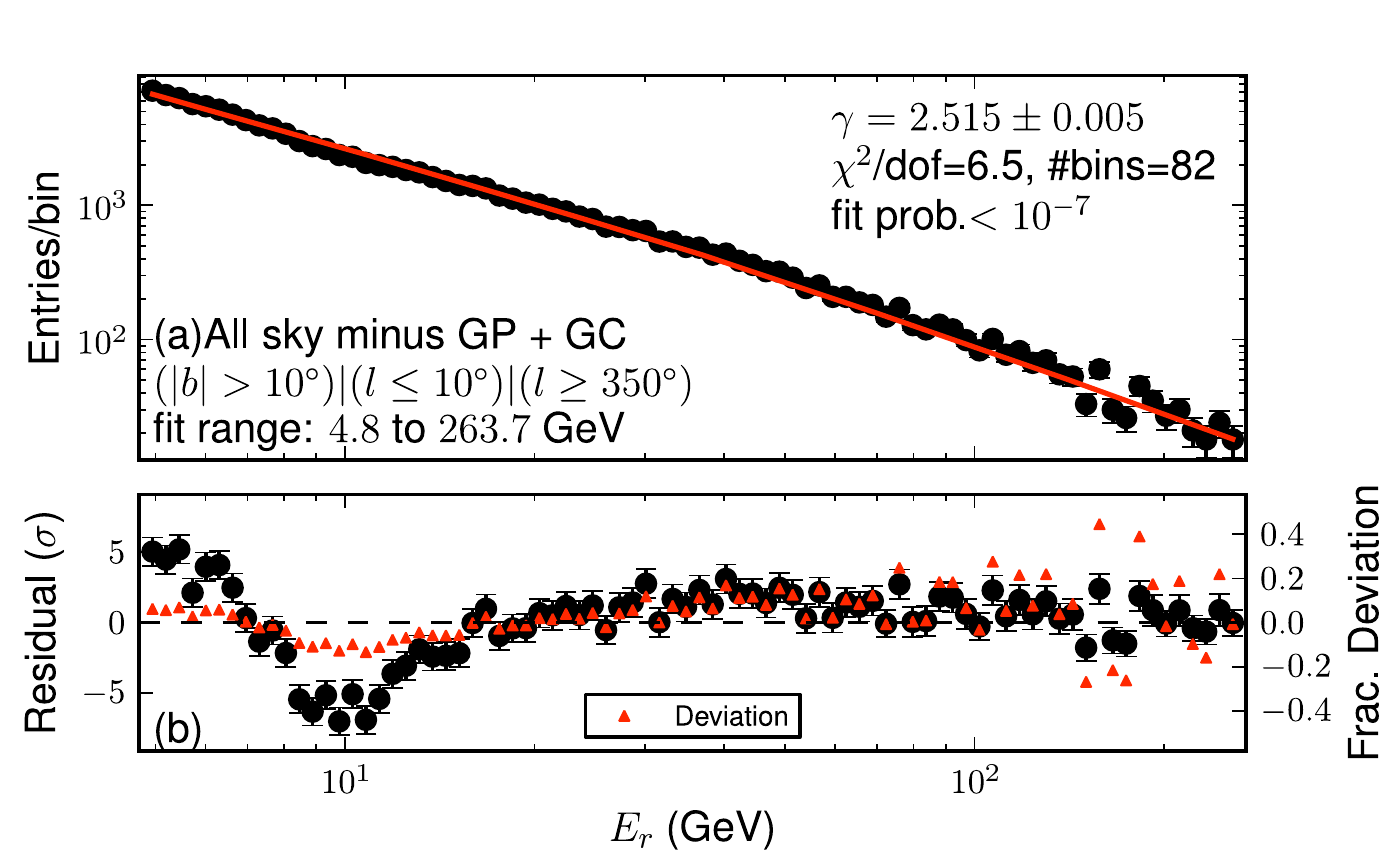}}
\hskip 2mm
\subfigure{\includegraphics[width=0.49\textwidth]{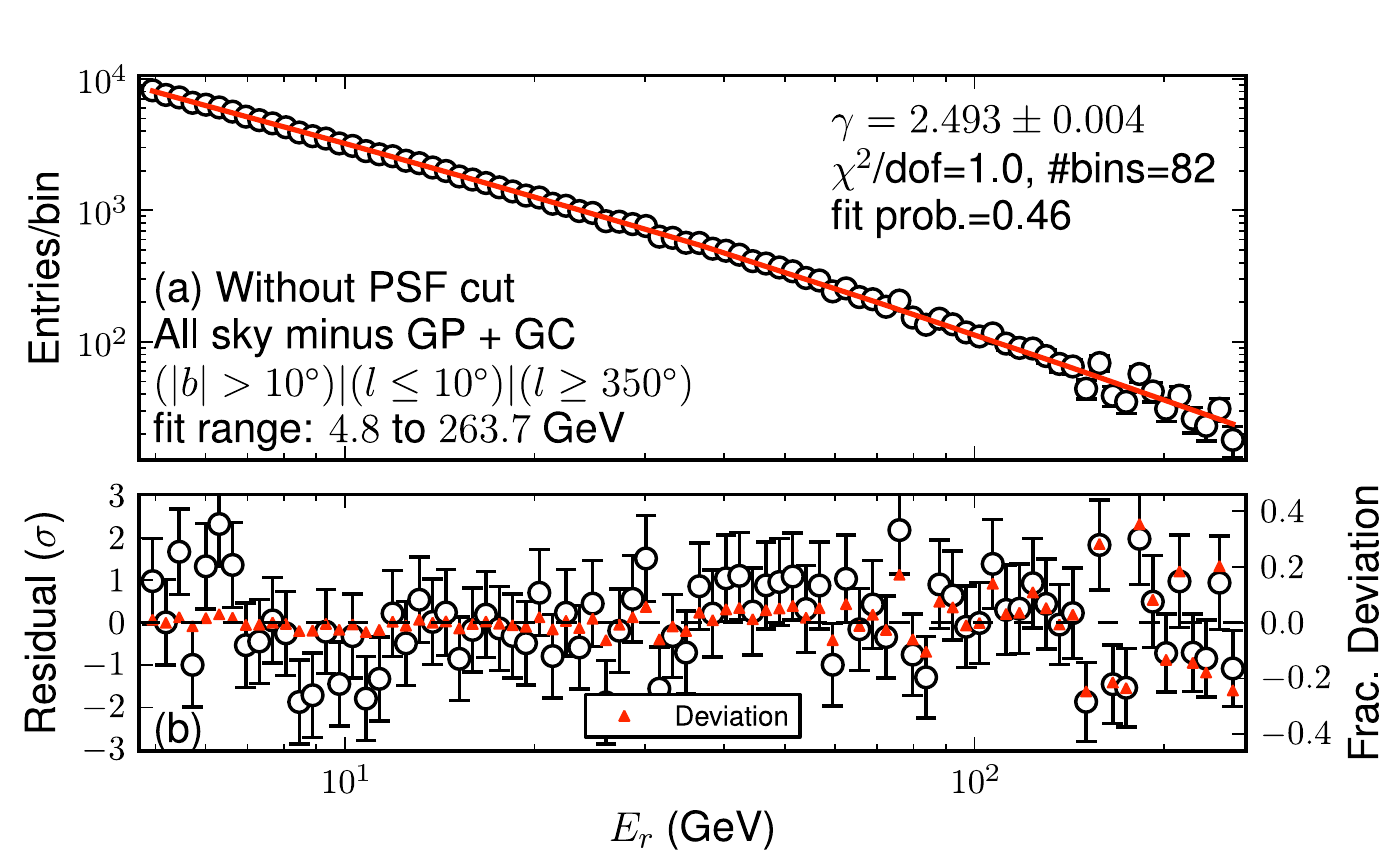}}
\end{center}
\caption{The inclusive photon counts spectrum (circles) of the expected signal region.  
The left spectrum includes photons from the Pass 6 DATACLEAN sample, while the right spectra also includes those photons which failed to pass the ``PSF cut'', which induces systematic 
effects below $\sim 13$ GeV, as shown in Fig.~\ref{fig:albedo_inverseROI} with control datasets.
A power-law fit (top, line), normalized fit residuals (bottom, circles), and 
the fractional deviation of the data from the power-law fit (bottom, small triangles) are shown 
as in Fig.~\ref{fig:albedo_inverseROI}.  
We search for dark matter induced gamma-ray lines above $\sim 13$ GeV using the left dataset, 
and below $\sim 13$ GeV using the right dataset.  For dark matter contributions to the 
inclusive spectrum, we use the left spectrum (for which the systematics 
from the PSF cut are not important) -- see \S\ref{sec:results line search}. 
}\label{fig_range_spec_0to82}
\end{figure*}

In Fig.~\ref{fig:albedo_inverseROI}, we show the control data sets, both with (top) and without (bottom) 
the PSF cut.  The removal of the PSF cut clearly mitigates the systematic effects.  
Understanding the systematics that can lead to line-like structures (bump and dips) is 
of course particularly important for the line search, and will be discussed in \S\ref{sec:linesearch}.
These systematics have been mitigated in Pass 7 LAT data releases \cite{Pass7}.  
Fig.~\ref{fig_range_spec_0to82} left (right) shows the photon counts spectrum from 4.8 to 264 GeV for the line 
dataset with the PSF cut included (removed).  

\subsection{Spectral Fitting}\label{sec:spectral_fitting}

Although the unbinned line dataset (photon list) is used for fits, a binned dataset is used to define the goodness-of-fit, calculate residuals, and show the intensity.  We use 5\% energy bins for the binned dataset, with boundaries given by $E_k=1.05^k$ GeV, $k=0,1, \ldots, 82$.  The line search uses photons from 4.8 to 251 GeV.  Photons with energy up to 264 GeV are included in the counts spectrum and inclusive flux plots to avoid a partially-filled last bin.

We assume the photon intensity spectrum integrated over the entire selection region is described by a power-law spectrum with spectral index $\gamma$,
\begin{equation}
F\left(E,\gamma\right)=E^{-\gamma}\left[\text{photons cm}^{-2}\text{s}^{-1} {\rm sr}^{-1}\right].
\end{equation}
The probability density function (PDF) fit to the counts spectrum is the product of the intensity spectrum, $F\left(E,\gamma\right)$, and the exposure, $\varepsilon\left(E\right)$, to take into account the LAT acceptance.   
The exposure is calculated over the line dataset energy range at 5\% intervals in log-space 
and interpolated to obtain $\varepsilon\left(E_i\right)$. 
The likelihood function is given by
\begin{equation}
\mathscr{L}\left(\gamma\right)={\displaystyle\prod_{i=1}^{N} F(E_i,\gamma) \times \varepsilon(E_i)},
\end{equation}
where $N$ is the number of photons.  We use the ROOT RooFit module \citep{Verkerke2003} 
to perform the unbinned likelihood analysis. 

Figs~\ref{fig:albedo_inverseROI} and \ref{fig_range_spec_0to82} 
show the counts spectrum and the fit to $F\left(E,\gamma\right)\times\varepsilon\left(E\right)$ for the full 
energy range 4.8 to 264 GeV.  
For each energy bin, 
the residuals and fractional deviation are given by $(n_{\text{DATA}}-n_{\text{FIT}})/\sqrt{n}_{\text{DATA}}$, and 
$(n_{\text{DATA}}-n_{\text{FIT}})/n_{\text{FIT}}$, respectively, where $n_{\text{DATA}}$ ($n_{\text{FIT}}$) is 
the number of photons in the line dataset (calculated from the fit).  

The residuals show significant structure below $\sim$13 GeV, a broad peak at $\sim$5 GeV, and a broad dip at 
$\sim$10 GeV, which is mitigated by removing the PSF cut, as discussed above.  
Above $\sim$13 GeV, the residuals show no obvious structure, and  
thus the line search should not be significantly impacted by systematic effects above 12.8 GeV 
(a bin boundary).

\subsection{Flux Error Contributions to Inclusive Spectrum}\label{subsec:errors}

\begin{figure*}[t]
\begin{center}
\includegraphics[width=0.7\textwidth]{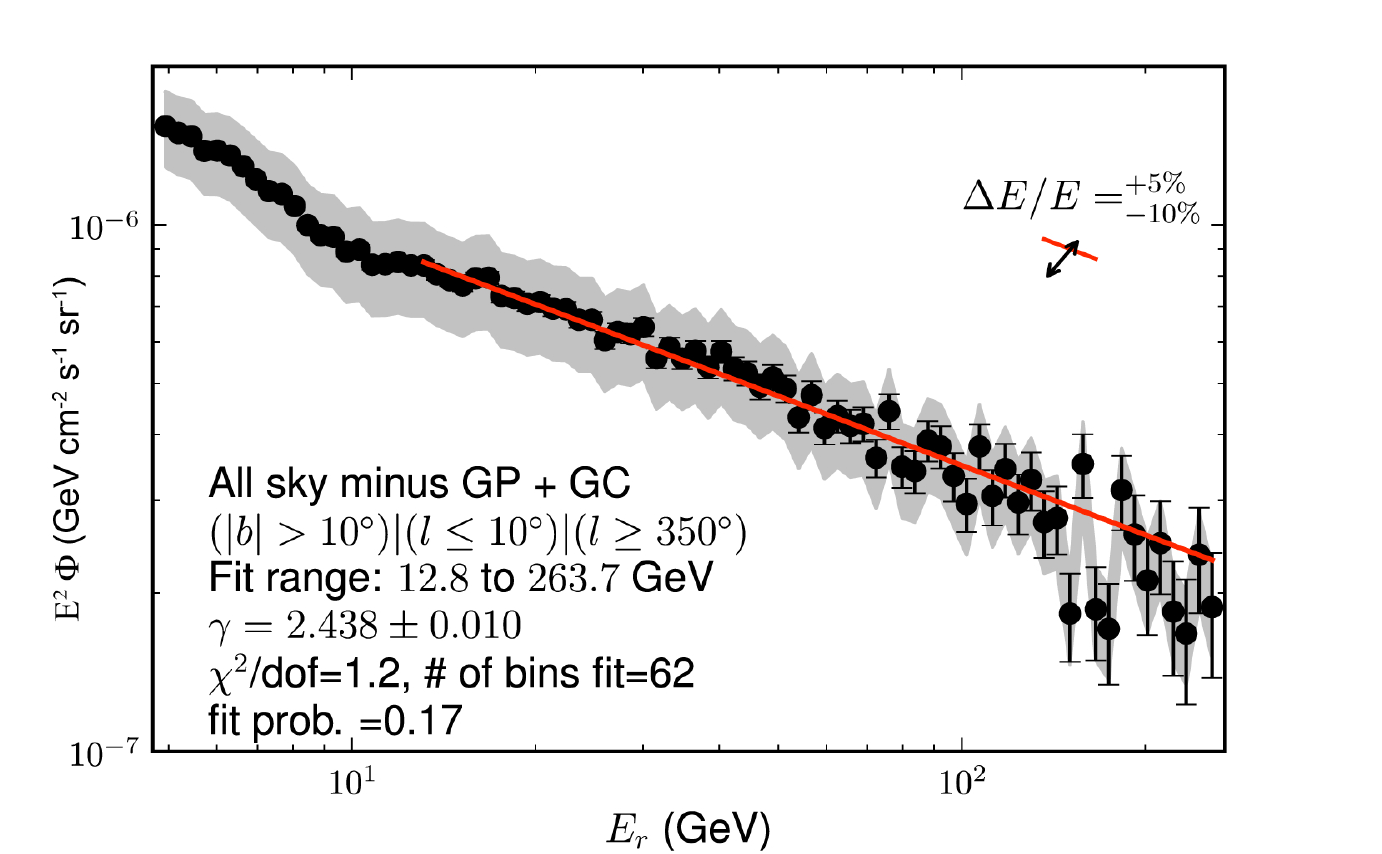}
\end{center}
\caption{$E^2\times\Phi$ with power-law fit.  The gray band shows the systematic error due to uncertainty in the effective area.  The arrow shows the shift due to uncertainty in the absolute energy scale.  
The intensities and energies shown here are tabulated in Table \ref{table_spectrum}. }
\label{esqur_dataclean_flux}
\end{figure*}

We define $E^2$ times the measured inclusive intensity, $\Phi$, by
\begin{equation}
(E^2 \times \Phi)_k = \left(\frac{E_k + E_{k+1}}{2}\right)^2 \times 
\frac{n_k}{(E_{k+1}-E_k)\, \bar{\epsilon}_k \, \Delta\Omega} \,\, [{\rm GeV}^{-1}\,{\rm cm}^{-2} \,{\rm s}^{-1}\, {\rm sr}^{-1}]\,,
\end{equation}
where for each energy bin $k$, $n_k$ is the number of photons and 
$\bar{\epsilon}_k$ is the exposure.  In Fig.~\ref{esqur_dataclean_flux}, we show this quantity 
overlaid with the best-fit power law ($\gamma=2.44$) from the energy range 12.8 to 264 GeV for the line dataset.  We use the dataset with the full analysis and PSF cuts for the flux to be consistent with the LAT IRFs which are defined for the standard Pass 6 DATACLEAN photon selection.  
Table \ref{table_spectrum} lists the energy range, number of photons, and 
$\Phi$ followed by its uncertainties.

\begin{turnpage}
\begingroup
\squeezetable
\begin{table*}
\begin{center}
\begin{tabular}{crr|crr|crr|crr}
Energy & \multicolumn{1}{c}{n} & \multicolumn{1}{c}{$\Phi$} &  Energy & \multicolumn{1}{c}{n} & \multicolumn{1}{c}{$\Phi$} &  Energy & \multicolumn{1}{c}{n} & \multicolumn{1}{c}{$\Phi$} &  Energy & \multicolumn{1}{c}{n} & \multicolumn{1}{c}{$\Phi$}\\
(GeV) & & \multicolumn{1}{c}{(GeV$^{-1}$s$^{-1}$cm$^{-2}$sr$^{-1}$)} & (GeV) & &  \multicolumn{1}{c}{(GeV$^{-1}$s$^{-1}$cm$^{-2}$sr$^{-1}$)} & (GeV) & & \multicolumn{1}{c}{(GeV$^{-1}$s$^{-1}$cm$^{-2}$sr$^{-1}$)} & (GeV) & & \multicolumn{1}{c}{(GeV$^{-1}$s$^{-1}$cm$^{-2}$sr$^{-1}$)}\\
\hline
\hline
4.8-5.1 & 7156 & $629\pm  7.4\pm102.8^{+90}_{-45}\cdot10^{-10}$ & 13.4-14.1 & 1625 & $424\pm 10.5\pm 84.9^{+61}_{-31}\cdot10^{-11}$ & 35.7-37.5 & 484 & $431\pm 19.6\pm 86.2^{+62}_{-31}\cdot10^{-12}$ & 
99.4-104.3 & 83 & $285\pm 31.2\pm 56.9^{+41}_{-20}\cdot10^{-13}$ \\ 

5.1-5.3 & 6638 & $554\pm  6.8\pm 91.8^{+80}_{-40}\cdot10^{-10}$ & 14.1-14.8 & 1519 & $375\pm  9.6\pm 75.0^{+54}_{-27}\cdot10^{-11}$ &37.5-39.3 & 429 & $364\pm 17.6\pm 72.8^{+52}_{-26}\cdot10^{-12}$ & 
104-110 & 101 & $332\pm 33.0\pm 66.4^{+48}_{-24}\cdot10^{-13}$\\ 

5.3-5.6 & 6295 & $495\pm  6.2\pm 83.4^{+71}_{-36}\cdot10^{-10}$ & 14.8-15.6 & 1418 & $333\pm  8.8\pm 66.5^{+48}_{-24}\cdot10^{-11}$ &  
39.3-41.3 & 437 & $354\pm 16.9\pm 70.7^{+51}_{-25}\cdot10^{-12}$ & 
110-115 & 77 & $242\pm 27.6\pm 48.5^{+35}_{-17}\cdot10^{-13}$\\

5.6-5.9 & 5684 & $421\pm  5.6\pm 72.0^{+61}_{-30}\cdot10^{-10}$ & 15.6-16.3 & 1396 & $311\pm  8.3\pm 62.2^{+45}_{-22}\cdot10^{-11}$ &  41.3-43.4 & 386 & $298\pm 15.2\pm 59.6^{+43}_{-21}\cdot10^{-12}$  & 115-121 & 82 & $247\pm 27.3\pm 49.5^{+36}_{-18}\cdot10^{-13}$ \\
5.9-6.2 & 5477 & $383\pm  5.2\pm 66.4^{+55}_{-28}\cdot10^{-10}$ & 16.3-17.2 & 1337 & $283\pm  7.7\pm 56.6^{+41}_{-20}\cdot10^{-11}$ & 43.4-45.5 & 360 & $265\pm 14.0\pm 52.9^{+38}_{-19}\cdot10^{-12}$ &
 121-127 & 67 & $194\pm 23.7\pm 38.8^{+28}_{-14}\cdot10^{-13}$\\
6.2-6.5 & 5162 & $340\pm  4.7\pm 59.8^{+49}_{-24}\cdot10^{-10}$ & 17.2-18.0 & 1183 & $237\pm  6.9\pm 47.4^{+34}_{-17}\cdot10^{-11}$ & 45.5-47.8 & 323 & $226\pm 12.6\pm 45.3^{+33}_{-16}\cdot10^{-12}$\\
6.5-6.8 & 4744 & $294\pm  4.3\pm 52.5^{+42}_{-21}\cdot10^{-10}$ & 
18.0-18.9 & 1122 & $213\pm  6.4\pm 42.6^{+31}_{-15}\cdot10^{-11}$  & 
47.8-50.2 & 320 & $214\pm 12.0\pm 42.8^{+31}_{-15}\cdot10^{-12}$ &
127-133 & 70 & $194\pm 23.2\pm 38.8^{+28}_{-14}\cdot10^{-13}$ \\

6.8-7.1 & 4317 & $252\pm  3.8\pm 45.6^{+36}_{-18}\cdot10^{-10}$ & 18.9-19.9 & 1048 & $188\pm  5.8\pm 37.7^{+27}_{-14}\cdot10^{-11}$ & 50.2-52.7 & 289 & $185\pm 10.9\pm 36.9^{+27}_{-13}\cdot10^{-12}$ &
133-140 & 55 & $146\pm 19.7\pm 29.3^{+21}_{-11}\cdot10^{-13}$\\

7.1-7.5 & 3944 & $217\pm  3.5\pm 39.9^{+31}_{-16}\cdot10^{-10}$ & 
19.9-20.9 & 1012 & $172\pm  5.4\pm 34.4^{+25}_{-12}\cdot10^{-11}$  
& 52.7-55.3 & 242 & $148\pm  9.5\pm 29.5^{+21}_{-11}\cdot10^{-12}$ & 
140-147 & 53 & $135\pm 18.6\pm 27.0^{+19}_{-10}\cdot10^{-13}$\\

7.5-7.9 & 3744 & $194\pm  3.2\pm 36.2^{+28}_{-14}\cdot10^{-10}$ & 
20.9-21.9 & 942 & $152\pm  4.9\pm 30.3^{+22}_{-11}\cdot10^{-11}$  &
 55.3-58.1 & 253 & $148\pm  9.3\pm 29.5^{+21}_{-11}\cdot10^{-12}$ &
 147-154 & 33 & $81\pm 14.1\pm 16.1^{+12}_{-6}\cdot10^{-13}$\\
 
7.9-8.3 & 3416 & $167\pm  2.9\pm 31.6^{+24}_{-12}\cdot10^{-10}$ & 
21.9-23.0 & 901 & $137\pm  4.6\pm 27.4^{+20}_{-10}\cdot10^{-11}$ &
58.1-61.0 & 208 & $116\pm  8.0\pm 23.2^{+17}_{-8}\cdot10^{-12}$ &
154-162 & 60 & $141\pm 18.2\pm 28.2^{+20}_{-10}\cdot10^{-13}$\\

8.3-8.7 & 3012 & $140\pm  2.5\pm 26.7^{+20}_{-10}\cdot10^{-10}$ & 
23.0-24.1 & 825 & $119\pm  4.1\pm 23.8^{+17}_{-9}\cdot10^{-11}$   & 
61.0-64.1 & 208 & $111\pm  7.7\pm 22.2^{+16}_{-8}\cdot10^{-12}$ &
162-170 & 30 & $677\pm123.6\pm135.4^{+97}_{-49}\cdot10^{-14}$\\

8.7-9.1 & 2763 & $121\pm  2.3\pm 23.5^{+17}_{-9}\cdot10^{-10}$ & 
24.1-25.4 & 791 & $108\pm  3.8\pm 21.6^{+15}_{-8}\cdot10^{-11}$  & 
64.1-67.3 & 189 & $96\pm  7.0\pm 19.3^{+14}_{-7}\cdot10^{-12}$ & 
170-178 & 26 & $564\pm110.5\pm112.7^{+81}_{-41}\cdot10^{-14}$\\

9.1-9.6 & 2630 & $109\pm  2.1\pm 21.4^{+16}_{-8}\cdot10^{-10}$ & 
25.4-26.6 & 694 & $89\pm  3.4\pm 17.9^{+13}_{-6}\cdot10^{-11}$  & 
67.3-70.6 & 181 & $88\pm  6.6\pm 17.6^{+13}_{-6}\cdot10^{-12}$ &
178-187 & 45 & $94\pm 14.0\pm 18.7^{+13}_{-7}\cdot10^{-13}$\\

9.6-10.0 & 2372 & $93\pm  1.9\pm 18.5^{+13}_{-7}\cdot10^{-10}$   & 
26.6-27.9 & 690 & $84\pm  3.2\pm 16.8^{+12}_{-6}\cdot10^{-11}$  & 
70.6-74.2 & 148 & $690\pm 56.7\pm137.9^{+99}_{-50}\cdot10^{-13}$ & 
187-197 & 35 & $70\pm 11.8\pm 14.0^{+10}_{-5}\cdot10^{-13}$\\

10.0-10.5 & 2301 & $85\pm  1.8\pm 17.0^{+12}_{-6}\cdot10^{-10}$   & 
27.9-29.3 & 653 & $76\pm  3.0\pm 15.1^{+11}_{-5}\cdot10^{-11}$  &
74.2-77.9 & 172 & $77\pm  5.8\pm 15.3^{+11}_{-6}\cdot10^{-12}$ & 
197-207 & 27 & $520\pm100.0\pm103.9^{+75}_{-37}\cdot10^{-14}$\\

10.5-11.1 & 2071 & $72\pm  1.6\pm 14.4^{+10}_{-5}\cdot10^{-10}$  &
29.3-30.8 & 644 & $71\pm  2.8\pm 14.1^{+10}_{-5}\cdot10^{-11}$ &
77.9-81.8 & 128 & $546\pm 48.2\pm109.1^{+78}_{-39}\cdot10^{-13}$ & 
207-217 & 30 & $555\pm101.3\pm110.9^{+80}_{-40}\cdot10^{-14}$\\

11.1-11.6 & 1994 & $655\pm 14.7\pm131.0^{+94}_{-47}\cdot10^{-11}$ & 
30.8-32.4 & 537 & $560\pm 24.1\pm111.9^{+80}_{-40}\cdot10^{-12}$  &
81.8-85.8 & 119 & $485\pm 44.5\pm 97.0^{+70}_{-35}\cdot10^{-13}$  & 
217-228 & 21 & $373\pm 81.4\pm 74.6^{+54}_{-27}\cdot10^{-14}$\\

11.6-12.2 & 1936 & $600\pm 13.6\pm120.1^{+86}_{-43}\cdot10^{-11}$ &  
32.4-34.0 & 538 & $532\pm 22.9\pm106.4^{+76}_{-38}\cdot10^{-12}$  &
85.8-90.1 & 129 & $503\pm 44.3\pm100.6^{+72}_{-36}\cdot10^{-13}$  & 
228-239 & 18 & $307\pm 72.4\pm 61.5^{+44}_{-22}\cdot10^{-14}$\\

12.2-12.8 & 1833 & $536\pm 12.5\pm107.3^{+77}_{-39}\cdot10^{-11}$ &  
34.0-35.7 & 490 & $460\pm 20.8\pm 92.0^{+66}_{-33}\cdot10^{-12}$ &
90.1-94.6 & 119 & $445\pm 40.8\pm 88.9^{+64}_{-32}\cdot10^{-13}$ & 
239-251 & 24 & $394\pm 80.4\pm 78.8^{+57}_{-28}\cdot10^{-14}$\\

12.8-13.4 & 1760 & $487\pm 11.6\pm 97.3^{+70}_{-35}\cdot10^{-11}$ &  
- & - &  - & 
94.6-99.4 & 99 & $354\pm 35.6\pm 70.9^{+51}_{-25}\cdot10^{-13}$ &
251-264 & 18 & $284\pm 66.9\pm 56.8^{+41}_{-20}\cdot10^{-14}$\\

\end{tabular}
\caption{\scriptsize Energy range, number of photons, and intensity.  The intensity is followed by its errors due to statistical fluctuations (1st error), effective area uncertainty (2nd error), and energy measurement 
absolute scale uncertainty (3rd asymmetric error).  See Table \ref{table_cr} for an estimate of the CR contamination, which is 2--8\% -- cosmic rays have not been subtracted for the results in this table.}
\label{table_spectrum}
\end{center}
\end{table*}
\endgroup
\end{turnpage}

The absolute energy scale, $s$, has uncertainty $\Delta s/s=+5\%,-10\%$ \cite{CRE2010}.  For a spectrum described by a power law, the propagation of uncertainty in the energy scale to the measured intensity is the same as that used for the analysis of CR electrons \cite{CRE2010} in the LAT data. 
It is given by 
\begin{equation}
\Delta \Phi/\Phi = \left(\gamma-1\right)\Delta s/s,
\end{equation}
where $\Delta s/s \ll 1$. 
For $\gamma=2.44$, the shift in the flux is +7.2\%, -14.4\% (double-sided arrow in 
Fig.~\ref{esqur_dataclean_flux}).  
The shift for $\Phi$ and $E^2\times\Phi$ is the same because the energy bin boundaries do not change.

Abdo \emph{et al.} \cite{Ackermann2010} evaluate the uncertainty in the effective area for the Pass 6 DATACLEAN photon selection.
The dataclean effective area uncertainty was evaluated using cut efficiencies for data and MC observations of the Vela gamma-ray pulsar.  The systematic error in the normalization of the effective area is $\pm$5\% at 560 MeV, and increases to $\pm$20\% at 10 GeV with a linear dependence on $\log{E}$.  The error is $\pm16\%$ at 5 GeV, and remains constant at $\pm20\%$ above 10 GeV (gray band in Fig.~\ref{esqur_dataclean_flux}).  
Extending the power-law fit of Fig.~\ref{esqur_dataclean_flux} to 4.8 GeV, the fit is within the gray band systematic error.

We define the fractional CR contamination as the ratio of the CR intensity (i.e., the rate of residual cosmic rays that pass the gamma-ray selection cuts) to the gamma-ray intensity.  The LAT and CR intensities, and percent CR contamination, from 3.2 to 104.2 GeV, are given in Table \ref{table_cr} \cite{Ackermann2010}.  Above 100 GeV, the spectral indices for protons and electrons is $\sim$2.6 and $\sim$2.9, respectively \cite{Edmonds2011}.  We assume a conservative estimate of $\sim$2.6 for the total CR spectrum since the electron spectrum is significantly softer than the proton spectrum.  We estimate the total CR contamination at 200 GeV by scaling the percent CR contamination at 10 GeV (8\%) using the spectral indices of the CR (2.6) and photon (2.44) flux.  At 200 GeV, the CR contamination 
is $5\%$, and is less above this energy.  This contamination thus adds a negligible amount to the systematic error. 

\begin{table*}[t]
\begin{center}
\begin{tabular}{lccccc}
Energy (GeV) & 3.2-6.4 & 6.4-12.8 & 12.8-25.6 & 25.6-51.2 & 51.2-102.4 \\
Scale factor & $\times10^{-8}$ & $\times10^{-9}$ & $\times10^{-9}$ & $\times10^{-9}$ & $\times10^{-10}$ \\
\hline
CR intensity (cm$^{-2}$s$^{-1}$sr$^{-1}$) & $0.8\pm0.4$ & $6.3\pm3.0$ & $1.4\pm0.8$ & $0.6\pm0.4$ & $0.9\pm0.9$ \\
LAT intensity (cm$^{-2}$s$^{-1}$sr$^{-1}$) & $25.3\pm4.5$ & $81.3\pm16$ & $28.3\pm5.7$ & $10.6\pm2.1$ & $37.9\pm7.7$ \\
Percent contamination & $3\%$ & $8\%$ & $5\%$ & $6\%$ & $2\%$ \\
\end{tabular}
\caption{CR and LAT intensity for P6V3 dataclean photons for the ROI $|b|\geq10^{\circ}$ from \cite{Ackermann2010}. The percent contamination for the line dataset is given by the ratio of CR to the LAT intensity.  Above 100 GeV the CR contamination is $\lesssim 5\%$.}
\label{table_cr}
\end{center}
\end{table*}

\section{Line Search}\label{sec:linesearch}

\subsection{Statistical Analysis}
\label{stat_analysis}

\begin{figure*}[t]
\begin{center}
  \includegraphics[width=0.7\textwidth]{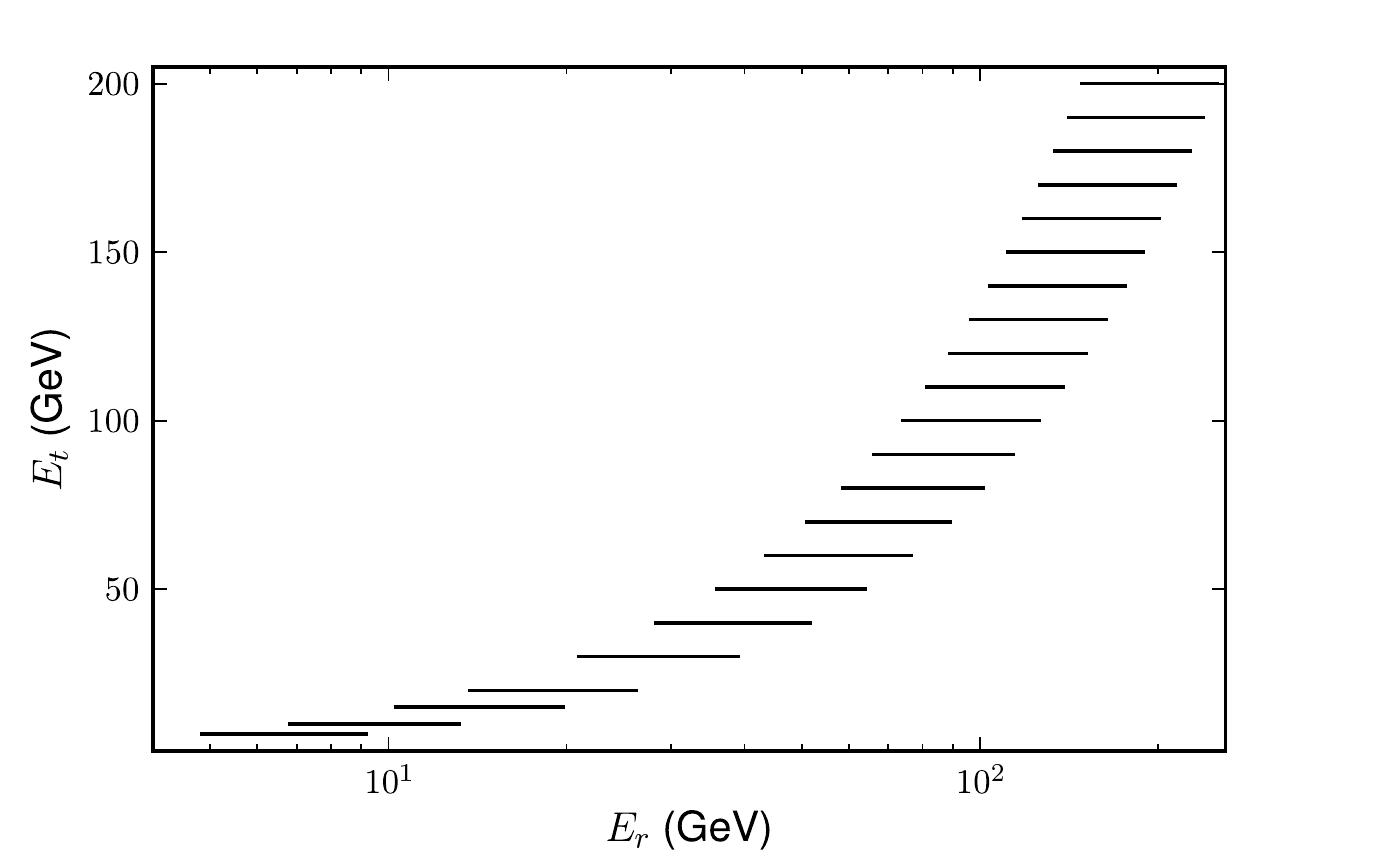}
  \caption{Spectral line search energy ranges.}  
  \label{ranges}
  \end{center}
\end{figure*}

We use an unbinned maximum likelihood signal-plus-background fit to search the counts spectrum for spectral lines with energy from 7 to 200 GeV (with trials at 7, 10, 15, and 20 to 200 GeV at 10 GeV increments).  The search energy range for a spectral line of true energy $E_j$ depends on the instrument resolution at $E_j$, where $j$ is the trial number.  The line resolution and normalized signal PDF, $S_j(E)$, are determined using the parameterization of the MC energy dispersion described in \S \ref{line_irfs_sec}; see \cite{Edmonds2011} for details.  
To determine the signal PDF parameters as a function of energy, the MC energy dispersion fit parameters are linearly interpolated, then transformed to reconstructed-energy space using $\sigma_{k,j}=\tilde{\sigma}_{k,j}E_j$ and $\mu_{k,j}=\left(\tilde{\mu}_{k,j}+1\right)E_j$  \cite{Edmonds2011}.
The search energy range for a line at $E_j$ is 
$E_{j} (1 \pm 4 \overline{\sigma}_{j} )$, where $\overline{\sigma}_j=E_j(\tilde\sigma_{1,j}+\tilde\sigma_{2,j}+\tilde\sigma_{3,j})/3$.  Fig.~\ref{ranges} shows the search regions.  The area of $S_j(E)$ integrated over each range is $>98\%$.  For spectral lines with true energies from 7 to 200 GeV, the inclusive energy range is 4.8 to 251.4 GeV. 
Furthermore, the residual plot for each fit region was examined by eye to exclude significant signals; see Appendix B in \cite{Edmonds2011}, where the residual plots for all fits are shown.

A power law is sufficient to fit the full spectrum above 12.8 GeV; however, we allow the power-law index to vary between search ranges to allow for broad features.  The background function is
\begin{equation}
B_j(E,\Gamma_j)=E^{-\Gamma_j},
\end{equation}
where $\Gamma_j$ is free.  The composite signal-plus-background unbinned likelihood function for the $j^{th}$-search energy range with $N_j$ photons is
\begin{equation}
\label{ml_method}
\mathscr{L}_j\left(f_j,\Gamma_j\right)={\displaystyle\prod_{i=1}^{N_j} 
\Big(f_jS_j(E_i) + (1-f_j)B(E_i,\Gamma_j)}\Big),
\end{equation}
where $f_j$ is the signal fraction, and $S_j(E)$ and $B_j(E,\Gamma_j)$ are normalized to 1.  For the line search, the signal fraction may be negative, $-1<f_j<1$.  When constructing upper limits, the signal is constrained to be non-negative ($0\leq f_j<1$) to obtain physical flux limits.  In the fits, the position of the signal is fixed by the line energy $E_j$, whereas $f_j$ and $\Gamma_j$ are free parameters.

A maximum likelihood estimation of the signal fraction ($\widehat{f}_j$) and background index ($\widehat{\Gamma}_j$) is performed.  Use of the maximum likelihood instead of the extended maximum likelihood has negligible effects on our fitting results because the signal fraction is extremely small \cite{Lyons1986}.  
Since the effective area is varying slowly over these energy ranges, and we are looking for sharp spectral features, it is reasonable to fit for signal fraction without including the effective area.

The significance of the estimated signal is given by the number of standard errors ($N_{\sigma}$) corresponding to the difference in $\ln{\mathscr{L}}$ for the null hypothesis ($f_j=0$), and $\ln{\mathscr{L}}$ for the best fit ($f_j=\widehat{f}_j$),
\begin{equation}
(N_{\sigma})_j=\sqrt{-2\ln{\frac{\sup\mathscr{L}_j\left(0,\Gamma_j\right)}{\sup{\mathscr{L}_j\left(f_j,\Gamma_j\right)}}}}=\sqrt{T_j},
\end{equation}
where $T_j$ is the likelihood ratio test statistic \cite{Mattox:1996zz} 
and $\Gamma_j$ can be different in the numerator from the denominator.
The corresponding confidence level, $p$, is given by
\begin{equation}
\label{pval_eq}
1-p = \text{erf}\left(\frac{1}{\sqrt{2}}(N_{\sigma})_j\right).
\end{equation}
When $f_j<0$, $N_{\sigma}$ is multiplied by -1 to emphasize that there is a deficit of signal photons compared to the background.

\begin{figure*}[t]
\begin{center}
\subfigure[]{\label{coverage_100}\includegraphics[width=0.49\textwidth]{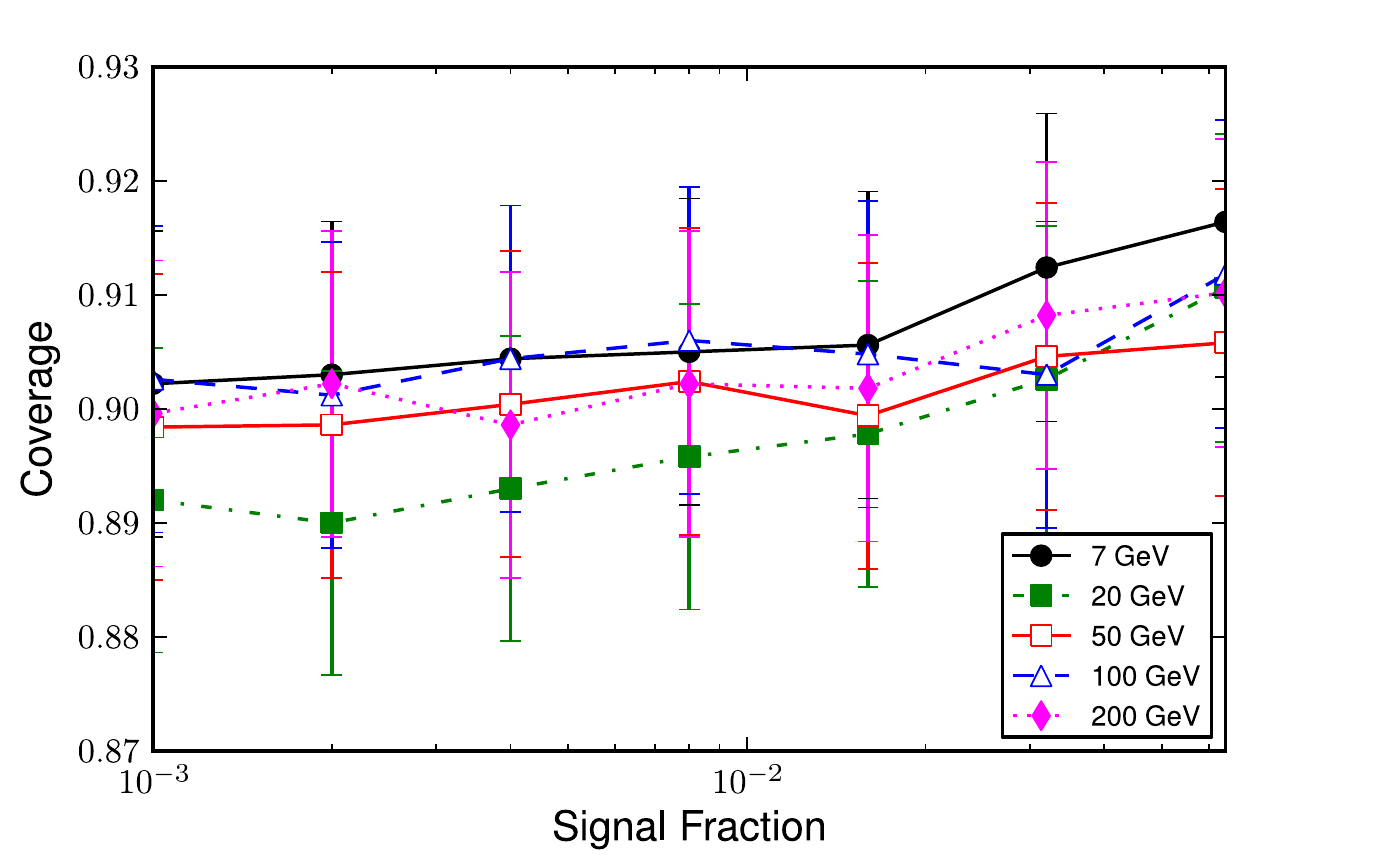}}
\subfigure[]{\label{power_100}\includegraphics[width=0.49\textwidth]{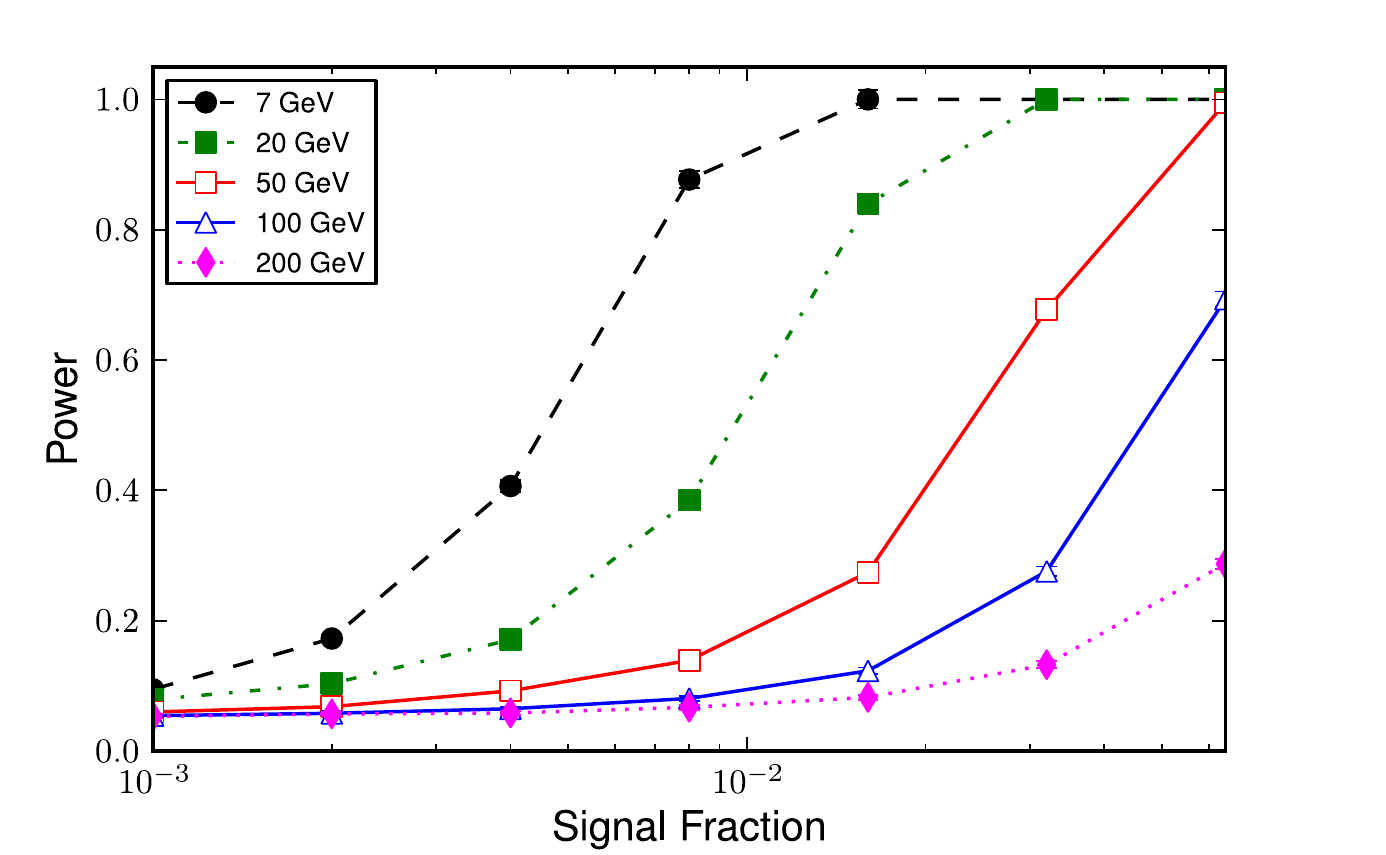}}
\end{center}
\caption{Coverage (a) and power (b) for 5000 MC trials for 7, 20, 50, 100, and 200 GeV spectral lines.  The error bars are statistical.  The background is simulated by a power-law with index $\Gamma=2.5$. 
In coverage plots, $f_j$ is not restricted to be greater than zero. Restricting to $f_j>0$ results 
in over-coverage at low signal fraction.}
\end{figure*}

To determine the upper limit to the signal fraction, $f_j$ is constrained to be non-negative and a confidence interval is constructed using the MINOS asymmetric error ($\delta^{+,-}_{f_j}$) with error level $\Delta\ln{\mathscr{L}_j}=1.35$ \cite{James2004b}.  The error level corresponds to the 90\% (95\%) coverage probability for the estimation of one Gaussian parameter in a two-sided (one-sided) confidence interval.  

\begin{figure*}[t]
\begin{center}
\includegraphics[width=0.7\textwidth]{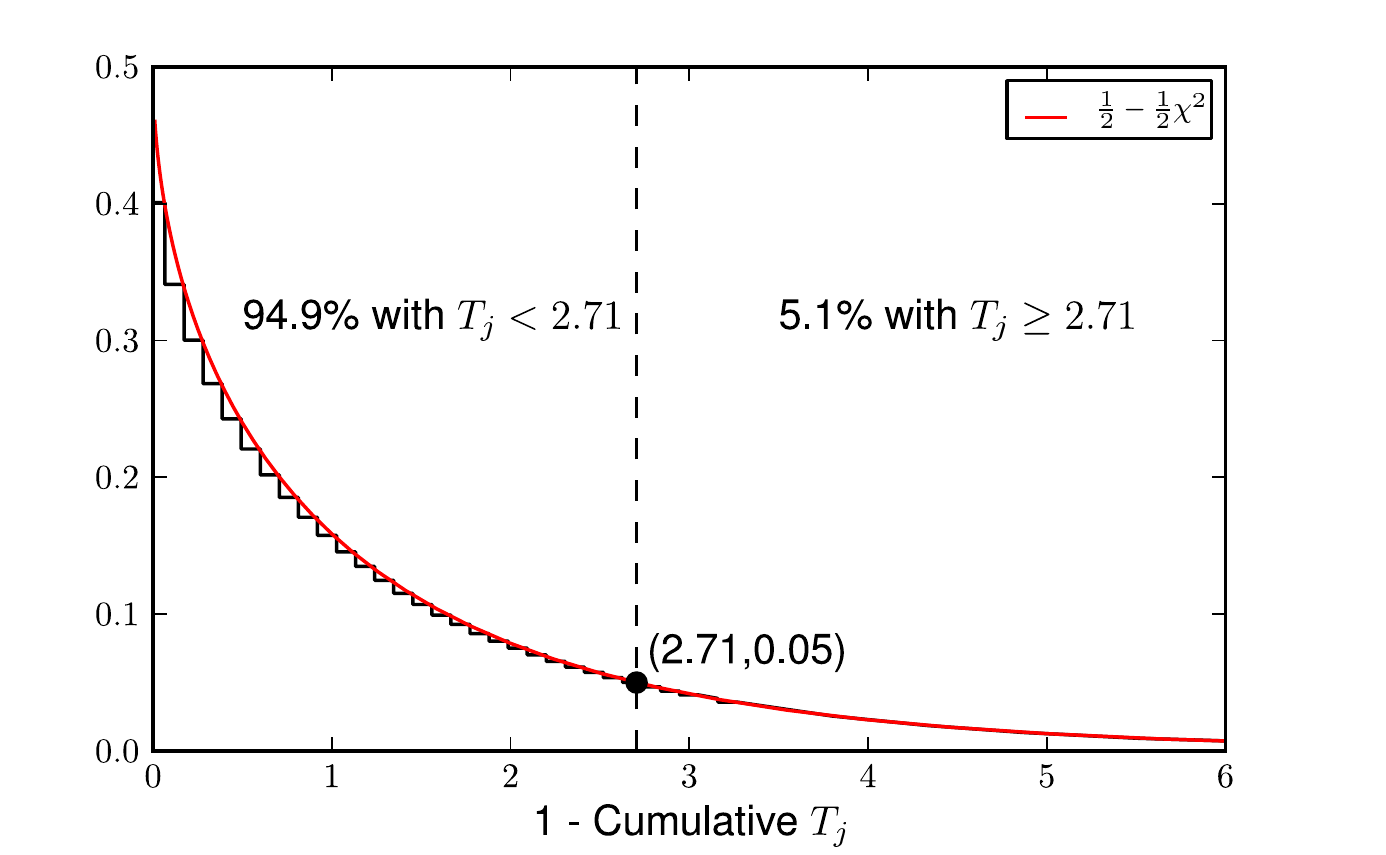}
\caption{The complement of the cumulative histogram of $T_j$ for the background only MC (black histogram).  
The red curve shows the $(1-\chi^2_{cdf})/2$ distribution from Chernoff's theorem~\cite{Chernoff1954}, where 
$\chi^2_{cdf}$ is the cumulative chi-squared distribution function.}
\label{fig_chernoff}
\end{center}
\end{figure*}

To test the coverage and power of the method, we simulate $\sim5000$ experiments with the background 
PDF $B(E,\Gamma=2.5)$ in the energy range 4.8 to 251 GeV.  
The power is defined as the fraction of experiments where a signal is detected.  
The number of background photons in each 
MC simulation is equal to the number of photons in the actual data.  
Figs.~\ref{coverage_100} and \ref{power_100} show the coverage and power for the 
7, 20, 50, 100, and 200 GeV spectral line fits.  In coverage plots, $f_j$ is not restricted to be greater than zero.  The coverage plot shows the fraction of MC runs with the true signal fraction within the constructed 
confidence interval, and behaves as expected, giving $\sim90\%$ coverage.
When obtaining limits, we require $f_j>0$, which results in over-coverage at low signal fraction. 
The decreased power to detect a signal in the 100 GeV range compared to the 7 GeV 
range reflects the decrease in the number of photons within the fitting range and the 
$\sim 25\%$ decrease in line resolving power. 

We calculate the likelihood ratio test statistic distribution for $f_j=0$, and compare it to the expected behavior from Chernoff's theorem \cite{Chernoff1954} to show that our likelihood estimator is approximately Gaussian, and $\Delta\ln{\mathscr{L}_j}=1.35$ ($T_j=2.71$) corresponds to 95\% coverage probability for a single-sided confidence interval.  For the MC with background only, the distribution of the likelihood ratio test statistic should be
\begin{equation}
\text{PDF}\left(T_j\right)=\frac{1}{2}\delta\left(T_j\right)+\frac{1}{2}\chi^2\left(T_j\right) .
\end{equation}
In Fig.~\ref{fig_chernoff}, we show the curve $(1-\chi^2_{cdf})/2$, where $\chi^2_{cdf}$ is 
the cumulative chi-squared distribution function, and the complement of the cumulative histogram for $T_j>0$ for the MC simulated with background only, for all energy ranges (110,000 trials).  The percentage of photons 
with $T_j>2.71$ is 5.1\%, consistent with expectations.

\subsection{Results of the Line Search}\label{sec:results line search}

\begin{figure*}
\begin{center}
\includegraphics[width=0.65\textwidth]{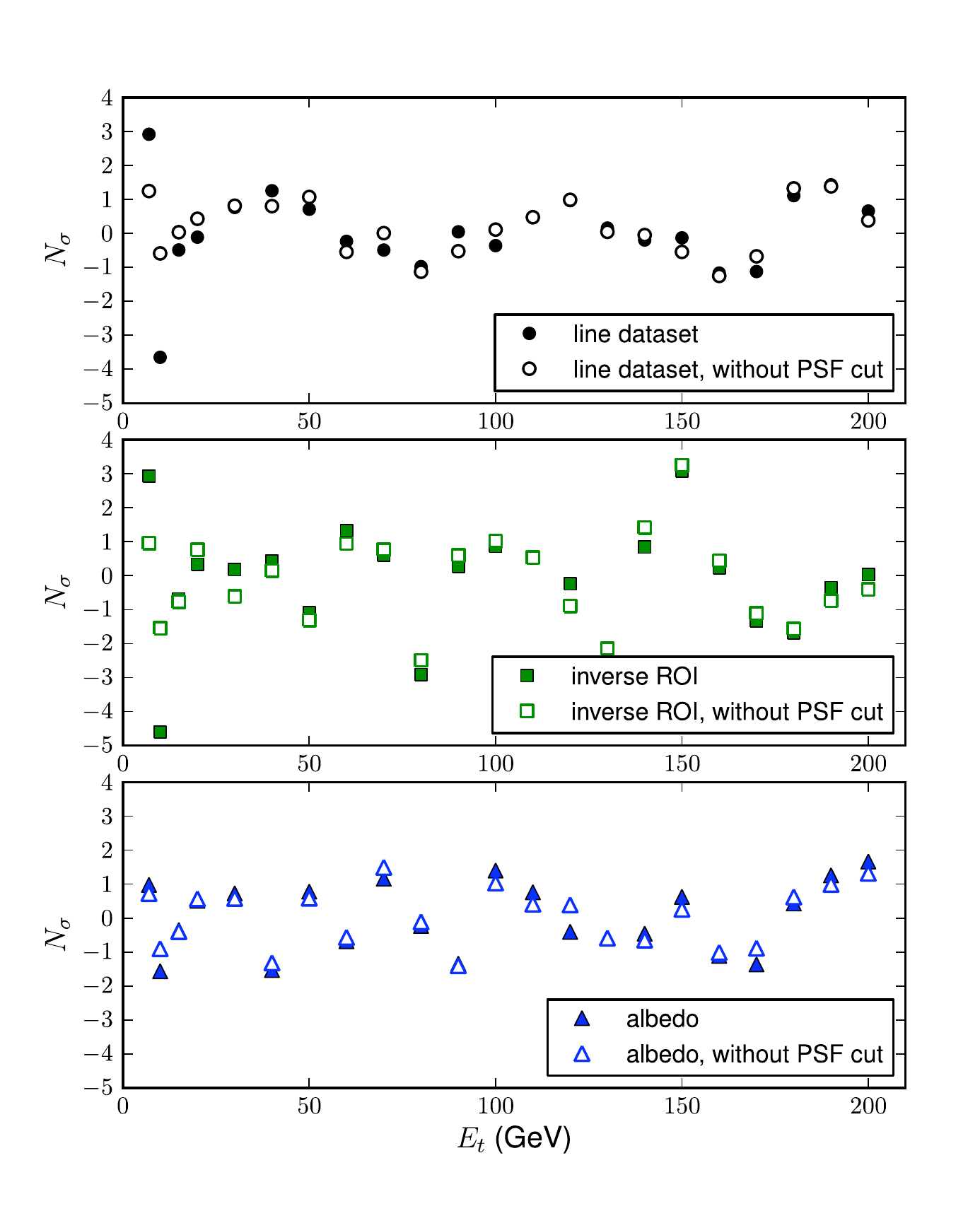}
\vskip -0.8cm
\caption{Signal significance for the line, inverse ROI, and albedo datasets with and without the PSF cut.  A trial factor is not applied to the signal significance.  All datasets show an excess at 7~GeV and a deficit at 10~GeV.  In the line dataset and inverse ROI datasets, the 7 and 10 GeV features have magnitude of significance $\geq3\sigma$.  Removing the PSF cut decreases the magnitude of their significance, particularly in the line dataset and inverse ROI.} \label{sigsignif}
\end{center}
\end{figure*}

The spectral line search was performed for the  full photon selection, inverse ROI, and albedo datasets with and without the PSF cut.  In Fig.~\ref{sigsignif}, we show the signal significance ($N_{\sigma}$) from fitting a spectral line in each energy range for the datasets with and without the PSF cut.  A trials factor is not applied.  
Both the line analysis and inverse ROI datasets show features  
at 7 and at 10 GeV.  The Earth limb data set also shows structure at these energies. 
Given that these features appear in the inverse ROI and the Earth limb data set suggests that 
they are a systematic effect and not a real signal.  Moreover, removal 
of the PSF cut significantly reduces the  
significance of these features, see Fig.~\ref{sigsignif}. Tables of the fitting results for the line, inverse ROI, 
and albedo datasets are given in \cite{Edmonds2011}. 

\begin{figure*}
\begin{center}
\includegraphics[width=1.0\textwidth]{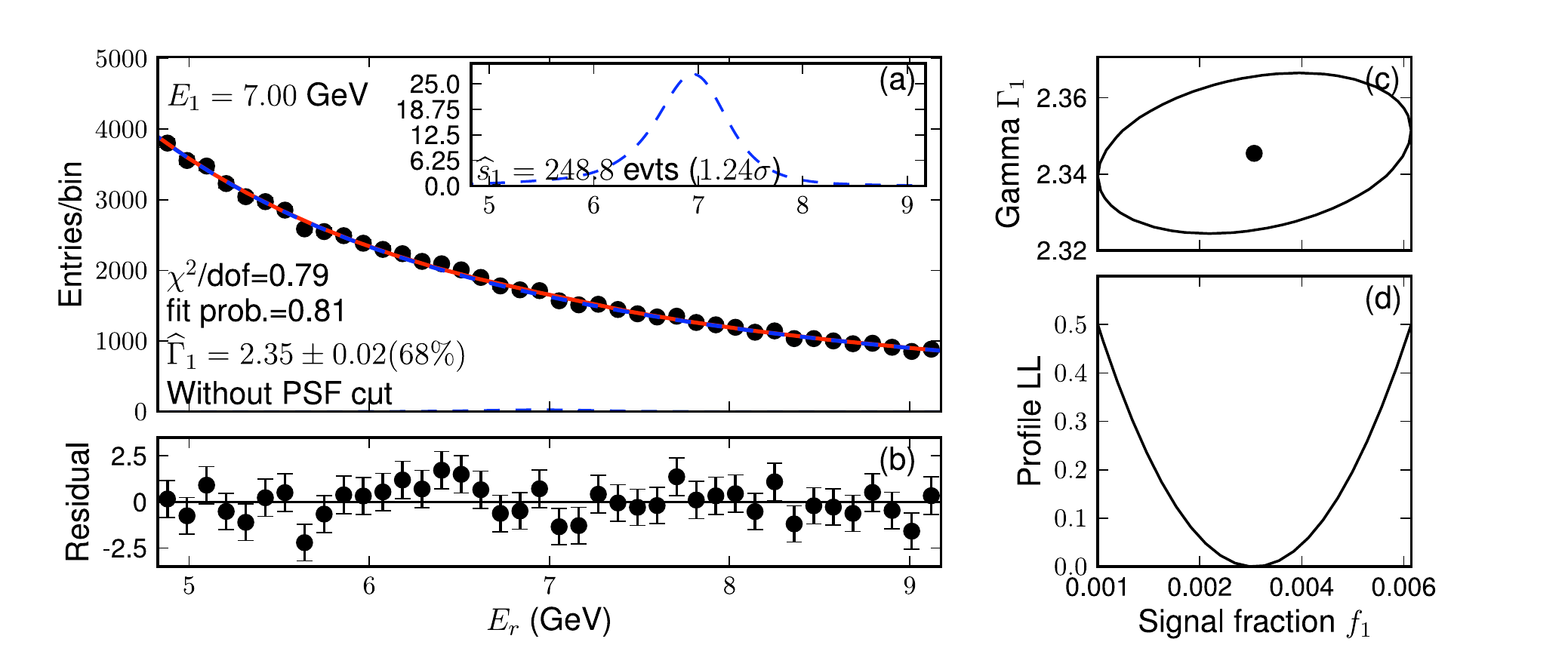}
\caption{Line dataset spectral line fitting results, excluding the PSF cut, for 7 GeV. Removing the PSF cut from the line dataset reduces the signal significance from $2.9\sigma$ to $1.2\sigma$, indicating that the feature is caused by a systematic effect. 
(a) Composite unbinned maximum likelihood fit (signal and background components in blue dashed, combined signal+background in solid red).  Signal pdf shown in inset (blue dashed line). (b) Normalized residuals for the composite fit. (c) Standard error ($1\sigma$) CL error ellipse. (d) Profile of the log-likelihood (LL) for the signal fraction, which is maximized with respect to the spectral index at each point.
Note that there are two blue lines in the top left figure, one for the background, which is largely indistinguishable from the red line, and one for a small simulated signal, which is barely distinguishable from the x-axis in this plot, but also shown in the inset.} 
\label{fig_line_dataset_wo_CTBCOREcut_7GeV}
\end{center}
\end{figure*}

\begin{figure*}
\begin{center}
\includegraphics[width=0.65\textwidth]{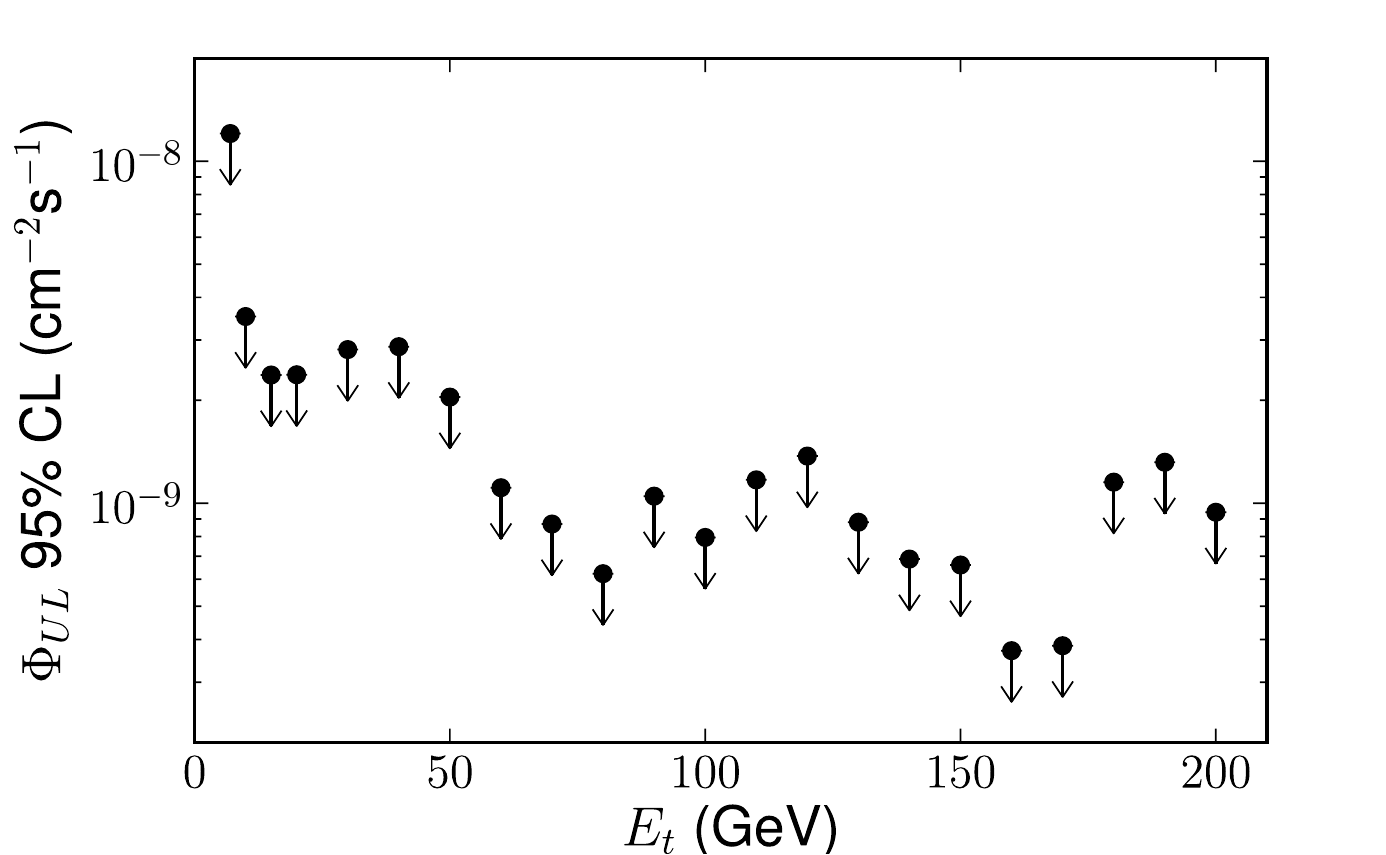}\\
\caption{95\% CL flux upper limit (integrated over the entire ROI) for photons from spectral lines.  The bin-to-bin correlations are due to the overlap of the energy ranges, which is shown in Fig.~\ref{ranges}.  There is a systematic uncertainty on these upper limits 
of 23\% for $E \le 130$ GeV and 30\% for $E > 130$ GeV (see text).  
}\label{fluxul}
\end{center}
\end{figure*}

Since the systematic effect of the PSF cut at 7 and 10 GeV results in line-like structures, we use the line dataset without the PSF cut selection for the 7 and 10 GeV lines.  Above 12.8 GeV (bin boundary) we 
use the line dataset with the PSF cut included.  Examination of the residuals (Fig.~\ref{fig_line_dataset_wo_CTBCOREcut_7GeV}) by eye 
suggests that a spectral line fit at $\sim$6.5 GeV would 
increase the signal significance and the fit probability, although only by an insignificant amount.  
To find the best position of the signal, we repeated the statistical analysis scanning from 6 to 7 GeV at 50 MeV increments.  The 6.5 GeV spectral line fit resulted in the largest signal, which had a significance of $2.6\sigma$.  The significance of the estimated signal is decreased by a ``look elsewhere effect,'' or trials factor, since the line search is performed multiple times.  An estimate of the number of independent trials is the number of half resolutions that fit in the full energy range.  For the line search from 4.8 to 251.4 GeV and $\sigma\sim$7\%, $N_t\sim30$, and the $2.6\sigma$ signal is degraded to $1.2\sigma$.

We find no detection of spectral lines from 7 to 200 GeV and calculate the 95\% CL flux upper limits to spectral lines, see Table \ref{tab:limits}. 
The 7 and 10 GeV flux limits using the line dataset without the PSF cut  
give better limits than including this cut, since the effective exposure is increased by 
25\% by removing the PSF cut.  We use the exposure as determined with the PSF cut included.  Thus, these limits are conservative.  We chose not to create new line shapes and P6\_V3 IRFs without the PSF cut.
Fig.~\ref{fluxul} shows the spectral line flux upper limits.  The uncertainty in the flux limits 
have been discussed in \S \ref{subsec:errors}.  
The cosmic ray contamination of the dataset varies from to 2\% to 8\%.  
There is an overall systematic error on these flux limits from the effective area.  
A conservative upper limit is obtained by multiplying the 95\% CL flux upper limit in Table \ref{tab:limits} by 1.23 for 
$E \le 130$ GeV and by $1.3$ for $E > 130$ GeV, which takes into account the systematic error on the exposure, a systematic 
error of 10\% on the energy resolution, and the systematic error for the energy dispersion distribution tails, all added in 
quadrature. 

For a recent analysis using public \emph{Fermi} data with the standard energy variable and cuts 
see \cite{Vertongen:2011mu}.  Our flux limits are slightly stronger (without our systematic error estimates), 
but cover a smaller energy range. We also minimized features in the response of the instrument that create 
significant line-like structures over our entire energy range, and particularly at low energy.

\section{Dark Matter Implications}\label{sec:DMimplications}

In this section, we discuss the implications for indirect dark matter searches of the absence of 
significant gamma-ray spectral lines as well as of the measurement of the inclusive photon spectrum.  

The differential photon flux from WIMP annihilation is 
\begin{equation}\label{eq:flux}
\frac{d\Phi_\gamma}{dE_\gamma} = \frac{1}{8\pi} \, \frac{\langle \sigma v\rangle}{m_\chi^2}\,
\frac{dN_\gamma}{d E_\gamma}\, r_\odot \rho_\odot^2\, J,
\end{equation}
with: 
\begin{equation}\label{eq:J}
J = \int db \int d\ell \int \frac{ds}{r_\odot} \, \cos b\, \Bigg(\frac{\rho(r)}{\rho_\odot}\Bigg)^2,
\end{equation}
where the integral is over the ROI, $\langle\sigma v\rangle$ is the annihilation 
cross section, $m_\chi$ is the WIMP mass, $dN_\gamma/dE_\gamma$ is the photon energy 
spectrum, $r_\odot \simeq 8.5$ kpc is the distance from the Sun 
to the GC \citep{Ghez:2008ms}, $\rho(r)$ is the WIMP halo profile, $\rho_\odot \simeq 0.4$ 
GeV cm$^{-3}$ is the WIMP halo density at the Sun \citep{Catena:2009mf}, 
$r = (s^2 + r_{\odot}^2 - 2 s r_\odot \cos\ell \cos b)^{1/2}$ is the Galactocentric distance, where $(\ell,b)$ are the Galactic 
longitude and latitude, respectively, and $s$ is the line of sight distance. 
For decays, the flux is given by substituting in Eq.~(\ref{eq:flux}),
$\langle\sigma v\rangle \rho_\odot^2/2m_{\chi}^2 \to \rho_\odot/\tau m_\chi$,
where $\tau$ is the WIMP lifetime, and $\rho^2/\rho_\odot^2\to \rho/\rho_\odot$ in Eq.~(\ref{eq:J}).
Note that since our ROI is very large, we do not integrate over the energy-dependent PSF and the 
photon energy spectrum when integrating over the DM halo profile and defining the $J$-factor. 

For halo profiles $\rho(r)$, we consider the Navarro-Frenk-White (NFW) profile,
\begin{equation}
\rho_{\rm NFW}(r) = \frac{\rho_s}{(r/r_s)(1+r/r_s)^2}
\end{equation}
with $r_s=20$ kpc  \cite{Navarro:1996gj}, the Einasto profile,
\begin{equation}
\rho_{\rm Einasto}(r) = \rho_s \exp\{ -(2/\alpha)[(r/r_s)^\alpha - 1]\}
\end{equation}
with $r_s=20$ kpc and $\alpha=0.17$~\cite{Einasto:1965,Navarro:2008kc},
and the isothermal profile
\begin{equation}
\rho_{\rm isothermal}(r) = \frac{\rho_s}{1+(r/r_s)^2}
\end{equation}
with $r_s=5$ kpc \citep{Bahcall:1980fb}.  
We take the maximum values for $r$ to be $\sim 150$ kpc for the Einasto and NFW profiles, 
and \mbox{$\sim 100$ kpc} for the isothermal profile, so that the Milky Way 
halo has a mass of $\sim 1.2 \times 10^{12} M_\odot$ (see, e.g., \citep{Wilkinson:1999hf,Xue:2008se}).
The value of $\rho_s$ is determined by requiring $\rho(r_\odot) = 0.4$ GeV cm$^{-3}$.  
Table \ref{tab:j_val} lists the resulting values for $J$ for our ROI.
For annihilations, the GC region contributes $\sim 36\%$ ($\sim 51\%$) for the NFW (Einasto) 
profiles, but a negligible fraction for the isothermal profile, while for 
decays, the GC region is always negligible.
The Einasto profile used here gives limits 40\% stronger than the NFW profile, while
the limits for the NFW profile are 40\% stronger than the isothermal profile.  
A very cuspy profile, such as the Moore profile (integrated to within $10^{-3}$ pc of the GC), 
can result in limits stronger than the Einasto profile by a factor $\sim6$.  A boost factor, the 
ratio of WIMP annihilation flux from substructure to the annihilation flux from the smooth 
profile for an Earth observer, can strengthen the limits by another factor of a few 
(at least, assuming a velocity-independent annihilation cross section).
We will ignore such substructure enhancement in this work.   
For decays, all profiles give similar limits.  

\begin{table}[t]
\begin{center}
\begin{tabular}{|l|c|c|c|c|c|c|}
  \hline
  \multirow{2}{*}{} & \multicolumn{2}{|c|}{NFW} & \multicolumn{2}{|c|}{Einasto} & \multicolumn{2}{|c|}{Isothermal} \\
  \hline
  & $|b|>10^{\circ}$ & GC & $|b|>10^{\circ}$ & GC & $|b|>10^{\circ}$ & GC \\
  \hline
  $J$ annihilation & 19.89 & 11.40 & 22.40 & 21.17 & 16.74 & 1.51 \\
  $J$ decay  & 20.69 &  1.19 & 20.7 & 1.48 & 21.02 & 0.68 \\
  \hline
\end{tabular}
\end{center}
\caption{Table of annihilation and decay $J$ values as defined in Eq.~(\ref{eq:J}),  
for the high latitude region $|b|>10^{\circ}$, and a $20^{\circ}\times20^{\circ}$ square at the GC, calculated for the NFW, Einasto and Isothermal dark matter profiles in the Milky Way.}
\label{tab:j_val}
\end{table}

For thermal WIMP annihilation to $\gamma X$ (where $X$ has mass $m_X$ and 
is another photon, $Z$ boson, or non-Standard Model particle), the photon energy spectrum, $dN_\gamma/dE_\gamma$, is
\begin{equation}
\frac{dN_\gamma}{dE_\gamma} = N_\gamma\, \delta\Big(E_\gamma-m_\chi\Big(1-\frac{m_X^2}{4 m_\chi^2}\Big)\Big)\,,
\end{equation}
where $N_\gamma=2~(1)$ when $X=\gamma$ ($X\ne \gamma$).
For the channels $W^+W^-$, $b\bar{b}$, $gg$, and $\tau^+\tau^-$, we use the photon spectra from DarkSUSY \cite{Gondolo:2004sc,darksusy}.  
For the direct annihilation channels into electrons or muons, 
the energy spectrum from final state radiation (FSR) is given by
\begin{equation}
\frac{dN_{\gamma}}{dy}
=
\frac{\alpha}{\pi}\left(\frac{1 + (1 - y)^2}{y}\right)\left(\ln\bigg(\frac{s (1 - y)}{m_{\ell}^2}\bigg) - 1 \right)
\label{equ:fsr1}
\end{equation}
\citep{Beacom:2004pe,Birkedal:2005ep}.
Here $\alpha\simeq 1/137$, $y=E_{\gamma}/m_{\chi}$ and $s=4 m_{\chi}^2$ for annihilation or 
$y=2E_{\gamma}/m_{\chi}$ and $s=m_{\chi}^2$ for decay, and $m_\ell$ is the 
electron or muon mass.
This formula holds in the collinear limit, where the photon is emitted collinearly with one
of the leptons and when $m_\ell$ is much less than the 
WIMP mass.  
This spectrum scales as $\sim 1/E_\gamma$, which is harder than the expected background spectrum, 
and has a sharp cut-off at the dark matter mass.  This feature may be clearly visible above backgrounds, 
especially for heavier dark matter masses as is suggested by models motivated by the PAMELA and 
\emph{Fermi} data~\cite{Cirelli:2008pk,ArkaniHamed:2008qn,Pospelov:2008jd,Cholis:2008qq}.  
The energy spectrum for FSR for the annihilation into $\phi\phi$ where $\phi$ 
is a new force mediator (scalar or vector) and decays to 
electrons or muons is given explicitly in \cite{Essig:2009jx}.  It is slightly softer than the previous 
case, but still harder than conventional backgrounds.  
If the final state includes $\mu$'s, we also include photons from the radiative decay of the 
muon, e.g.~$\mu^{-} \to e^- \nu_\mu \bar{\nu}_e \gamma$, using the formulas found in,  
e.g., \cite{Kuno:1999jp,Mardon:2009rc,Essig:2009jx}.
Including contributions to the signal from high-energy electrons or muons inverse Compton scattering 
(ICS) off starlight and the cosmic microwave background would significantly improve limits presented below, but 
also include more model dependence.  We postpone a discussion of the ICS contribution to a future 
publication.

\begin{table*}[t]
\begin{center}
\begin{tabular}{c|c|rrr|rrr}
$E_\gamma$ & 95\%CLUL & \multicolumn{3}{c|}{$\langle\sigma v\rangle_{\gamma\gamma}$ [$\gamma Z$] ($10^{-27}$ cm$^{3}$s$^{-1}$)} & \multicolumn{3}{c}{$\tau_{\gamma\nu}$ ($10^{28}$s)} \\
(GeV) & ($10^{-9}$ cm$^{-2}$s$^{-1}$) & NFW & Einasto & Isothermal & NFW & Ein. & Iso. \\
\hline
7&12.0&0.06[-]&0.04[-]&0.10[-]&10.8&11.0&10.7\\
10& 3.5&0.03[-]&0.02[-]&0.06[-]&26.0&26.3&25.8\\
15& 2.4&0.05[-]&0.04[-]&0.09[-]&25.7&26.0&25.5\\
20& 2.4&0.09[-]&0.07[-]& 0.2[-]&19.2&19.5&19.1\\
30& 2.8& 0.2[2.3]& 0.2[1.6]& 0.4[3.9]&10.8&11.0&10.7\\
40& 2.9& 0.4[ 2.7]& 0.3[ 1.9]& 0.8[ 4.6]& 8.0& 8.1& 7.9\\
50& 2.0& 0.5[ 2.3]& 0.4[ 1.7]& 0.8[ 4.0]& 8.9& 9.1& 8.9\\
60& 1.1& 0.4[ 1.5]& 0.3[ 1.1]& 0.7[ 2.6]&13.7&13.9&13.6\\
70& 0.9& 0.4[ 1.4]& 0.3[ 1.0]& 0.7[ 2.4]&15.0&15.2&14.9\\
80& 0.6& 0.4[ 1.2]& 0.3[ 0.9]& 0.7[ 2.1]&18.3&18.6&18.2\\
90& 1.0& 0.8[ 2.4]& 0.6[ 1.7]& 1.4[ 4.1]& 9.7& 9.8& 9.6\\
100& 0.8& 0.8[ 2.1]& 0.5[ 1.5]& 1.3[ 3.6]&11.5&11.7&11.4\\
110& 1.2& 1.4[ 3.6]& 1.0[ 2.6]& 2.3[ 6.1]& 7.1& 7.2& 7.0\\
120& 1.4& 1.9[ 4.8]& 1.4[ 3.5]& 3.2[ 8.3]& 5.5& 5.6& 5.5\\
130& 0.9& 1.4[ 3.5]& 1.0[ 2.5]& 2.4[ 6.0]& 8.0& 8.1& 7.9\\
140& 0.7& 1.3[ 3.1]& 0.9[ 2.2]& 2.2[ 5.3]& 9.5& 9.6& 9.4\\
150& 0.7& 1.4[ 3.3]& 1.0[ 2.4]& 2.4[ 5.7]& 9.2& 9.4& 9.2\\
160& 0.4& 0.9[ 2.1]& 0.7[ 1.5]& 1.6[ 3.6]&15.4&15.6&15.3\\
170& 0.4& 1.1[ 2.4]& 0.8[ 1.7]& 1.8[ 4.1]&14.0&14.2&13.9\\
180& 1.2& 3.6[ 8.0]& 2.6[ 5.8]& 6.1[13.8]& 4.4& 4.5& 4.4\\
190& 1.3& 4.6[10.1]& 3.3[ 7.3]& 7.8[17.4]& 3.6& 3.7& 3.6\\
200& 0.9& 3.6[ 7.9]& 2.6[ 5.7]& 6.2[13.6]& 4.9& 4.9& 4.8\\
\end{tabular}
\end{center}
\caption{
Flux, annihilation cross-section upper limits, and decay lifetime lower limits.  Spectral line energy and corresponding 95\% CLUL to spectral line flux, for $|b|>10^\circ$ plus a $20^\circ\times20^\circ$ square at the GC.  For each energy and flux limit, $\langle\sigma v\rangle_{\gamma\gamma}$ and $\langle\sigma v\rangle_{Z\gamma}$ upper limits, and $\tau_{\nu\gamma}$ lower limits are given for the NFW, Einasto, and isothermal DM distributions.  $\gamma Z$ limits below $E_\gamma < {\rm \ 30 \ GeV}$ are not shown; see text for explanation.  
A conservative upper limit is obtained by multiplying the 95\% CL flux upper limits by 1.23 for $E \le 130$ GeV and 1.3 for $E > 130$ GeV, which takes into account the systematic errors on the exposure, 
the energy resolution, and the energy dispersion distribution tails, all added in quadrature. 
}
\label{tab:limits}
\end{table*}

\subsection{Gamma-ray lines from Dark Matter Annihilation and Decay}\label{subsec:DMlines}

\begin{figure*}[t]
\begin{center}
\subfigure{\includegraphics[scale=0.58]{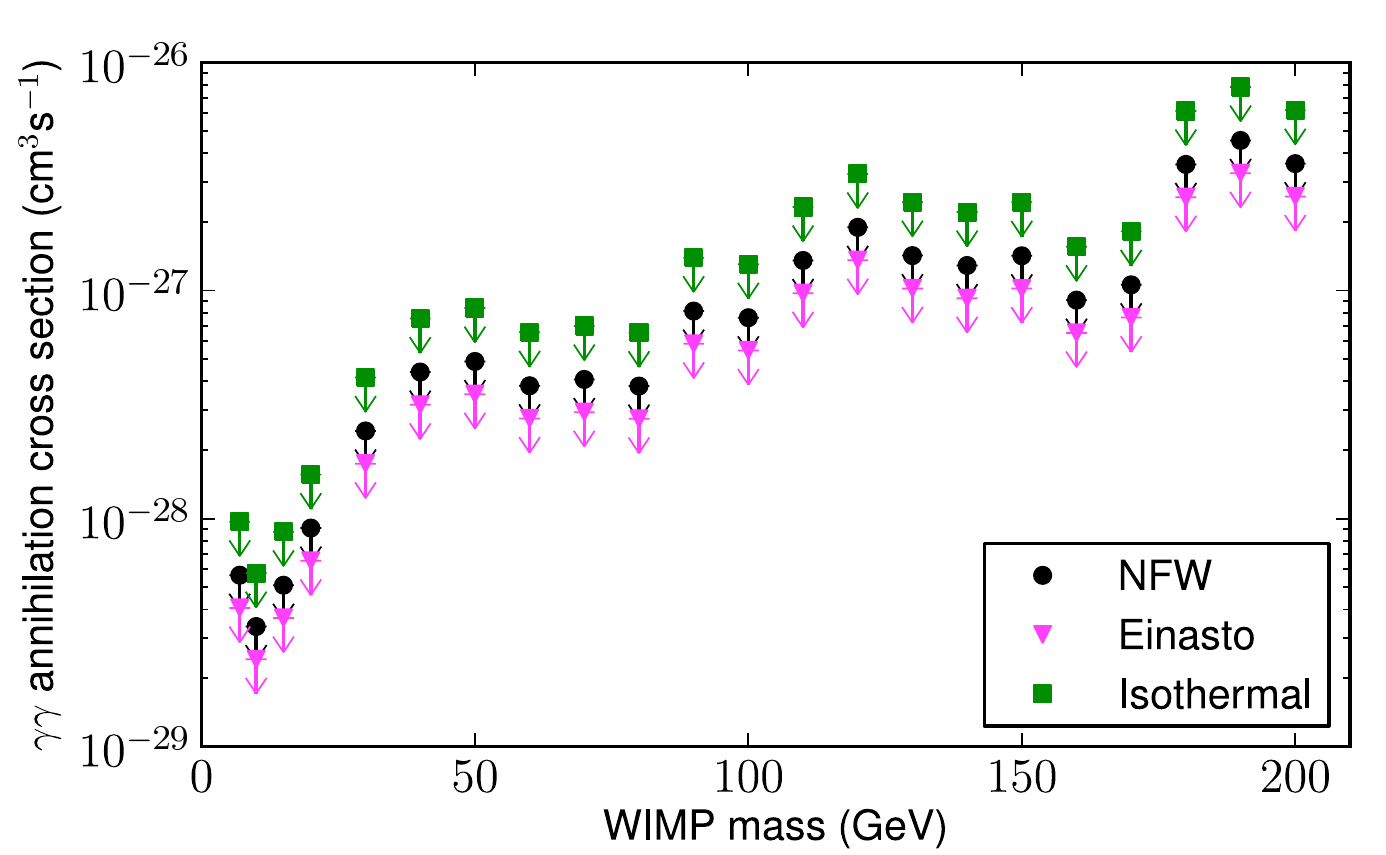}}
\hskip -0.1cm 
\subfigure{\includegraphics[scale=0.58]{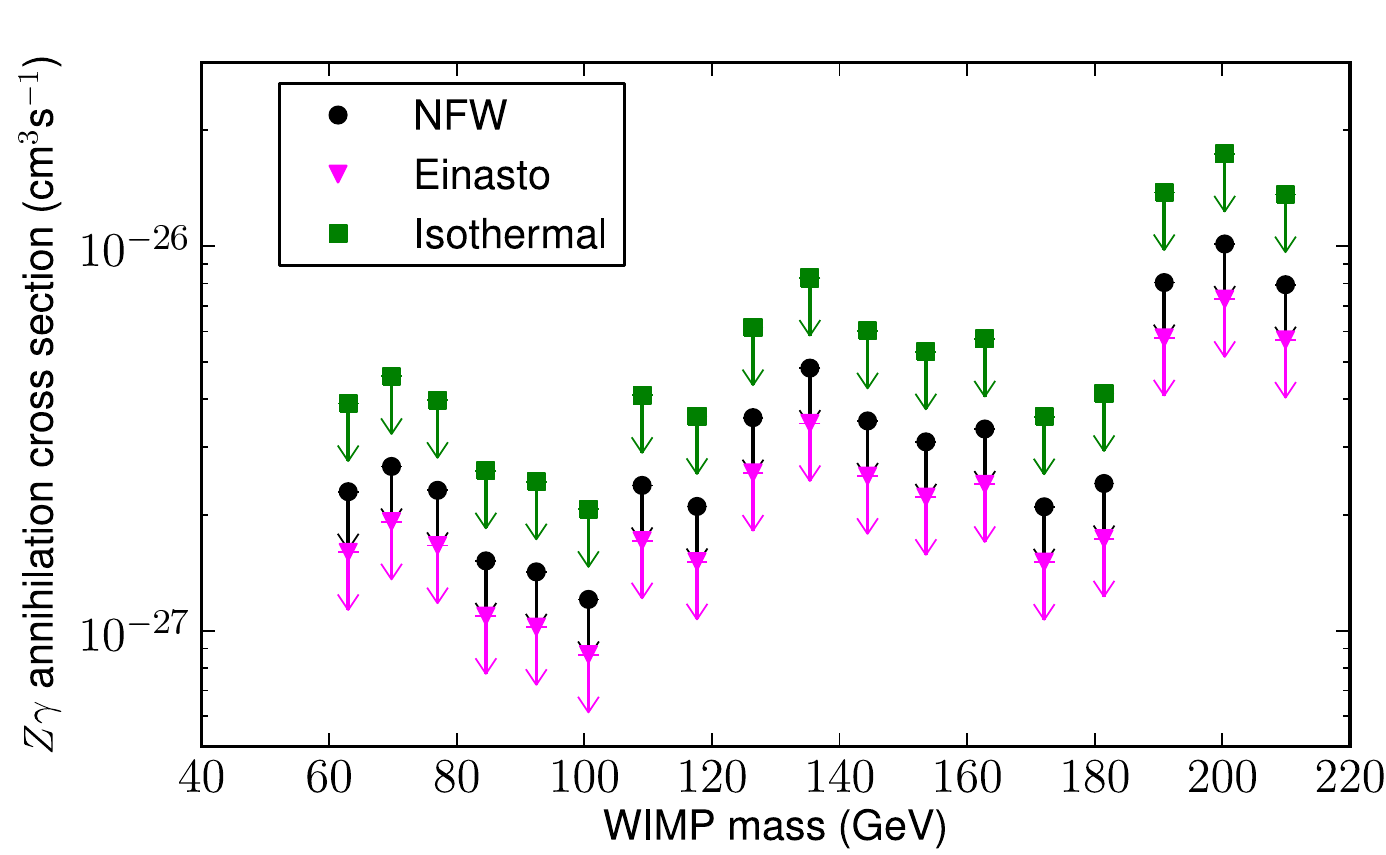}}\\
\subfigure{\includegraphics[scale=0.58]{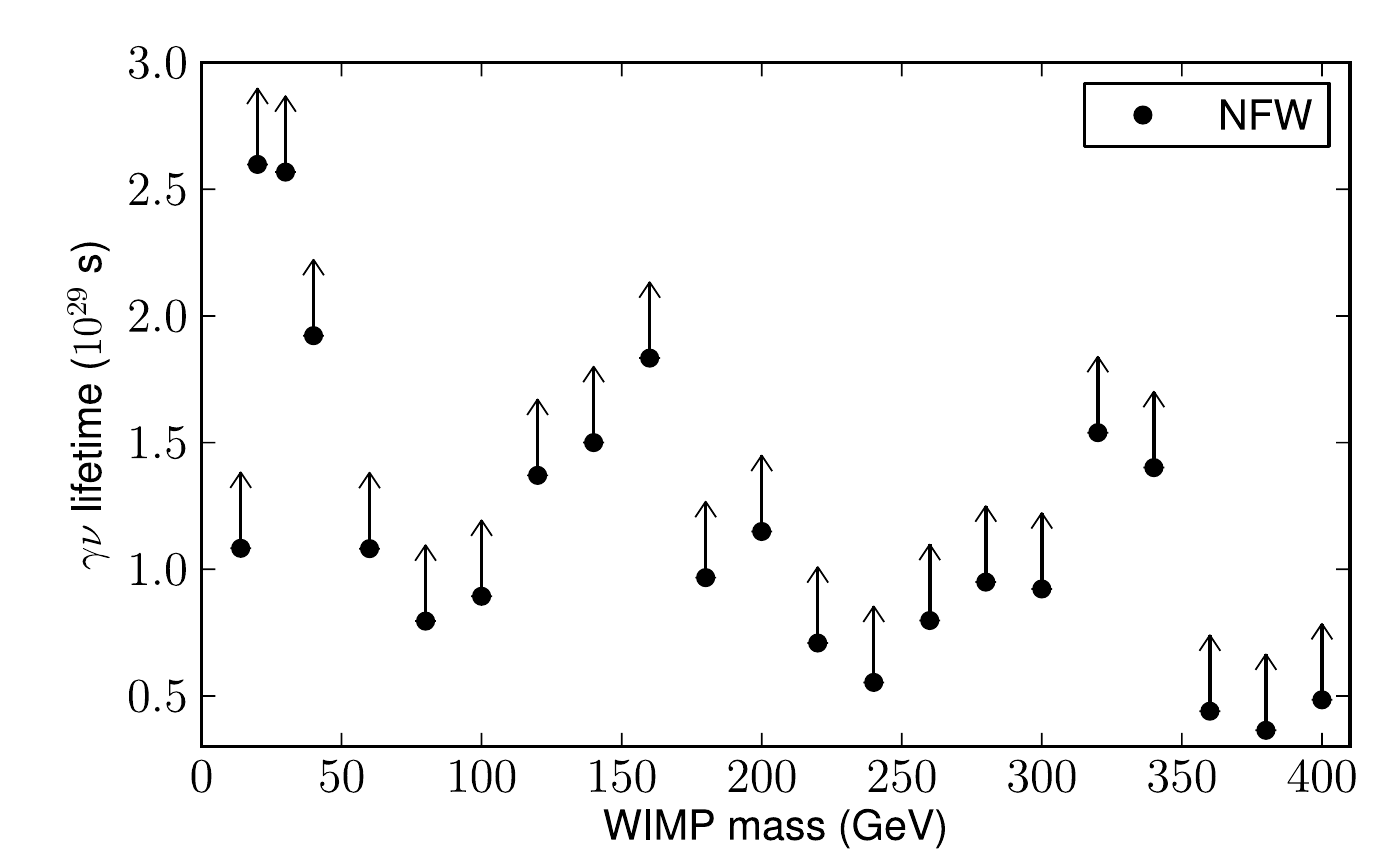}}
\caption{Top row: 
Dark matter annihilation 95\% CL cross section upper limits into $\gamma\gamma$ (left) 
and $Z\gamma$ (right) for the NFW, Einasto, and isothermal profiles for the region $|b|>10^\circ$ 
plus a $20^\circ\times20^\circ$ square at the GC. 
$\gamma Z$ limits below $E_\gamma < {\rm \ 30 \ GeV}$ are not shown; see text for explanation.
Bottom row: Dark matter decay 95\% CL lifetime lower limits into $\gamma\nu$ for the NFW profile 
and same ROI.  Systematic effects from the photon line flux upper limits are not included.  
}
\label{final_ulgg}
\end{center}
\end{figure*}

In this section, we calculate $\gamma\gamma$ and $Z\gamma$ annihilation cross section upper limits 
and $\gamma\nu$ lifetime lower limits for the NFW, Einasto, and isothermal density profiles using the above equations with $J$ values from Table \ref{tab:j_val}.  The limits are given by
\begin{equation}\label{cs_limit_eq}
\left<\sigma v\right>_{X\gamma,j}=5.99\times10^{-28}\frac{1}{N_\gamma}\left(\frac{\Phi_{UL,j}}{10^{-9}\text{ cm}^{-2}\text{s}^{-1}}\right)\left(\frac{m_\chi}{10\text{ GeV}}\right)^2\frac{1}{J_{ann}(\Delta\Omega)}\text{ cm}^{-3}\text{s}^{-1}\;\text{, and}
\end{equation}
\begin{equation}\label{cs_limit_eq2}
\tau_{X\gamma,j}=8.35\times10^{28} N_\gamma\left(\frac{10^{-9}\text{ cm}^{-2}\text{s}^{-1}}{\Phi_{UL,j}}\right)\left(\frac{10\text{ GeV}}{m_\chi}\right)\frac{1}{J_{decay}(\Delta\Omega)}\;\text{ s}\,.
\end{equation}
Table \ref{tab:limits} and Fig.~\ref{final_ulgg} give the spectral line flux upper limits, cross-section upper limits, and lifetime lower limits for various spectral line energies.

The $\gamma\gamma$ annihilation cross section $\langle\sigma v\rangle_{\gamma\gamma}$ upper limits are shown in Fig.~\ref{final_ulgg}.  The upper limits to $\langle\sigma v\rangle_{\gamma\gamma}$ using the NFW profile range from $\sim 3\times10^{-29}$ to $5\times10^{-27}$ cm$^{3}$s$^{-1}$ in the line (WIMP mass) energy range 7--200 GeV, while those for $\langle\sigma v\rangle_{Z\gamma}$ range from 
$\sim 10^{-27}$ to $10^{-26}$ cm$^3$s$^{-1}$ for WIMP masses 63--210 GeV.  
For decays to $\gamma Z$ with $M_\chi < 63 {\rm \ GeV}$ $(E_\gamma < 30 {\rm \ GeV})$, 
the finite width of the $Z$ boson introduces a $\gtrsim 20\%$ correction 
on the $\gamma Z$ limit compared to the case when the $Z$ width is 
approximated by zero (see e.g.~\cite{Gustafsson:2007pc}). Calculating this large correction to the
$\gamma Z$ limits due to the finite $Z$ width is beyond the scope 
of this paper. Thus all $\gamma Z$ limits for $E_\gamma < 30 {\rm \ GeV}$
are not shown in Table \ref{tab:limits} and Fig.~\ref{final_ulgg} (top right). 
The $\gamma Z$ limit for $E_\gamma = 30 {\rm \ GeV}$ $(M_\chi = 63 {\rm \ GeV})$ 
includes an approximate correction of 5.2\% due to the finite 
width of the $Z$, and this correction is well below our overall 
systematic error of 22\%. The correction for the $Z$ width on the 
$\gamma Z$ limits for points with $E_\gamma > 30{\rm \ GeV}$ is negligible. 

Fig.~\ref{final_ulgg} also shows the lower limits for the $\gamma\nu$ decay lifetime $\tau_{\gamma\nu}$ for the NFW profile versus WIMP mass.  The lower limits for $\tau_{\gamma\nu}$ range from $4\times10^{28}$ to $3\times10^{29}$ s for WIMP masses from 14 to 400 GeV.  
The Einasto and isothermal profiles give very similar results.

Theoretical predictions for gamma-ray line strengths are highly model dependent so that only some 
models are constrained by our results (see e.g.~\cite{Abazajian:2011tk}). 
For example, for a neutralino in the Minimal Supersymmetric Standard Model, the cross section to annihilate 
to two photons is generically well below our constraints, e.g.~\cite{Bergstrom:1997fh}.
Nevertheless, we are able to constrain some interesting models, including the one in \cite{Kane2009}, where the lightest supersymmetric particle (LSP) is a non-thermally produced $\sim 200$ GeV wino (supersymmetric partner of the $W$ boson) that may be able to explain the PAMELA satellite's 
CR data \cite{Adriani:2008zr}.  
While the dominant annihilation channel is into $W^+W^-$, which produces a continuous spectrum 
of photons (see \S \ref{subsec:inclusivelimits}),  
the wino can also annihilate into 
$\gamma\gamma$ and $\gamma Z$, with 
$\langle\sigma v\rangle_{\gamma\gamma}\sim2.3\times10^{-27}$ cm$^3$s$^{-1}$ and 
$\langle\sigma v\rangle_{Z\gamma}\sim1.4\times10^{-26}$ cm$^3$s$^{-1}$.  
The $\gamma Z$ annihilation produces a line at $E_\gamma \sim 160$ GeV, which limits 
the cross section for an NFW (isothermal) profile to be 
$\langle\sigma v\rangle_{Z\gamma}\lesssim 2.1 (3.6) \times10^{-27}$ cm$^3$s$^{-1}$, 
disfavoring the model by nearly a factor of 7 (4).  
In the more general ``wino'' LSP models with a higgsino component discussed in \cite{Acharya2011}, 
the mass can be in the range 
$\sim 140 - 155$ GeV with a 
$Z\gamma$ cross section of $(0.7-1.2)\times10^{-26}$ cm$^3$s$^{-1}$.  
These cross sections are also disfavored by our limits.

There are also other annihilation models that are partially constrained, 
including \cite{Gustafsson:2007pc,Mambrini:2009ad,Jackson2009},
while models that are only constrained assuming a much cuspier profile include \cite{Bertone:2009cb,Bringmann:2007nk}.
Using an effective field theory of WIMP interactions with Standard Model matter, the line constraints 
can be translated to limits on the cross section for dark matter scattering off 
nuclei in direct detection experiments.  This was done in \cite{Goodman2011} based on the 
eleven month \emph{Fermi} line search \cite{Bloom2010}.  The limits will now be slightly stronger 
and extend to lower masses. 
Our results on dark matter decay are able to constrain a subset of the lifetime range of interest for 
gravitinos decaying into monochromatic photons \cite{Ibarra:2007wg}.

\subsection{Inclusive Spectrum Limits} \label{subsec:inclusivelimits}

\begin{figure*}[t]
\begin{center}
\includegraphics[width=0.7\textwidth]{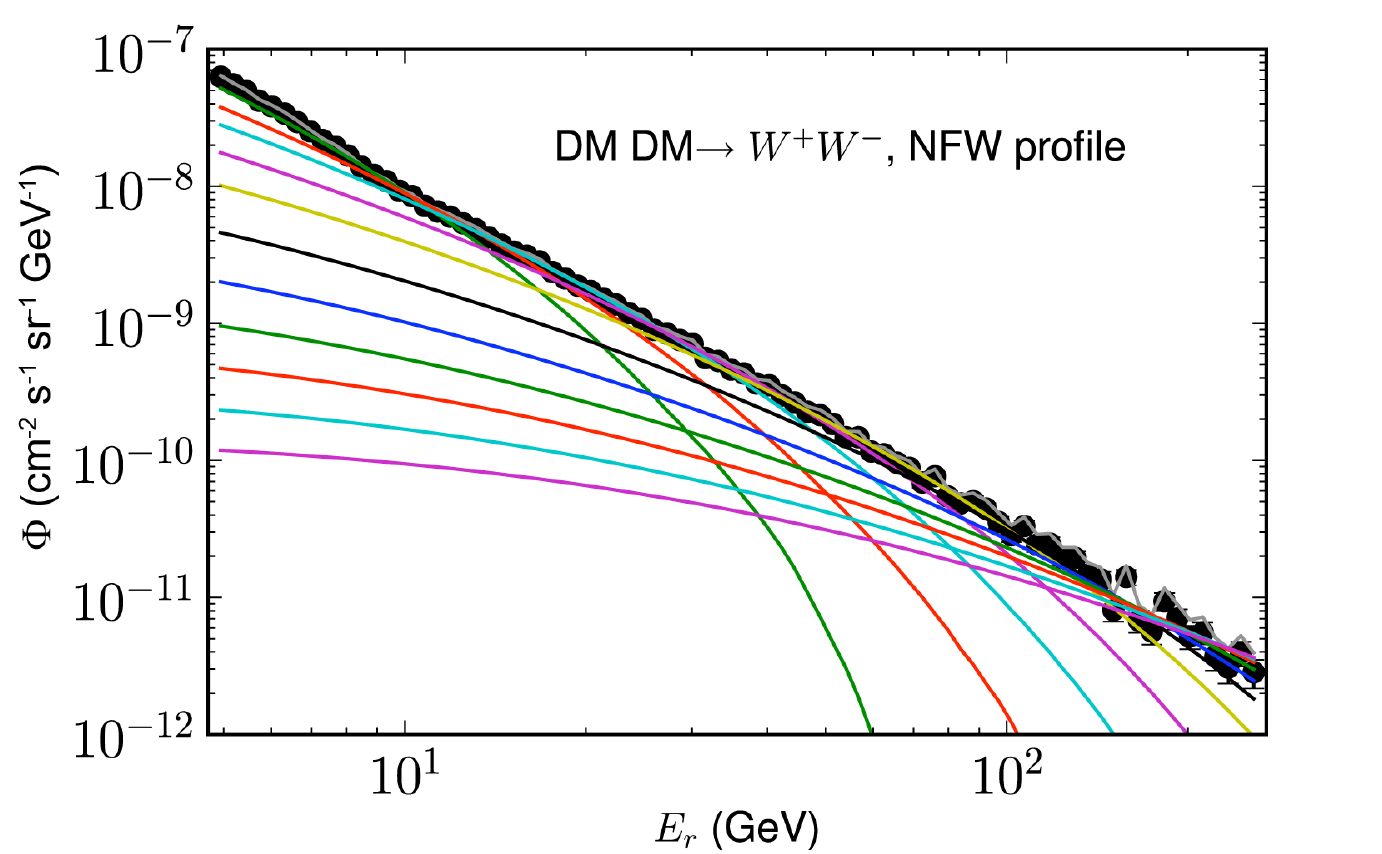}
\caption{Measured intensity (black) and largest allowed intensity (various colors) for WIMP 
annihilation to $W^+W^-$ for WIMP masses 
86, 139, 224, 360, 578, 923, 1495, 2405, 3867, 6219, and 10000 GeV, for the 
region $|b|>10^\circ$ plus a $20^\circ\times20^\circ$ square at the GC. 
}
\label{fig:all_flux_ann_wpwm}
\end{center}
\end{figure*}

In this section, we use the inclusive spectrum in Fig.~\ref{esqur_dataclean_flux} 
to calculate conservative upper limits on the annihilation 
cross section and lower limits on the decay lifetime for the channels $e^+e^-$, $\mu^+\mu^-$, 
$\tau^+\tau^-$, $W^+W^-$, $b\overline{b}$, $gg$ (gluons), $\phi\phi\to e^+e^-e^+e^-$, and 
$\phi\phi\to\mu^+\mu^-\mu^+\mu^-$.  
The models with final state leptons are motivated by a dark matter explanation to the 
PAMELA and \emph{Fermi} CR data \cite{Adriani:2008zr,Adriani:2010rc,FermiCRE2009}, 
and include models with a new force mediator coupled to 
dark matter \cite{Cirelli:2008pk,ArkaniHamed:2008qn,Pospelov:2008jd,Cholis:2008qq}.  

We note that the inclusive spectrum in our analysis' energy range is well fit by a simple power-law, making 
it a reasonable assumption that at least an $\mathcal{O}(1)$ fraction comes from astrophysical background 
sources and not from dark matter.  Nevertheless, we ignore any astrophysical background contribution to 
the photon spectrum, which makes the resulting limits very conservative.  

We compute the photon differential fluxes as described above for various WIMP annihilation and 
decay models.  To calculate the cross section upper limits, the WIMP model annihilation flux is 
compared to the 95\% CL upper limit to the measured flux in each energy bin.  
In energy bin $k$, the 95\% CL UL to the measured flux is 
given by $\Phi_k+1.64\sigma_k$, where $\sigma_k$ is the error due to statistical fluctuation (first 
error in Table \ref{table_spectrum}).  The value of $\sigma_k$ does not include the systematic errors, 
due to the effective area and absolute energy scale calibration uncertainty (also given in Table 
\ref{table_spectrum}).  The WIMP cross section is scaled until the model flux exceeds the measured 
flux in a single energy bin.  The WIMP lifetime lower limits are calculated in a similar way by scaling the 
decay lifetime.  

\begin{figure*}[t]
\begin{center}
\subfigure{\includegraphics[scale=0.58]{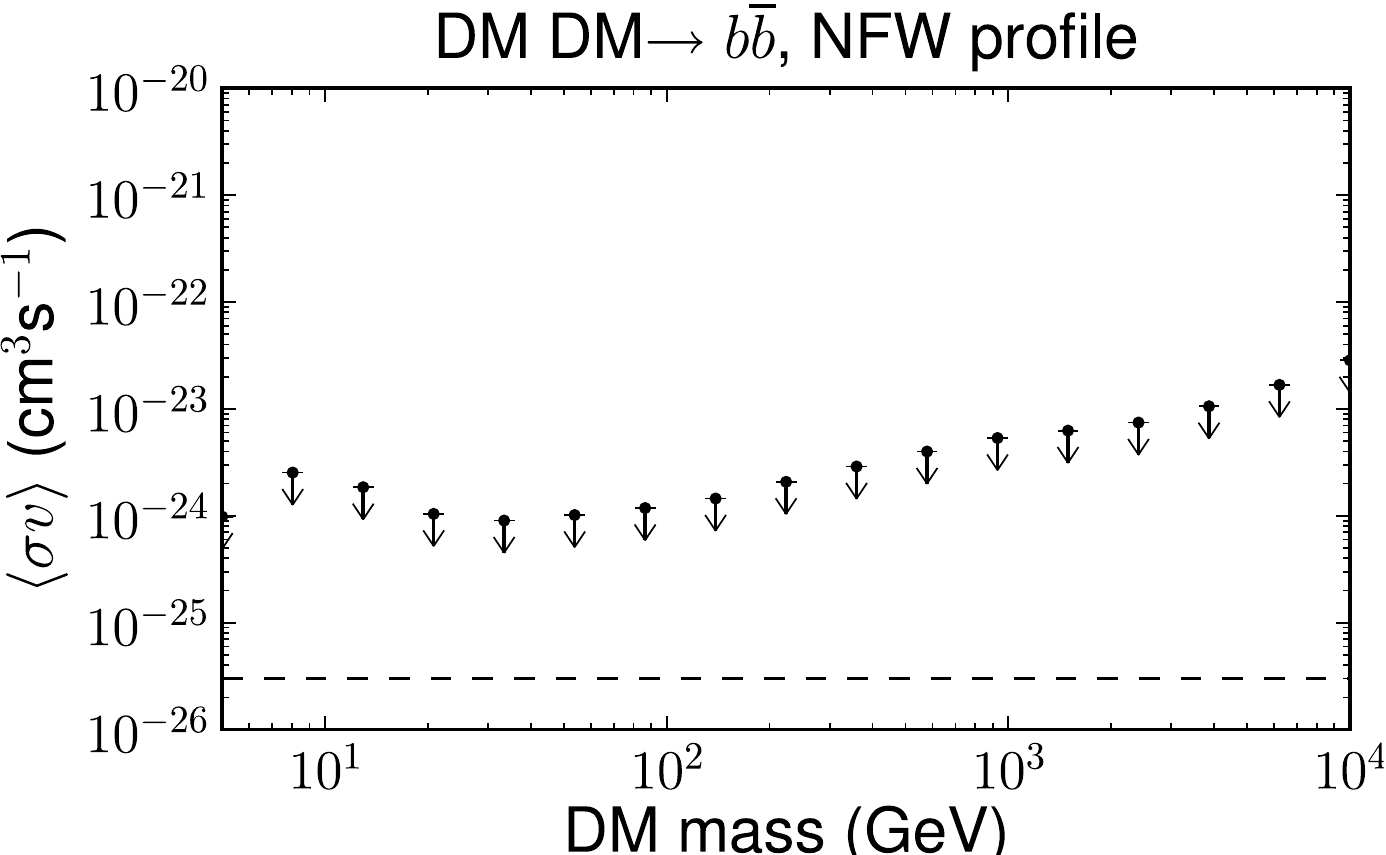}}
\subfigure{\includegraphics[scale=0.58]{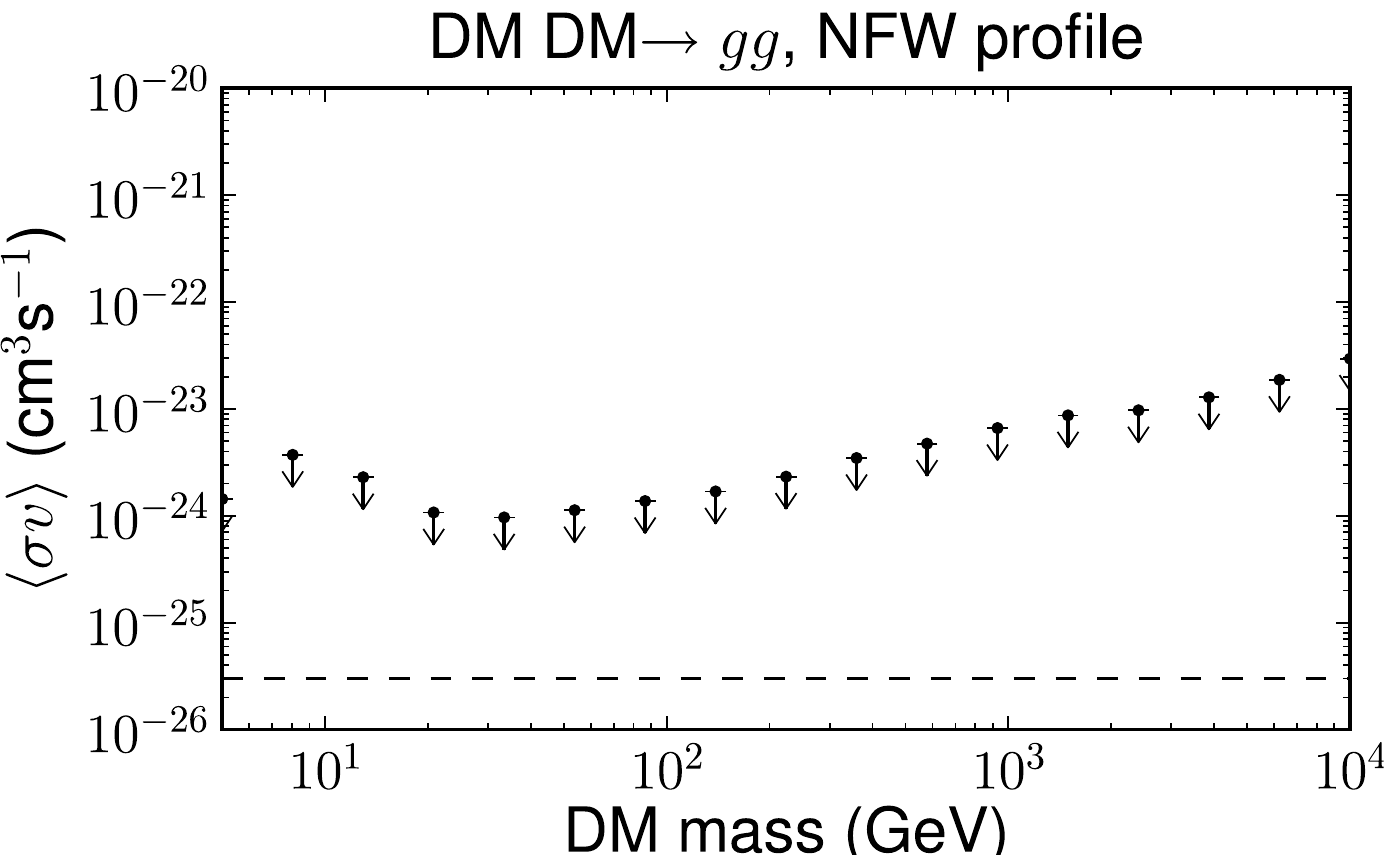}}\\
\vskip 3mm
\subfigure{\includegraphics[scale=0.58]{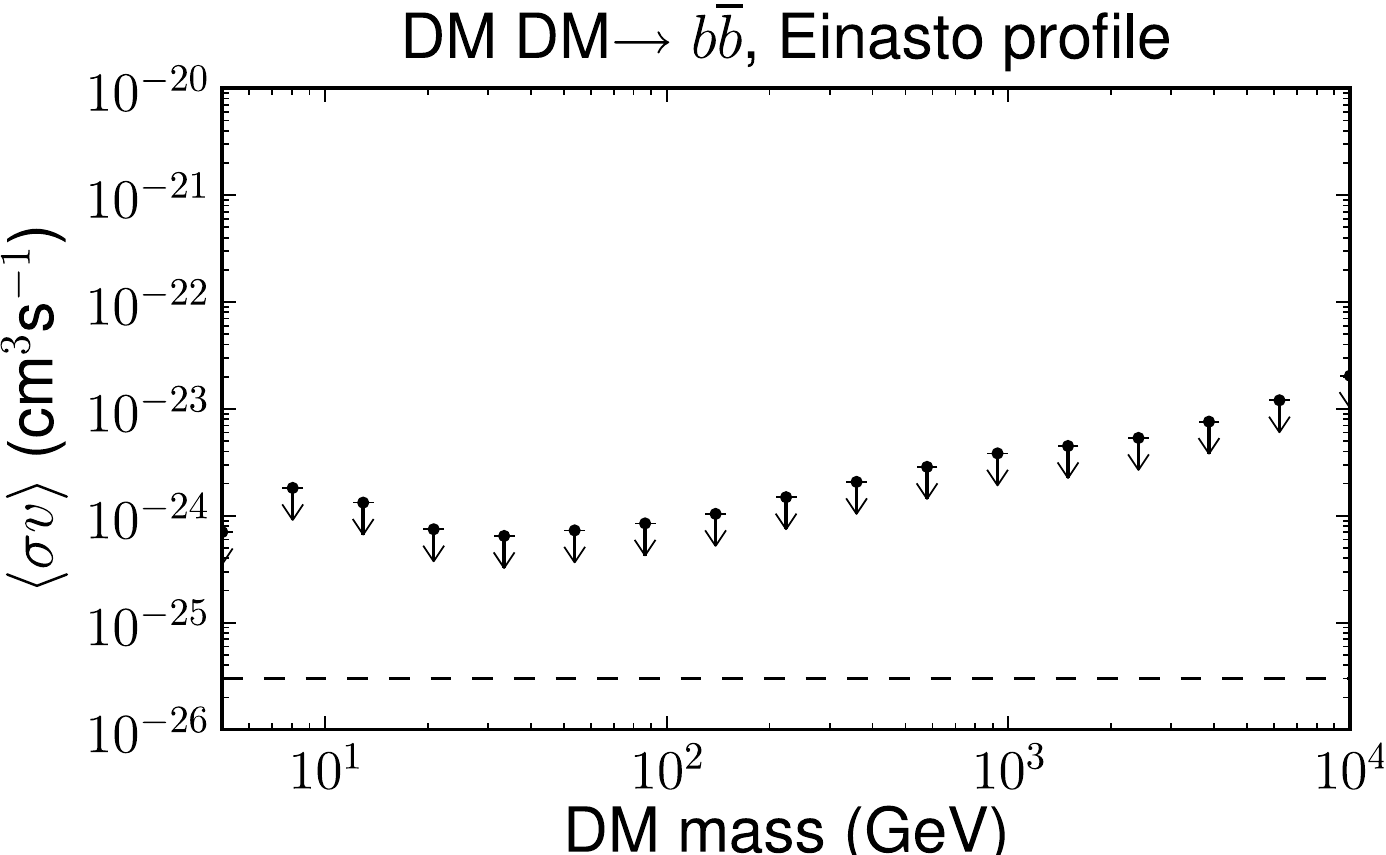}}
\subfigure{\includegraphics[scale=0.58]{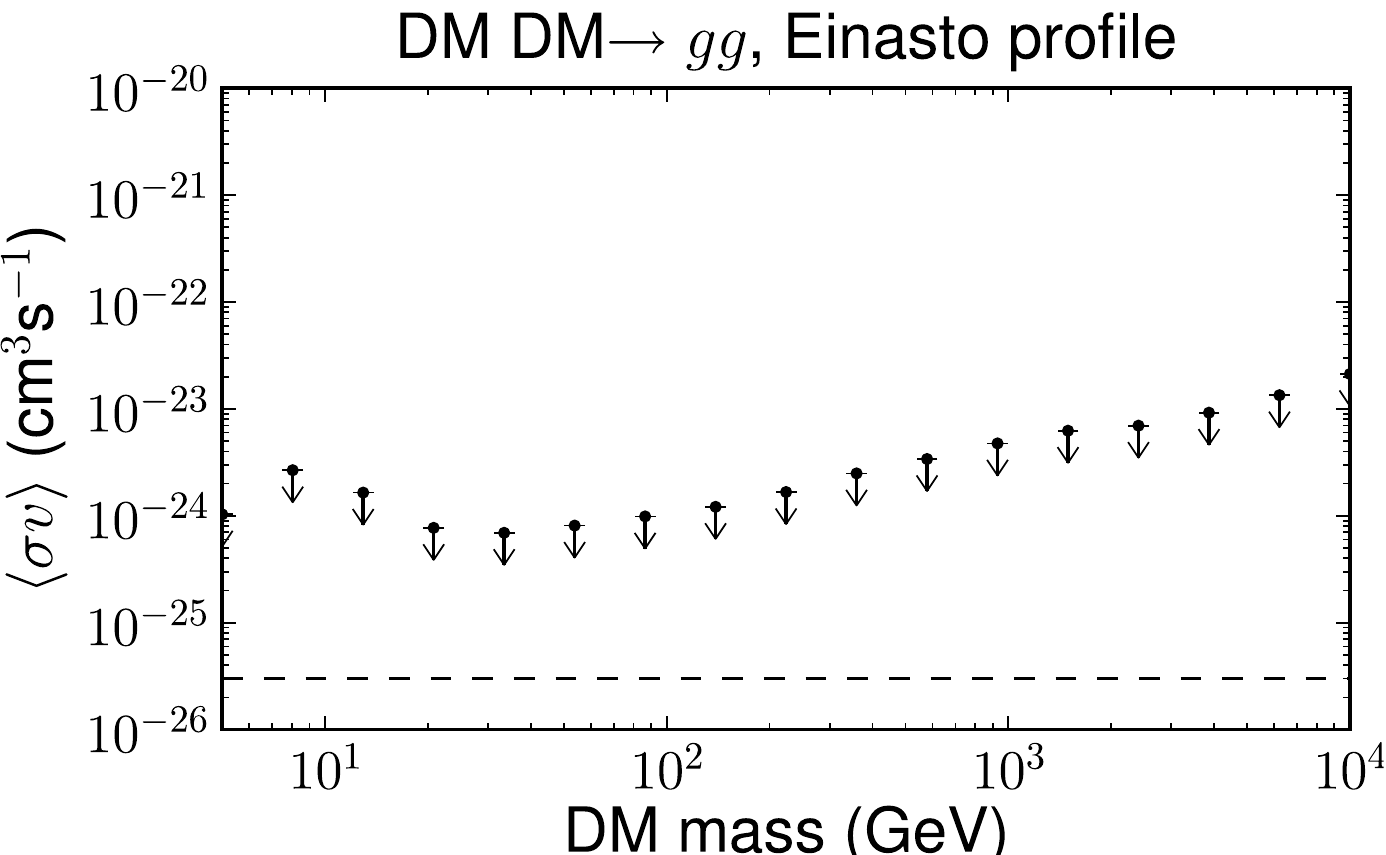}}\\
\vskip 3mm
\subfigure{\includegraphics[scale=0.58]{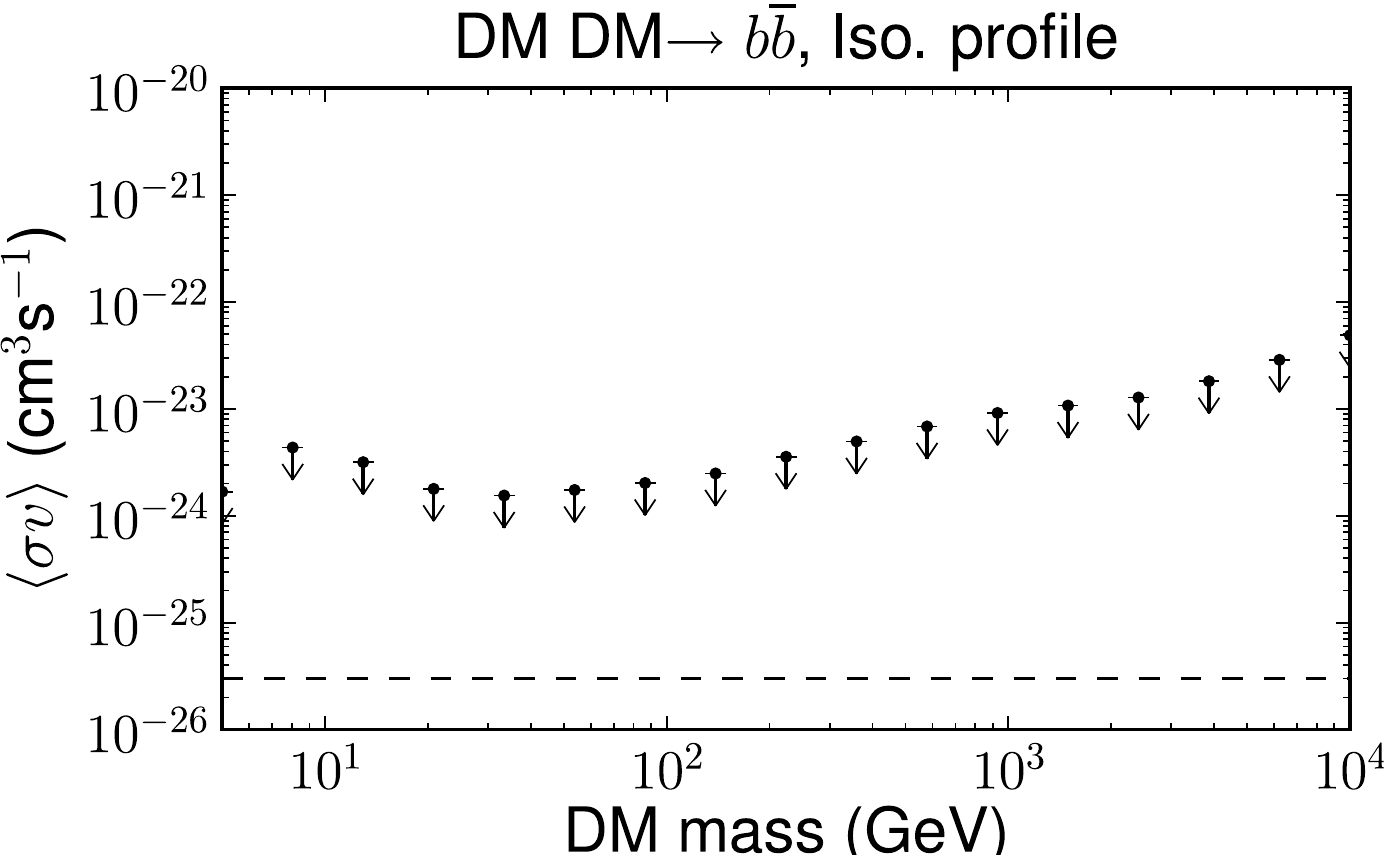}}
\subfigure{\includegraphics[scale=0.58]{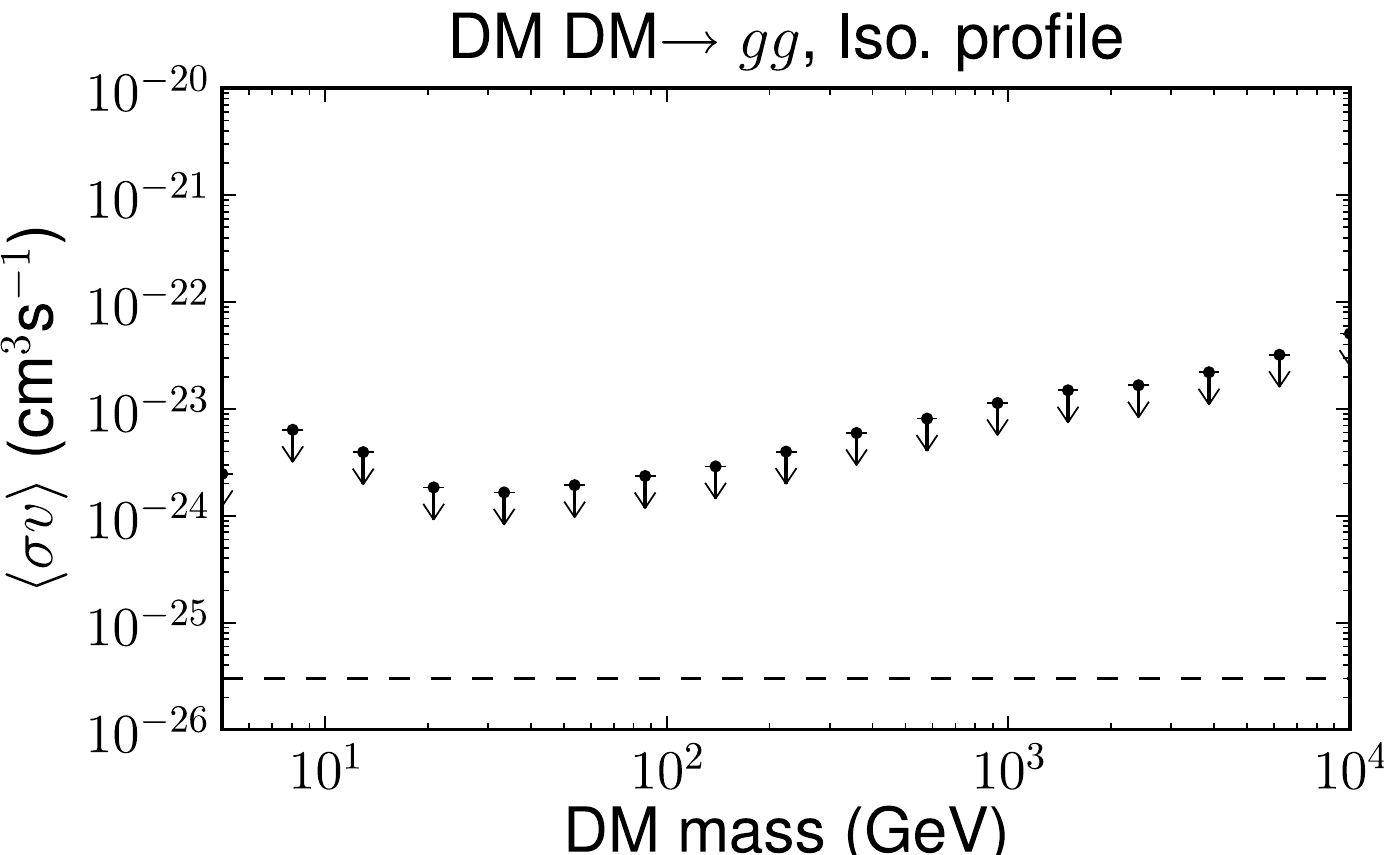}}
\end{center}
\caption{Cross section upper limits for dark matter annihilation to $b$-quarks 
or gluons from the diffuse gamma-ray background spectrum for the region $|b|>10^\circ$ plus a $20^\circ\times20^\circ$ at the GC, assuming the NFW, Einasto, or isothermal dark matter halo profiles.  
No photons from astrophysical background sources have been included, making these limits very conservative.  The horizontal dashed line is the cross section for a thermally produced WIMP 
($3\times 10^{-26}$ cm$^3$s$^{-1}$).
 }
\label{fig:flux_xsection_ul1}
\end{figure*}

\begin{figure*}[t]
\begin{center}
\subfigure{\includegraphics[scale=0.58]{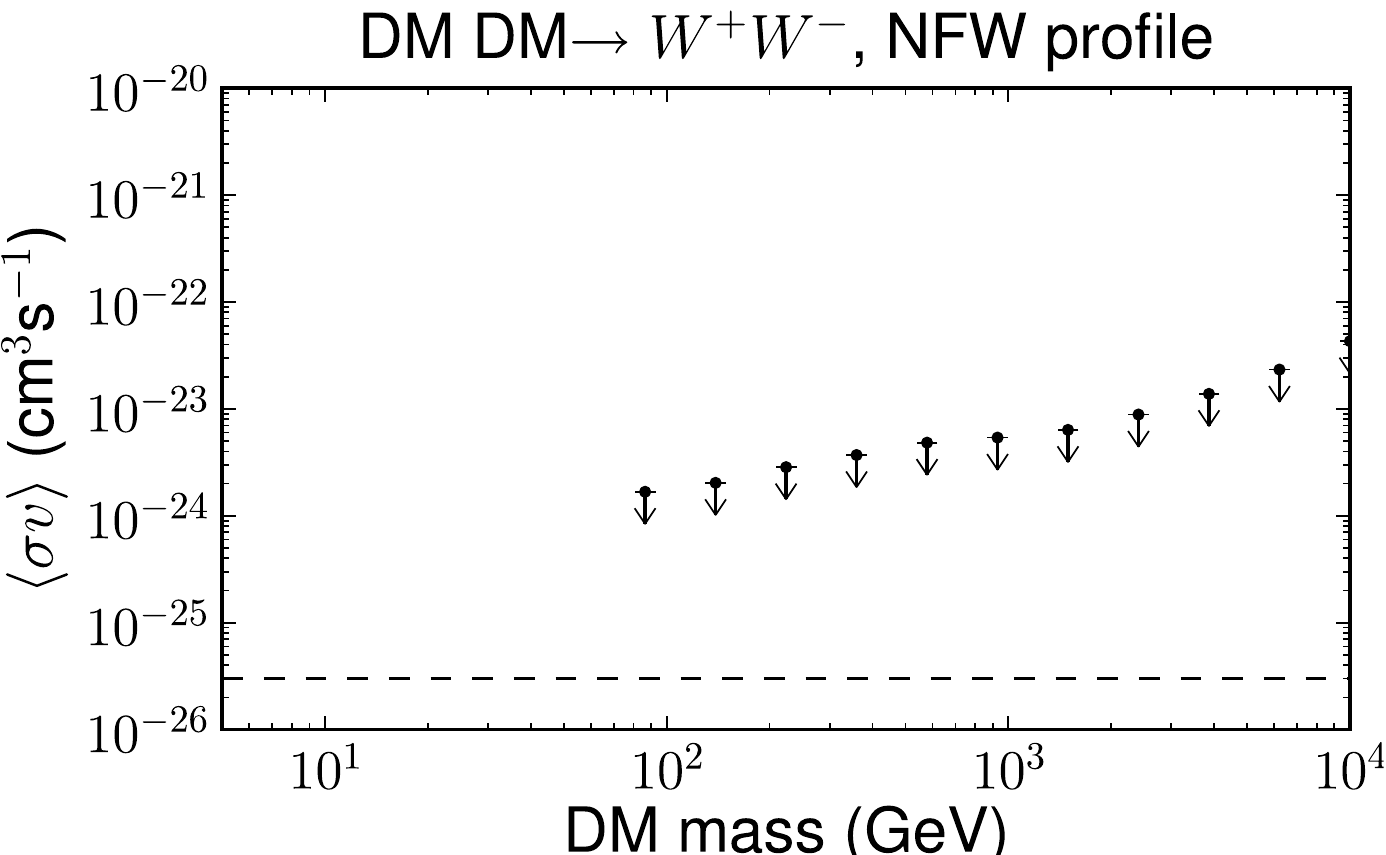}}
\subfigure{\includegraphics[scale=0.58]{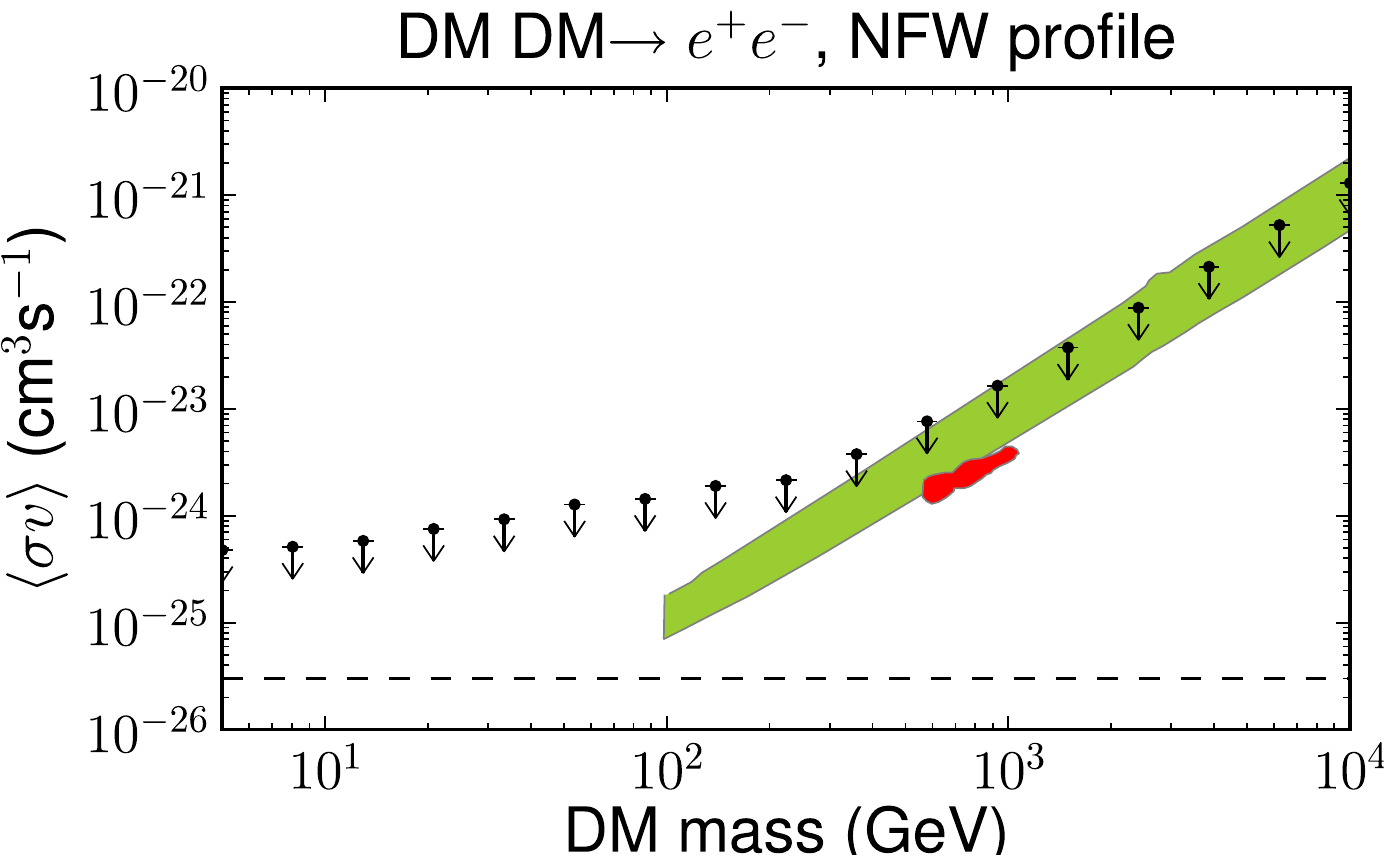}}\\
\vskip 3mm
\subfigure{\includegraphics[scale=0.58]{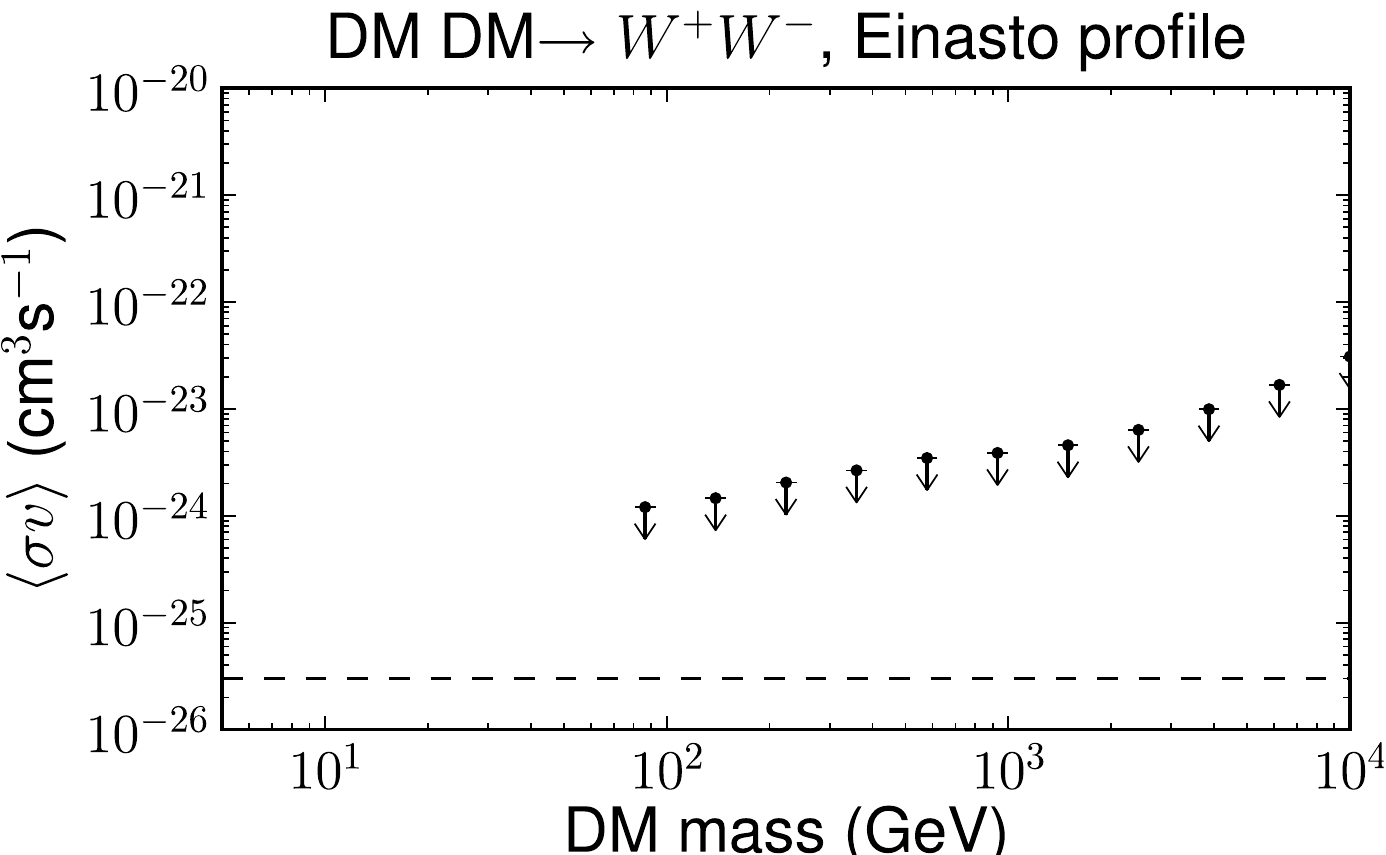}}
\subfigure{\includegraphics[scale=0.58]{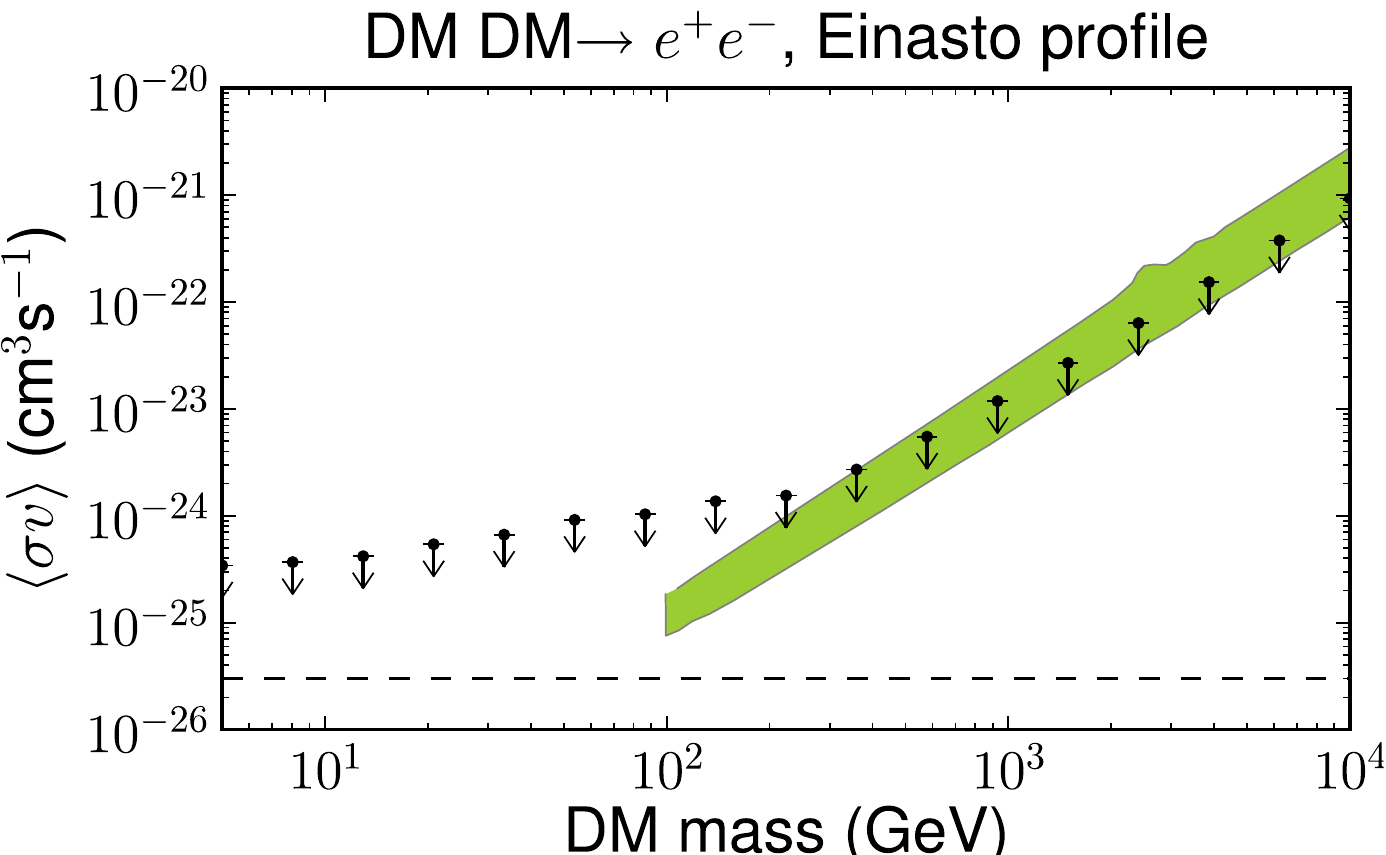}}\\
\vskip 3mm
\subfigure{\includegraphics[scale=0.58]{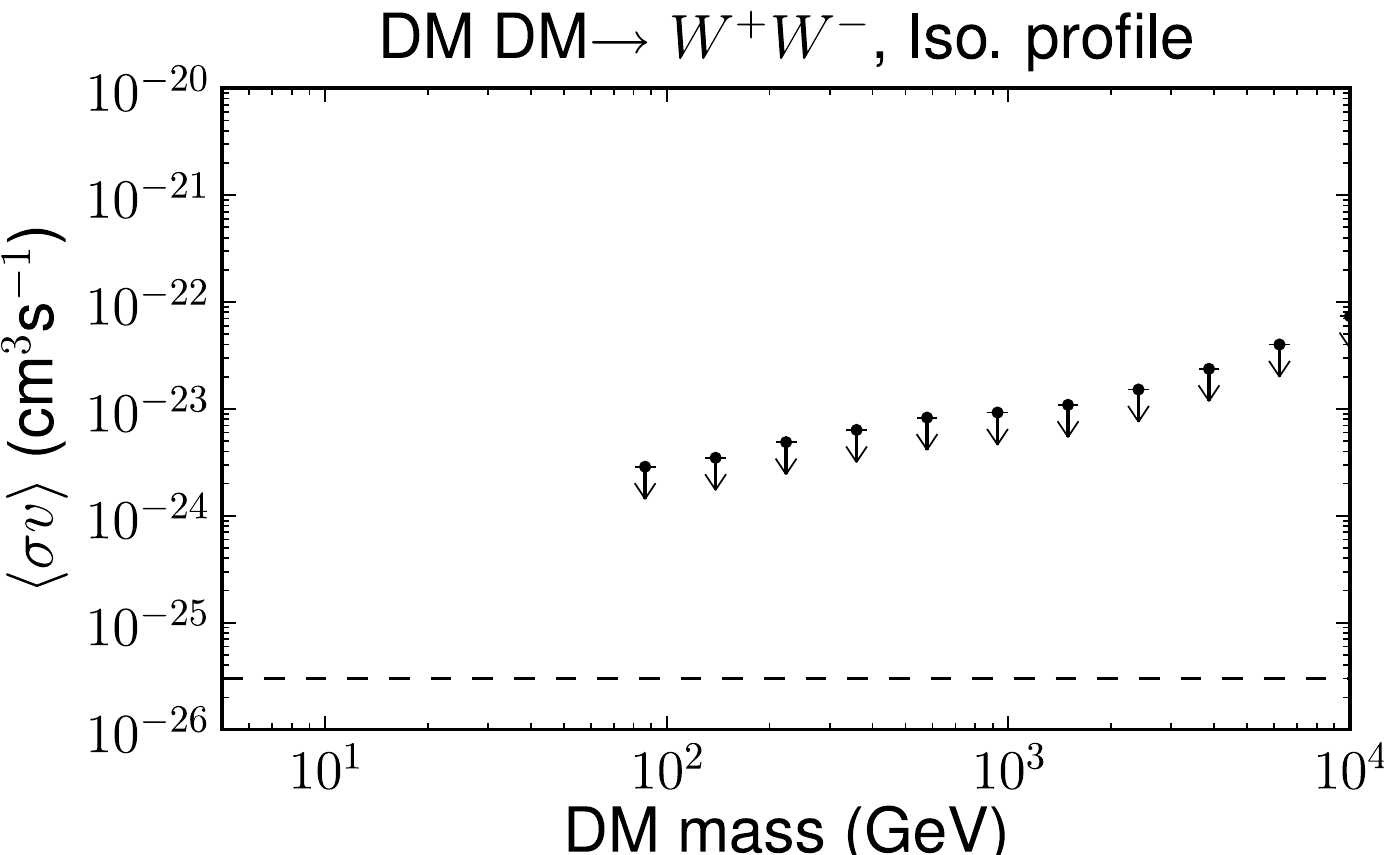}}
\subfigure{\includegraphics[scale=0.58]{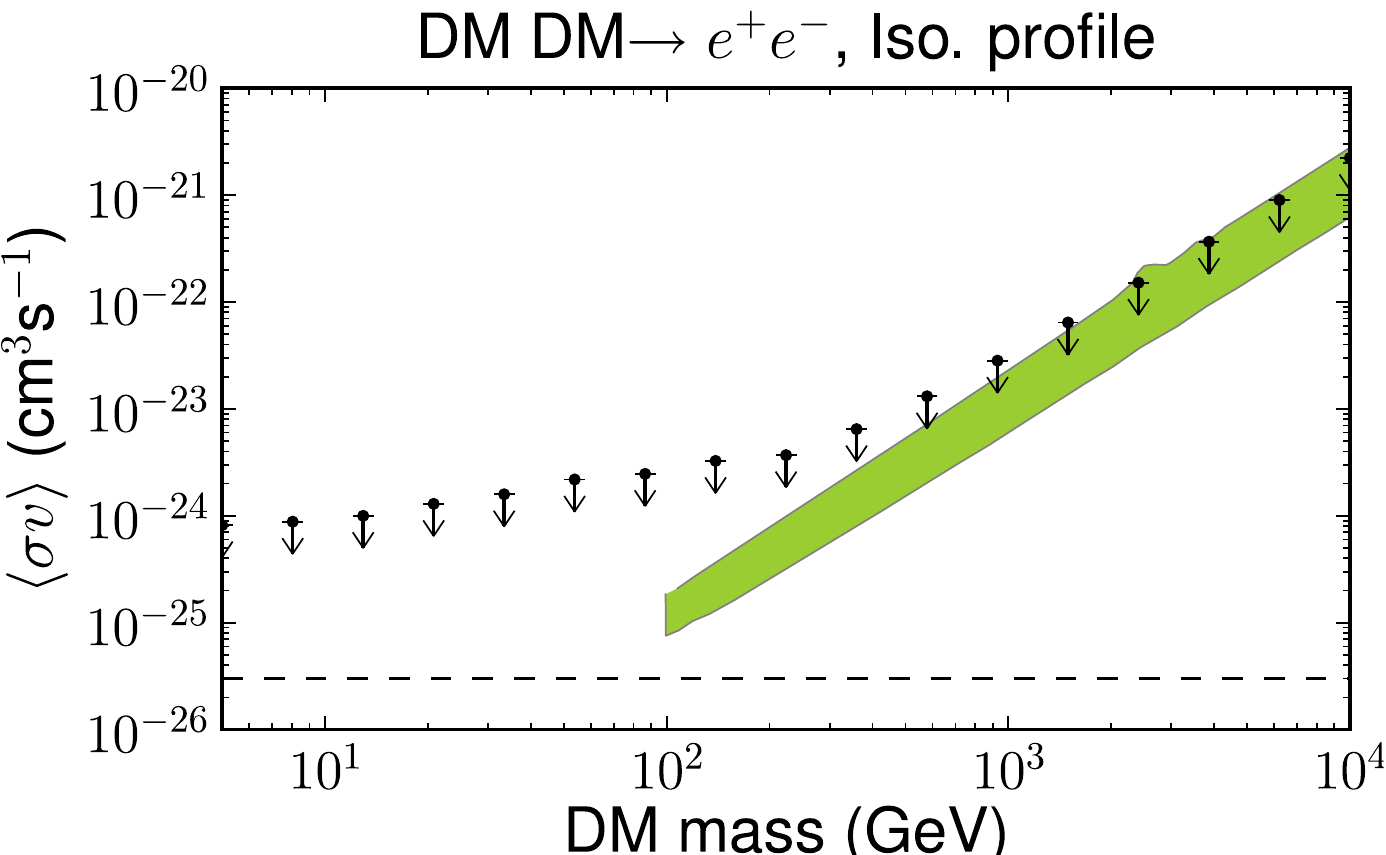}}
\end{center}
\caption{As in Fig.~\ref{fig:flux_xsection_ul1}, but for dark matter annihilation to $W$-bosons or 
$e^+e^-$.  
The annihilation to $e^+e^-$ only includes prompt photons from final states radiation and none from high-energy leptons inverse Compton scattering off starlight or Cosmic Microwave Background photons 
The green and red shaded regions can explain the PAMELA and \emph{Fermi} CR data, respectively. 
 They are taken from \citep{Meade2009}, but we have rescaled their regions by $(3/4)^2$ to a local 
 density of $\rho_{\odot}=0.4$ GeV cm$^{-3}$ from $0.3$ GeV cm$^{-3}$ used in \citep{Meade2009}.
 }
\label{fig:flux_xsection_ul2}
\end{figure*}

\begin{figure*}[t]
\begin{center}
\subfigure{\includegraphics[scale=0.58]{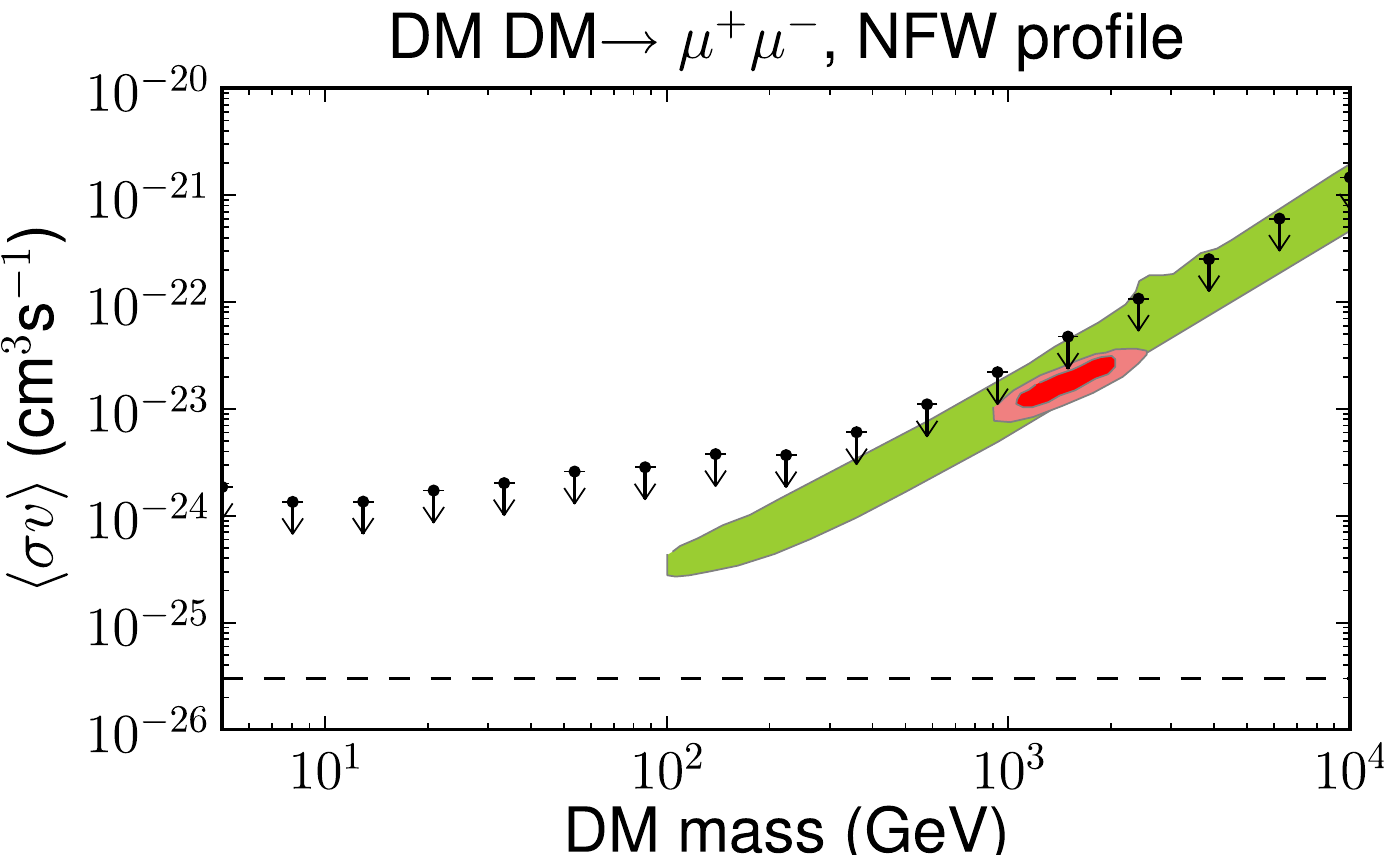}}
\subfigure{\includegraphics[scale=0.58]{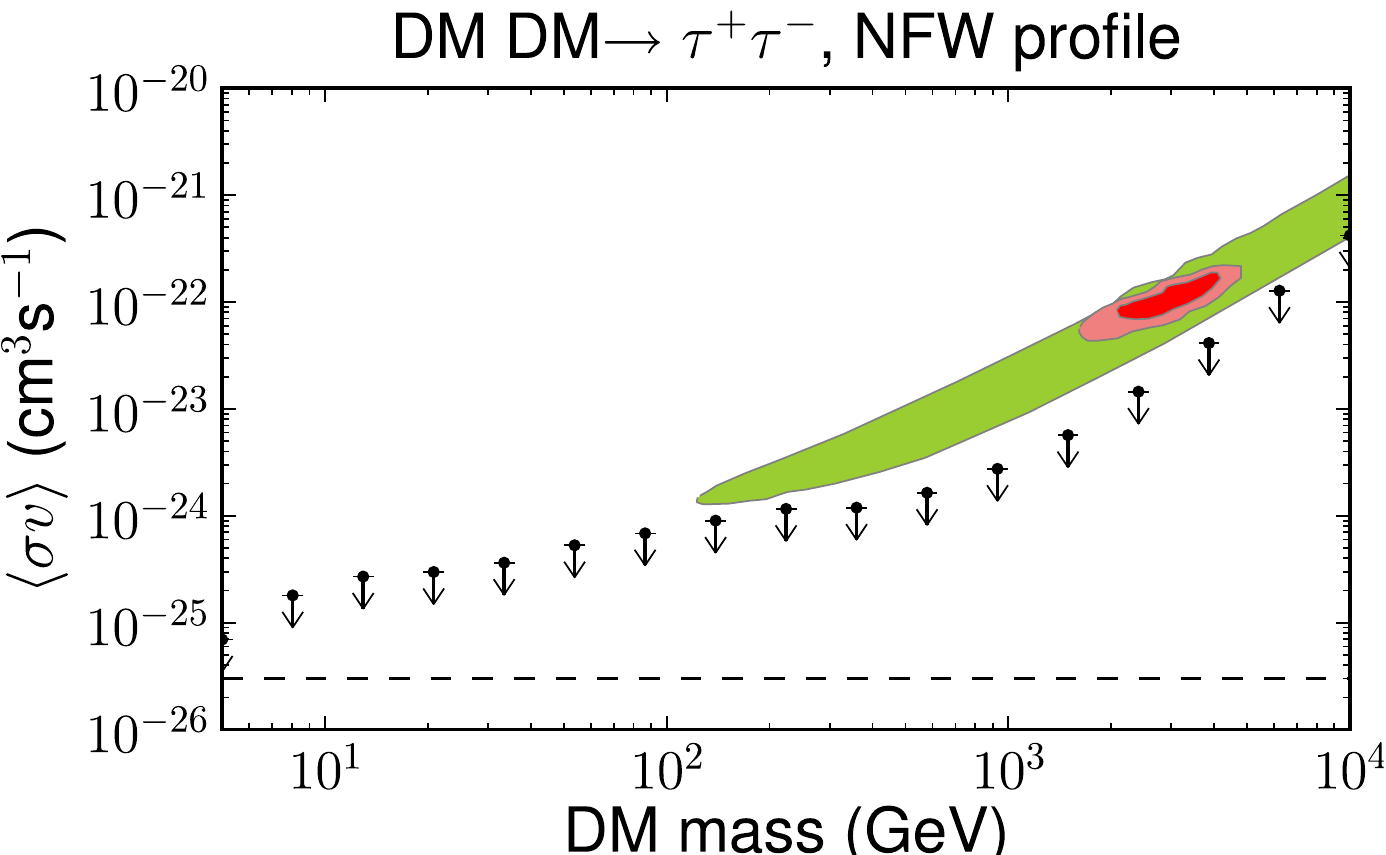}}\\
\vskip 3mm
\subfigure{\includegraphics[scale=0.58]{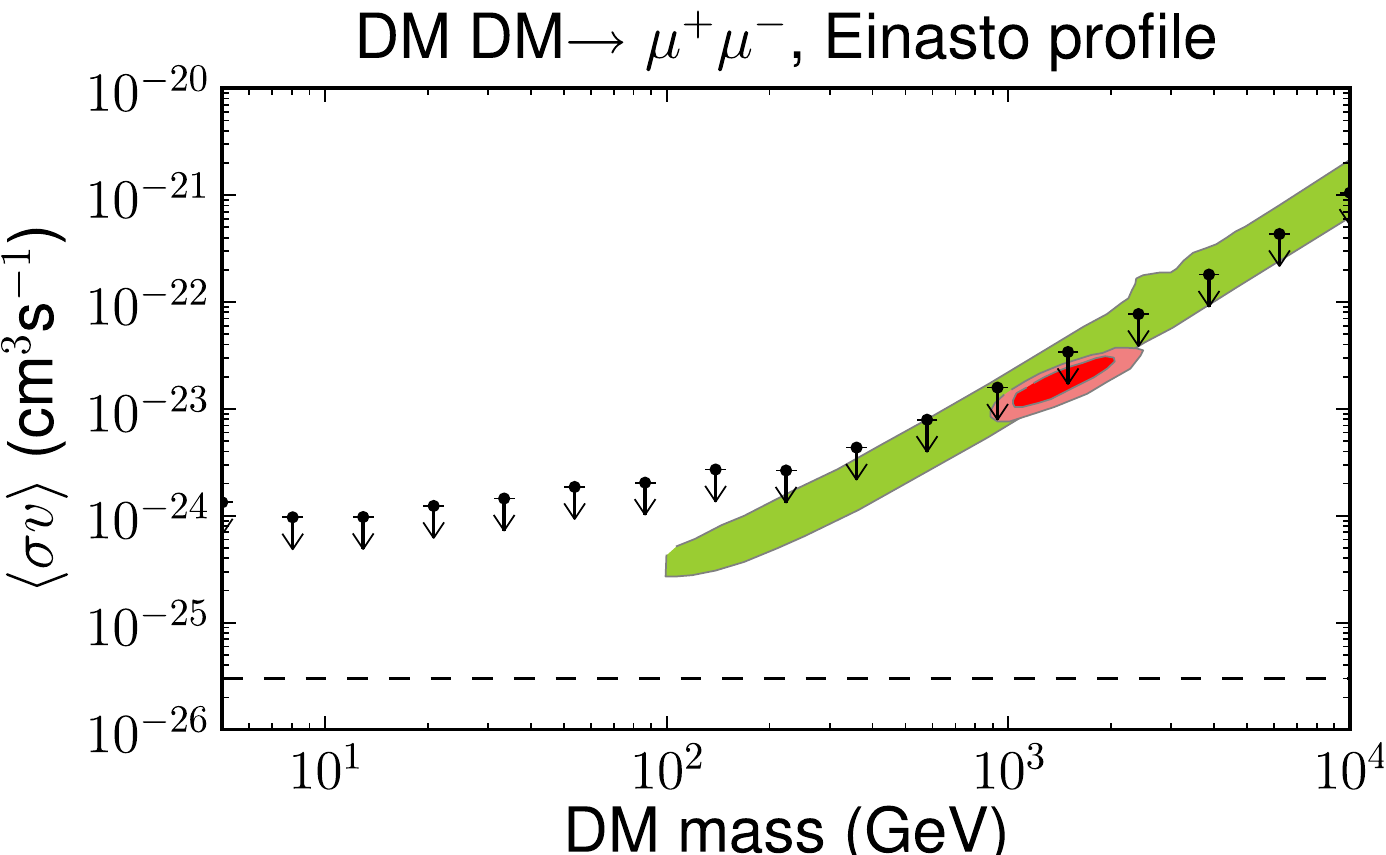}}
\subfigure{\includegraphics[scale=0.58]{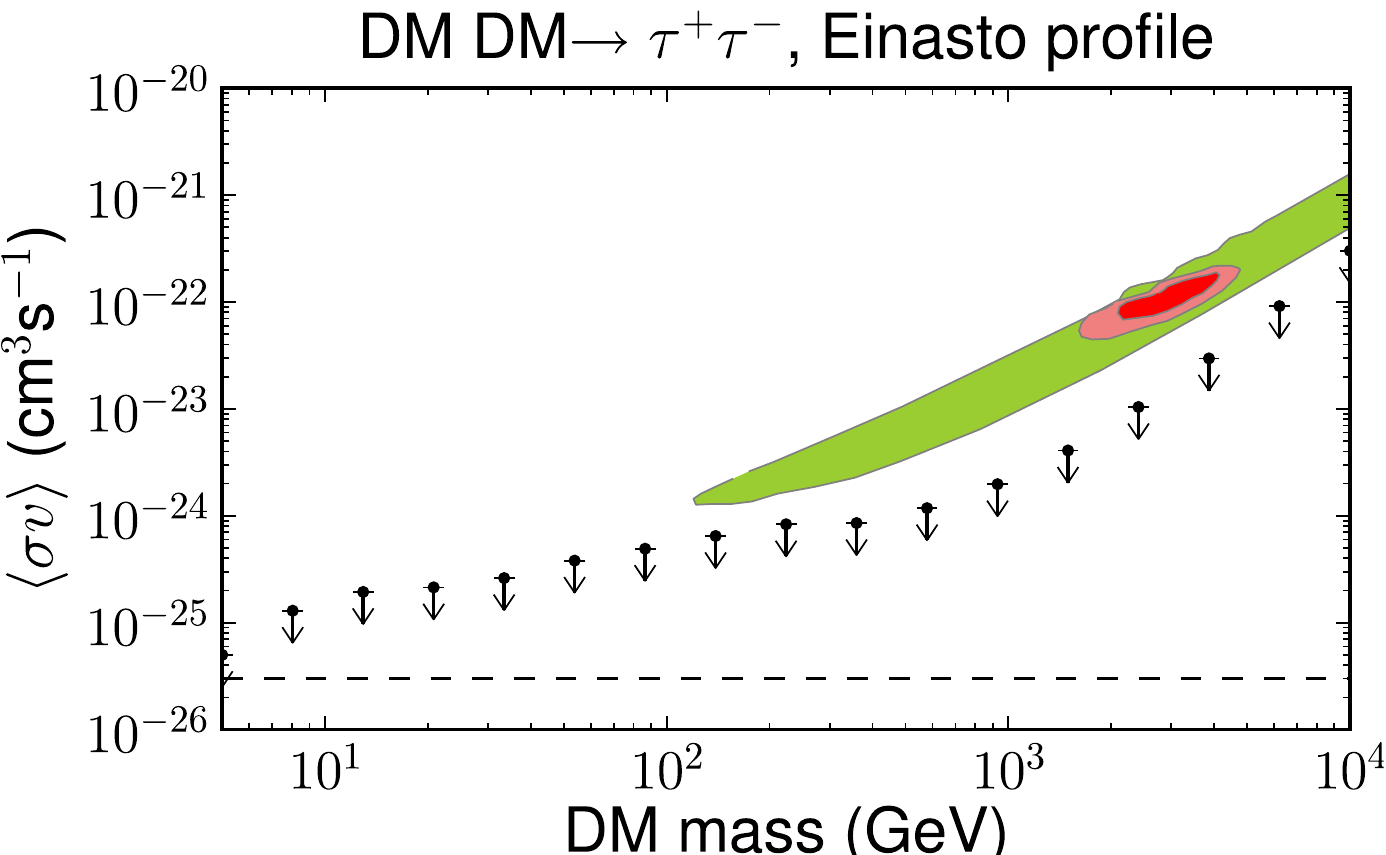}}\\
\vskip 3mm
\subfigure{\includegraphics[scale=0.58]{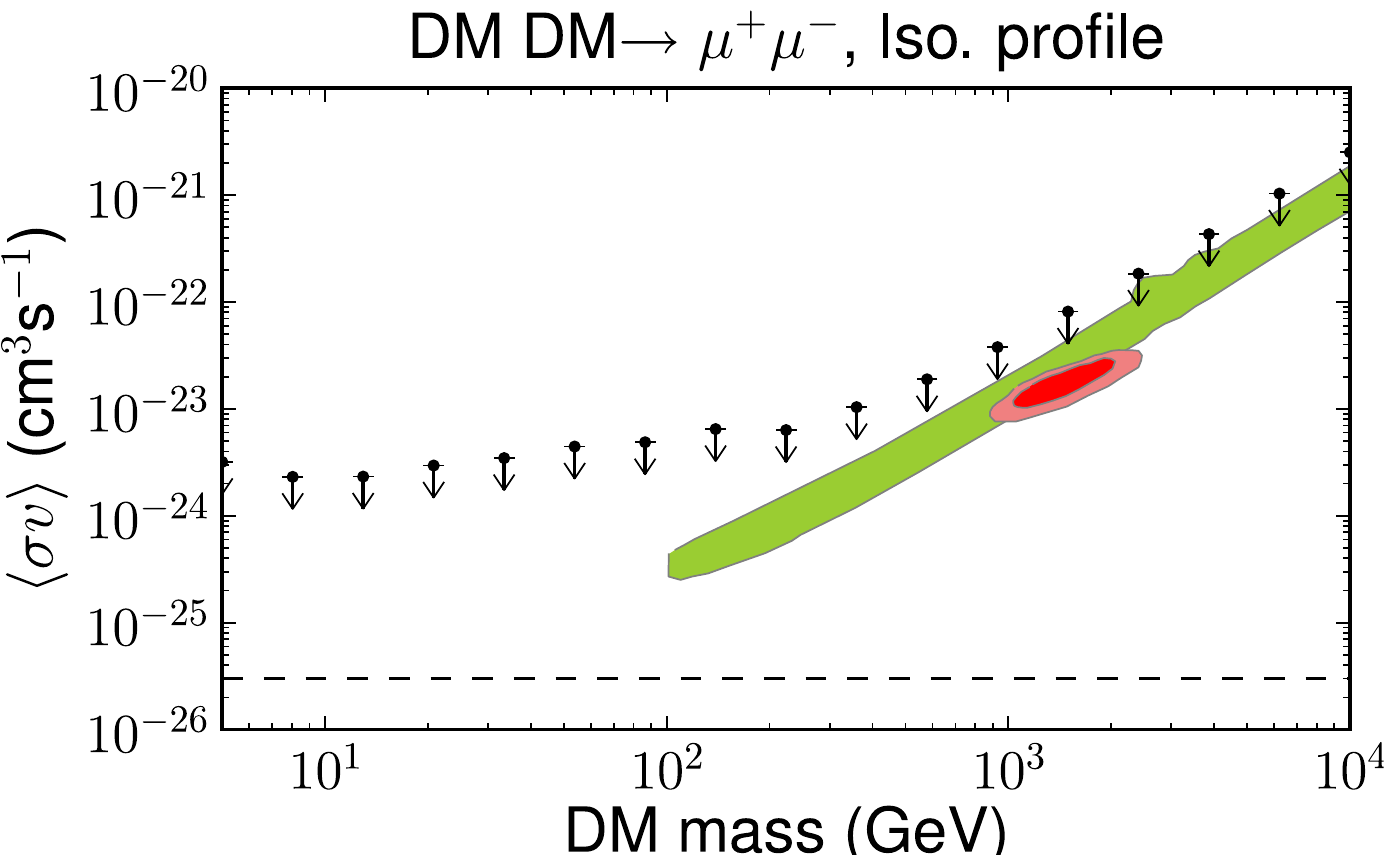}}
\subfigure{\includegraphics[scale=0.58]{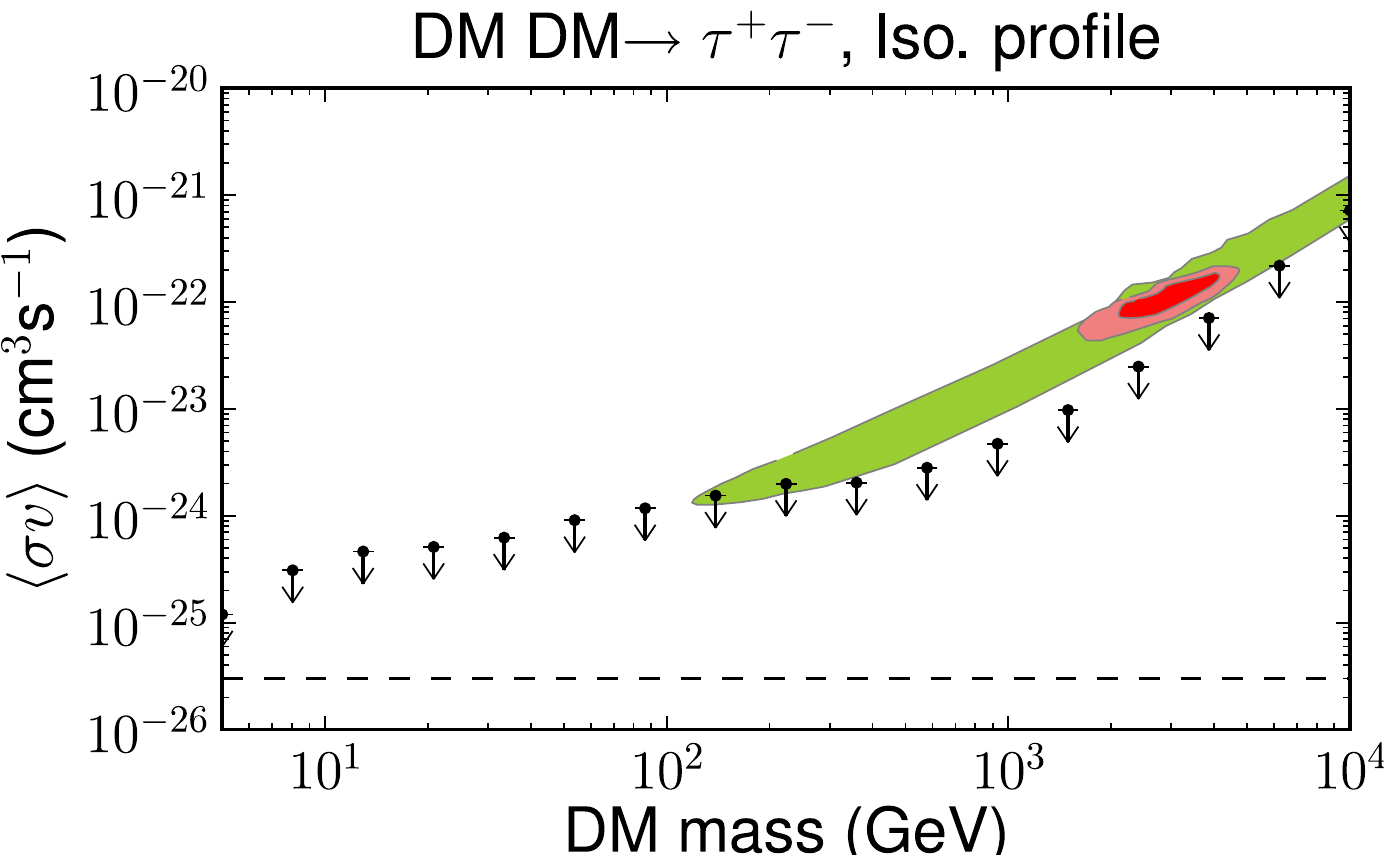}}
\end{center}
\caption{As in Fig.~\ref{fig:flux_xsection_ul1} and \ref{fig:flux_xsection_ul2}, but for dark 
matter annihilation to $\mu^+\mu^-$ or $\tau^+\tau^-$.  
 }
\label{fig:flux_xsection_ul3}
\end{figure*}

\begin{figure*}[t]
\begin{center}
\subfigure{\includegraphics[scale=0.58]{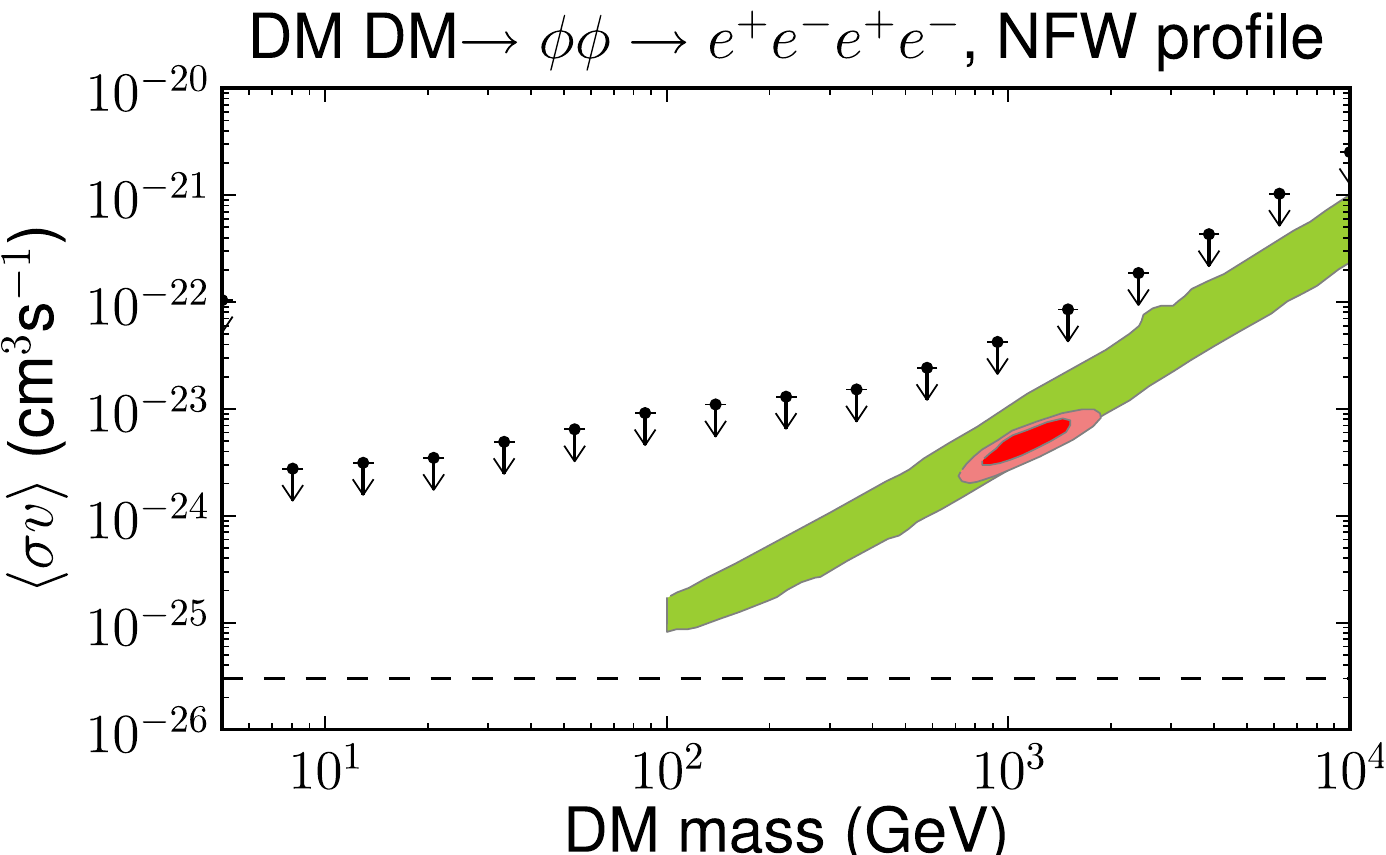}}
\subfigure{\includegraphics[scale=0.58]{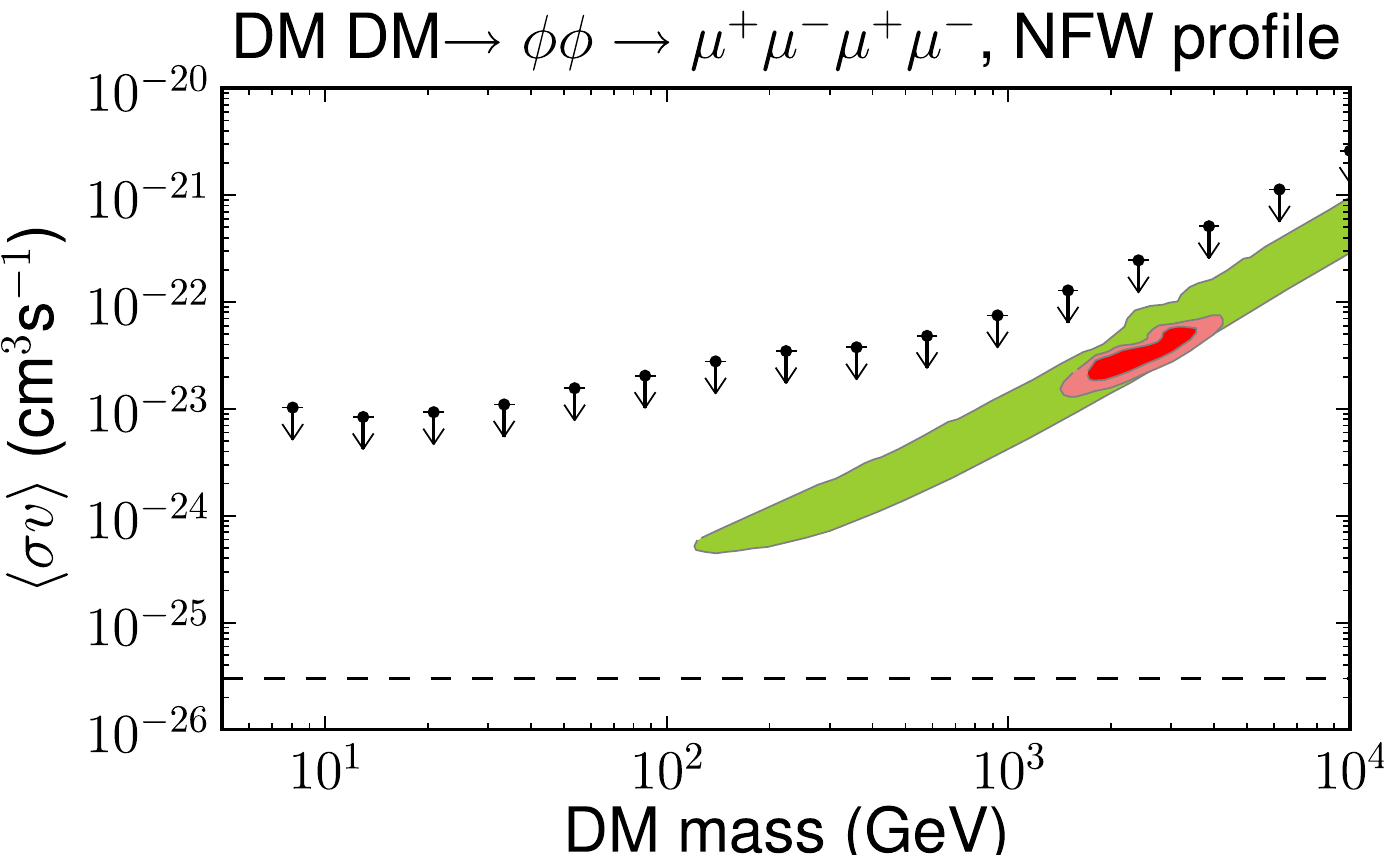}}\\
\vskip 3mm
\subfigure{\includegraphics[scale=0.58]{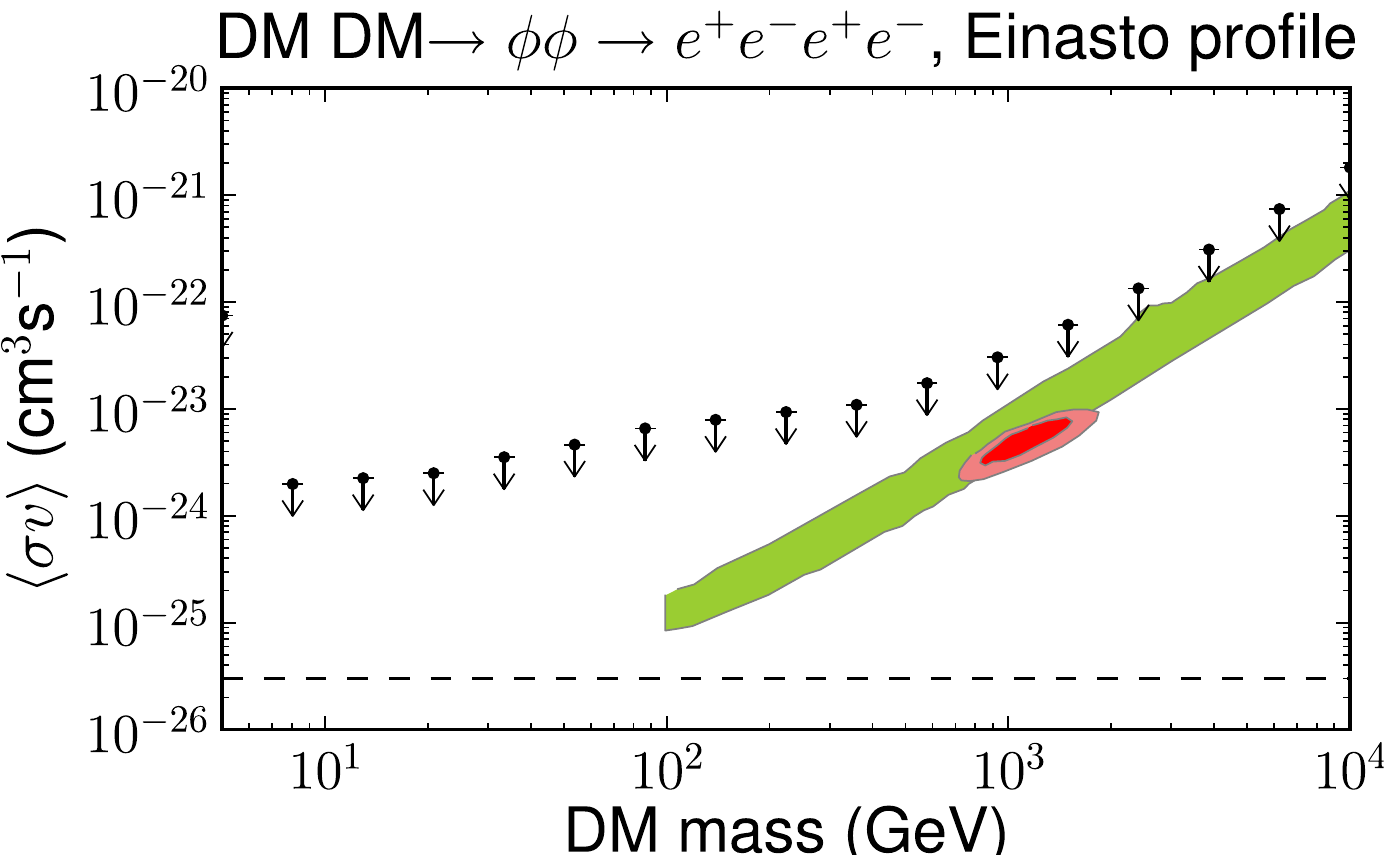}}
\subfigure{\includegraphics[scale=0.58]{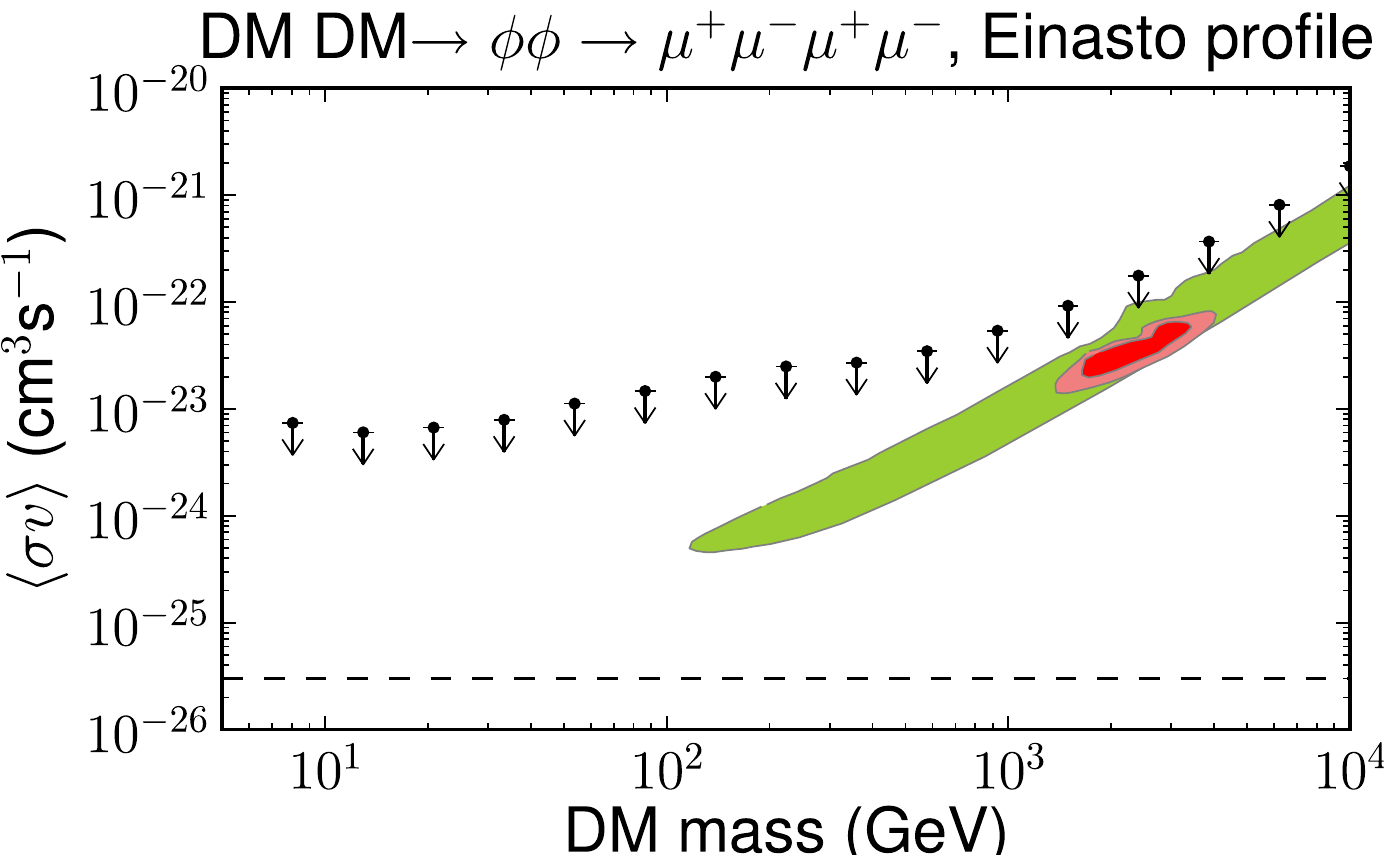}}\\
\vskip 3mm
\subfigure{\includegraphics[scale=0.58]{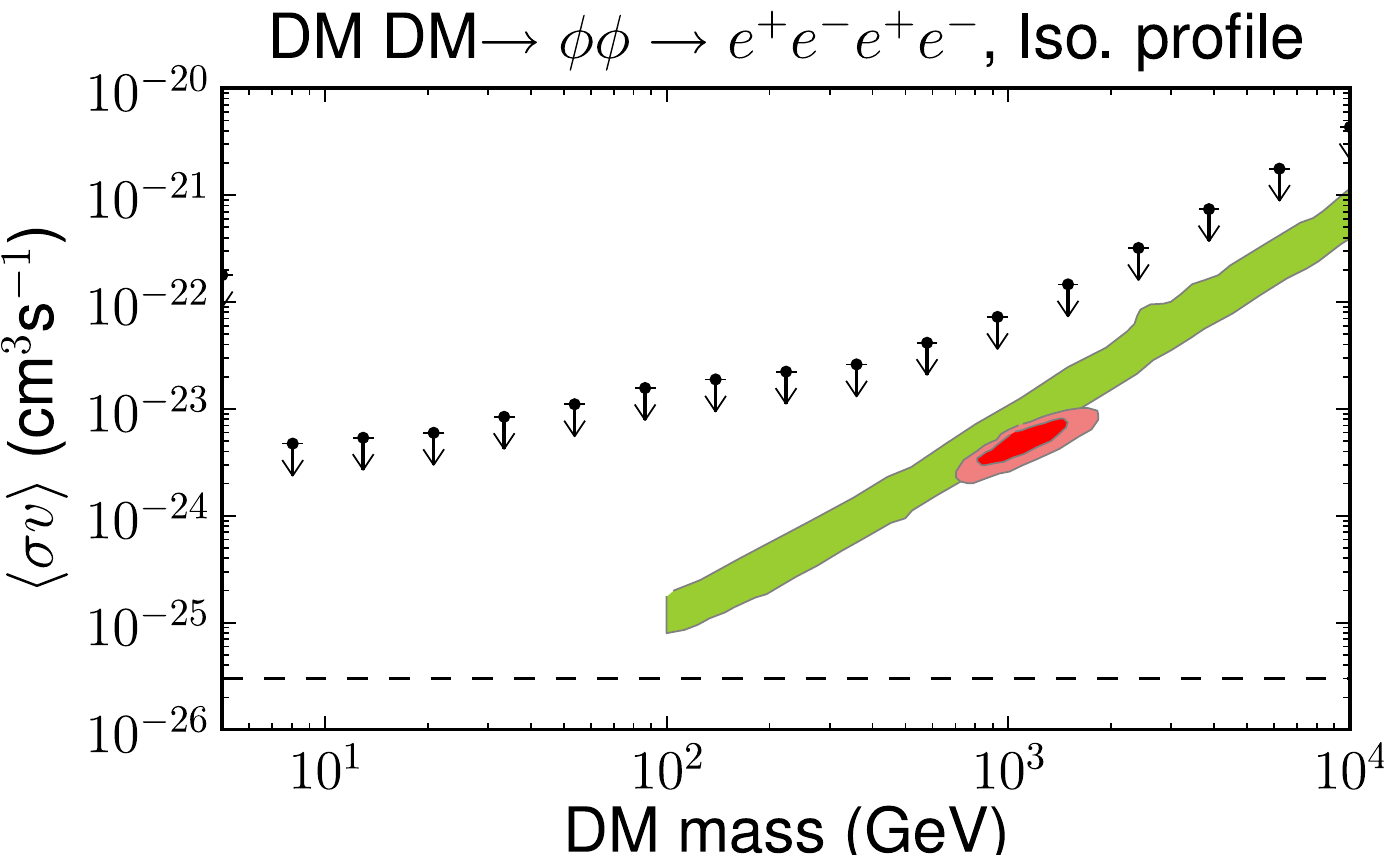}}
\subfigure{\includegraphics[scale=0.58]{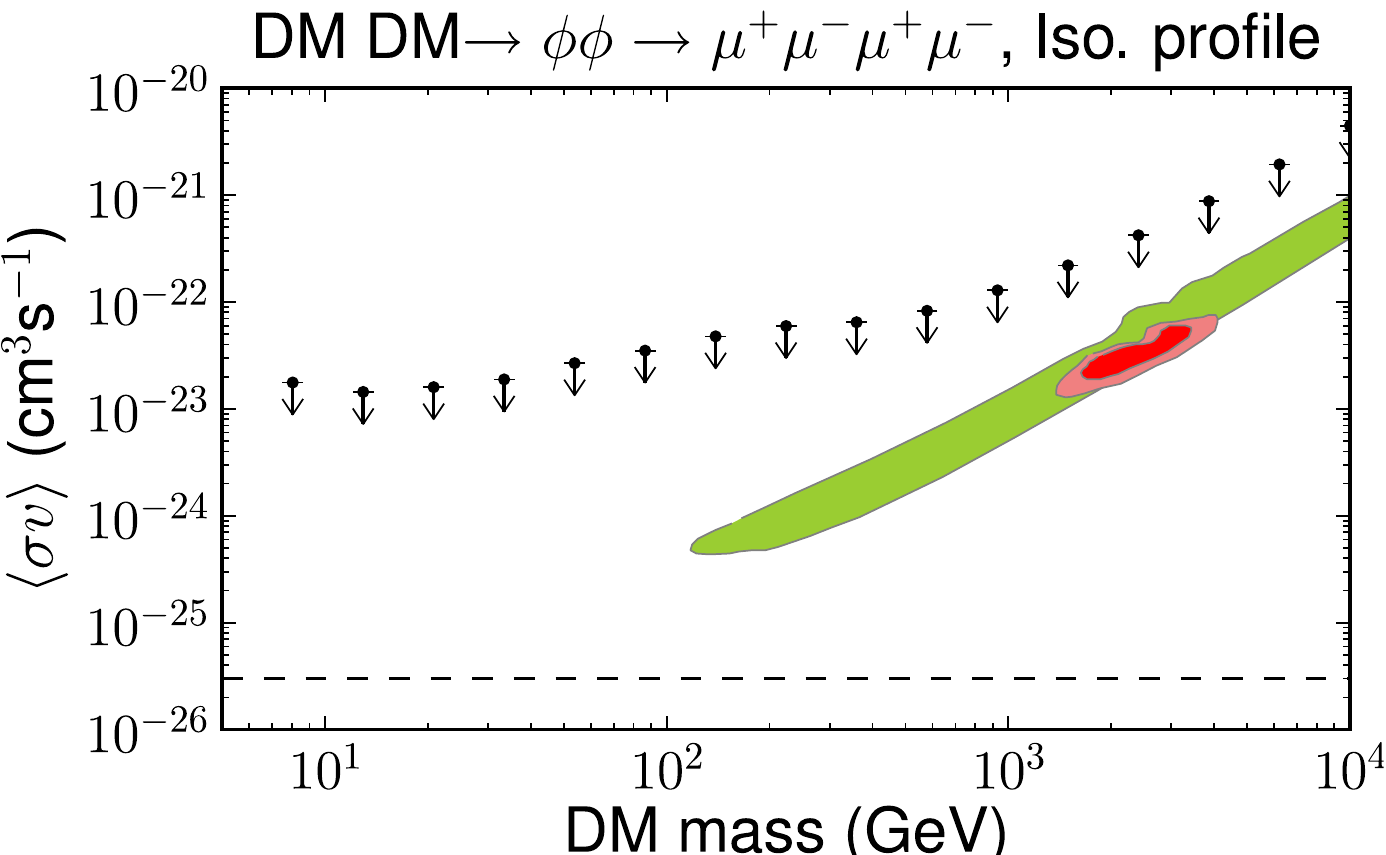}}
\end{center}
\caption{As in Fig.~\ref{fig:flux_xsection_ul1} and \ref{fig:flux_xsection_ul2}, but for dark 
matter annihilation to $\phi\phi$, where $\phi$ is a new low-mass force mediator with mass 
0.1 GeV (1 GeV), and decays to $e^+e^-$ ($\mu^+\mu^-$) in the figures on the left (right). 
 }
\label{fig:flux_xsection_ul4}
\end{figure*}

\begin{figure*}[t]
\begin{center}
\subfigure{\includegraphics[scale=0.58]{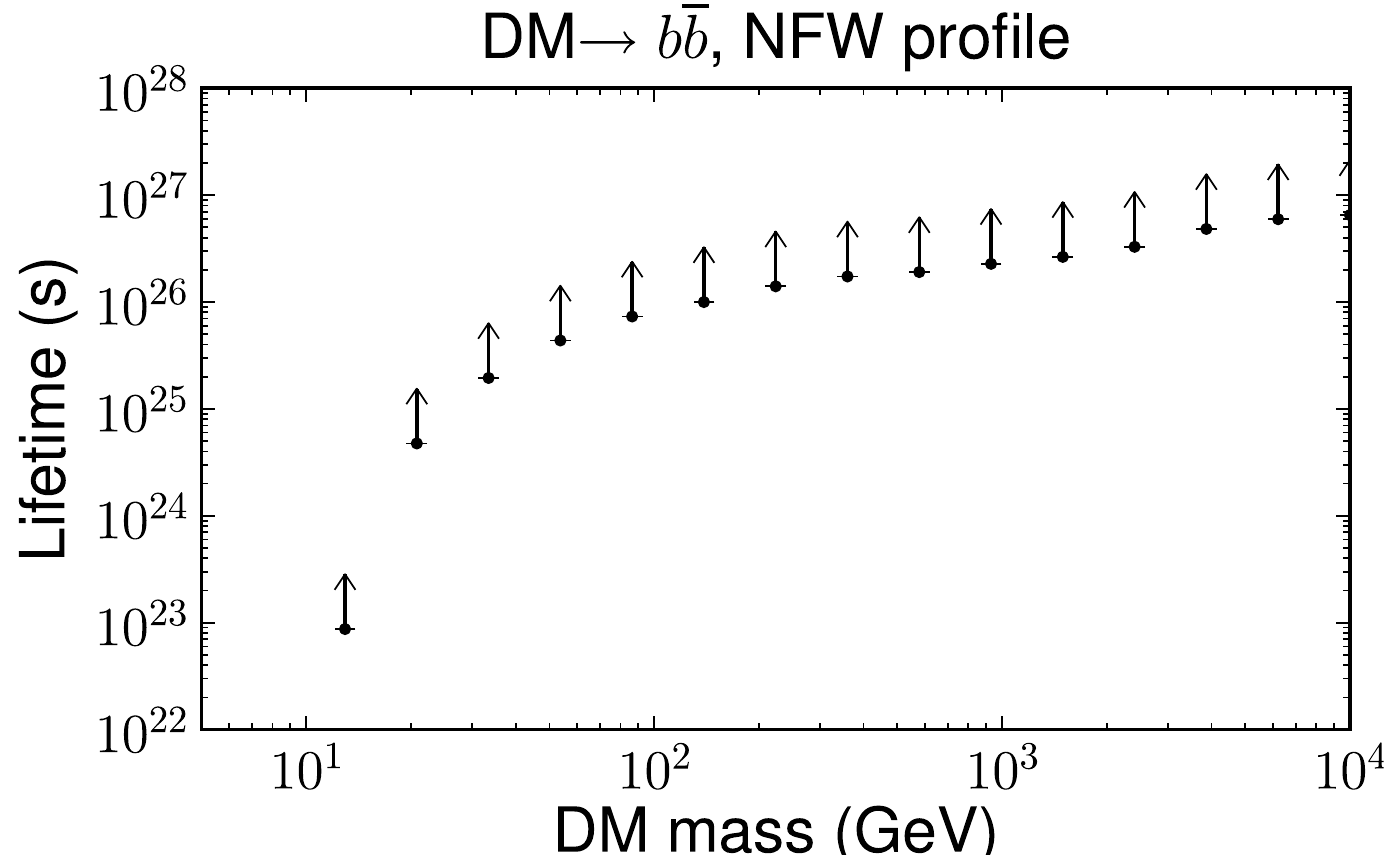}}
\subfigure{\includegraphics[scale=0.58]{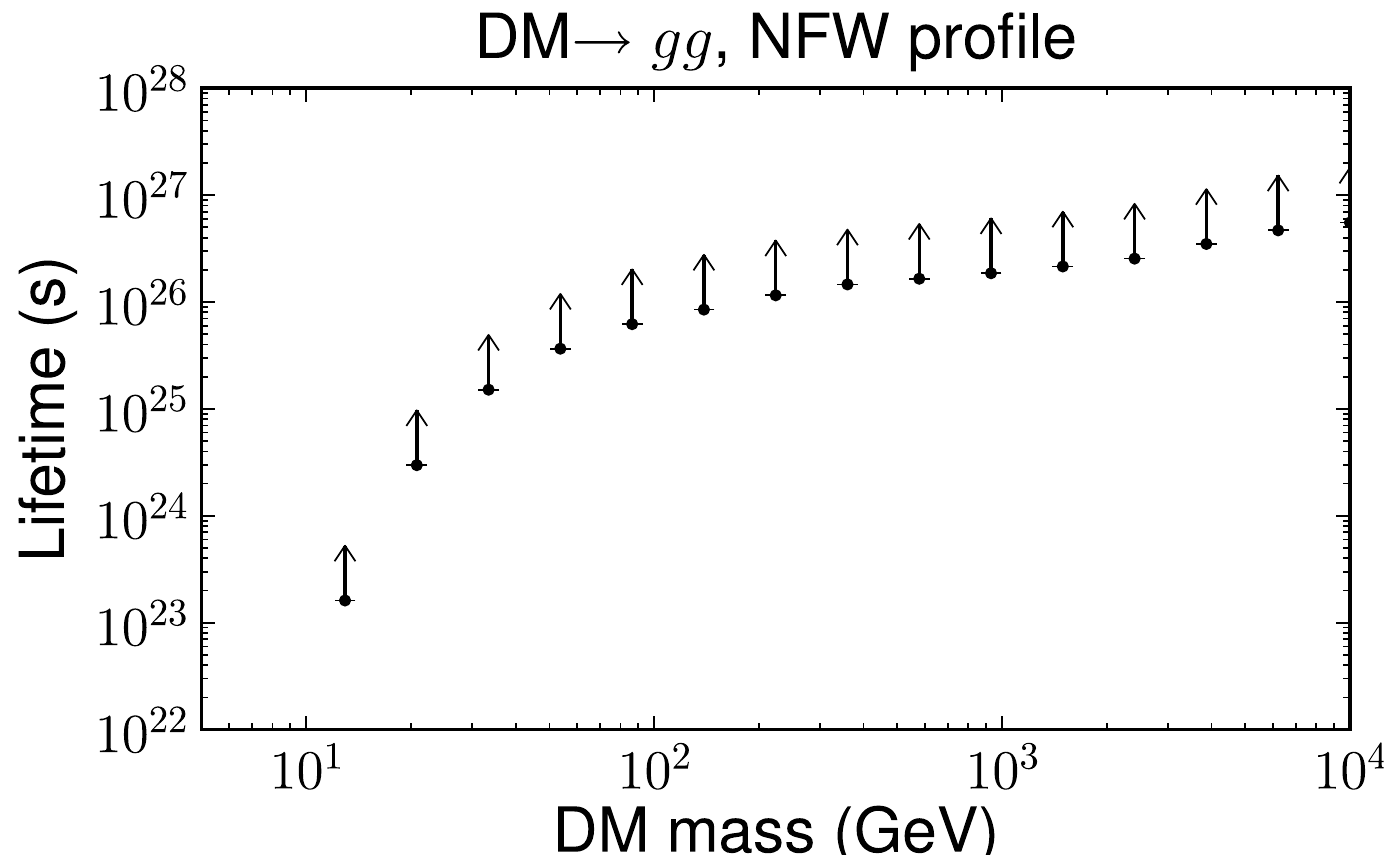}}\\
\vskip 3mm
\subfigure{\includegraphics[scale=0.58]{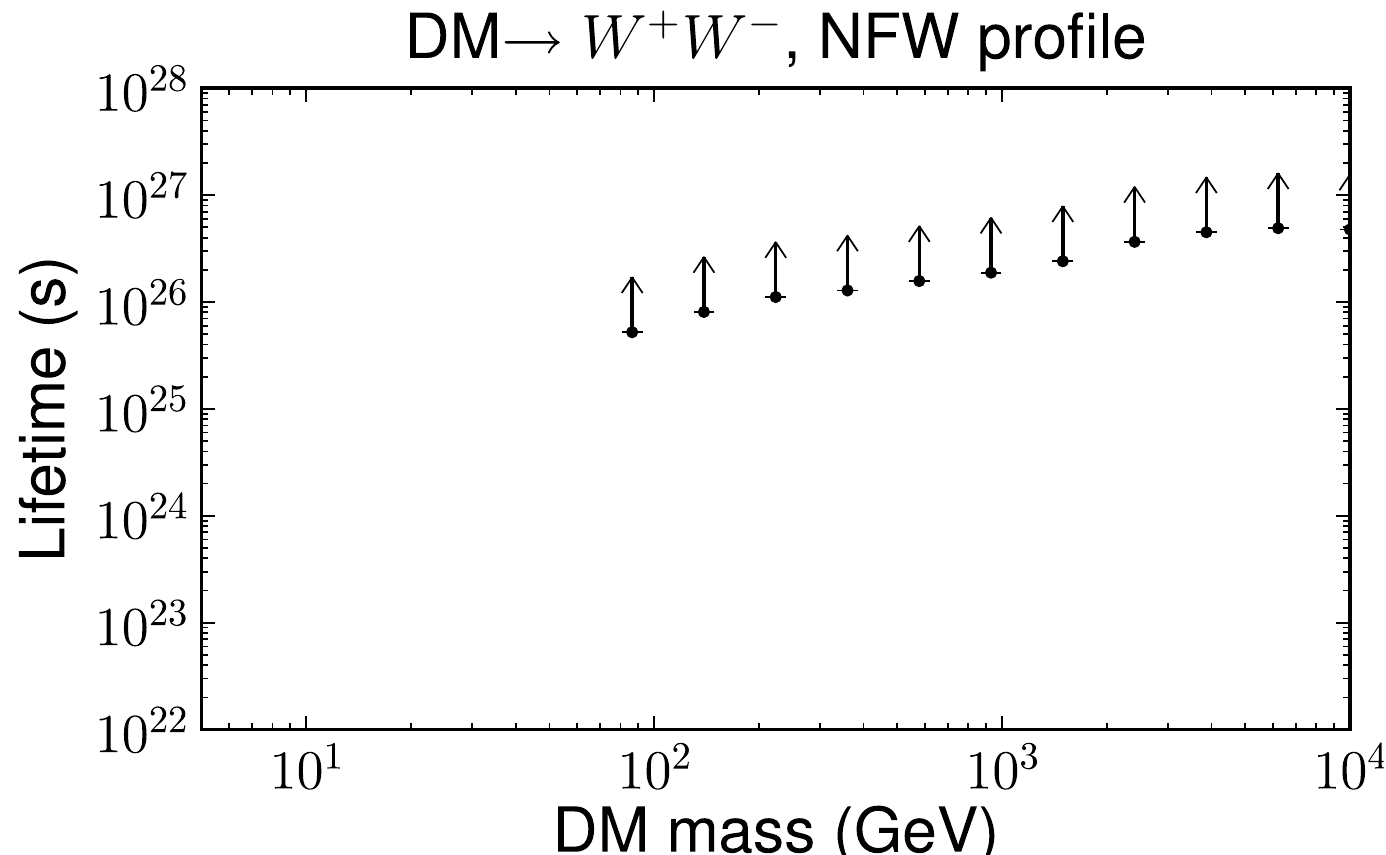}}
\subfigure{\includegraphics[scale=0.58]{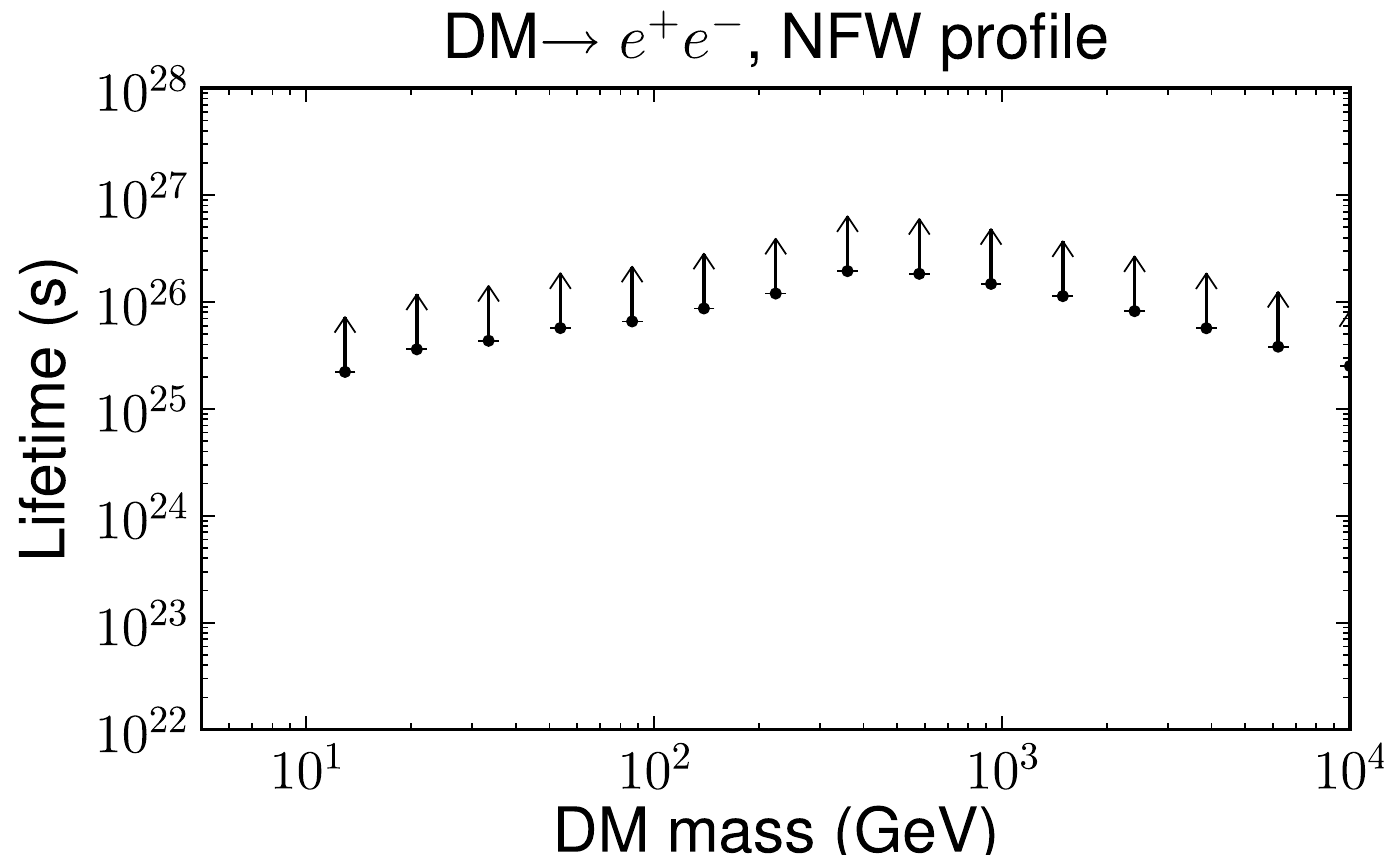}}
\end{center}
\caption{Lifetime lower limits for dark matter decay to $b$-quarks, gluons, $W$-bosons, or $e^+e^-$ 
from the diffuse gamma-ray background spectrum for the region 
$|b|>10^\circ$ plus a $20^\circ\times20^\circ$ at the GC, assuming the NFW dark 
matter halo profile (Einasto and isothermal give very similar constraints).  
No photons from astrophysical background sources have been included, making these limits very conservative.
 }
\label{fig:flux_lifetime_ll1}
\end{figure*}

\begin{figure*}[t]
\begin{center}
\subfigure{\includegraphics[scale=0.58]{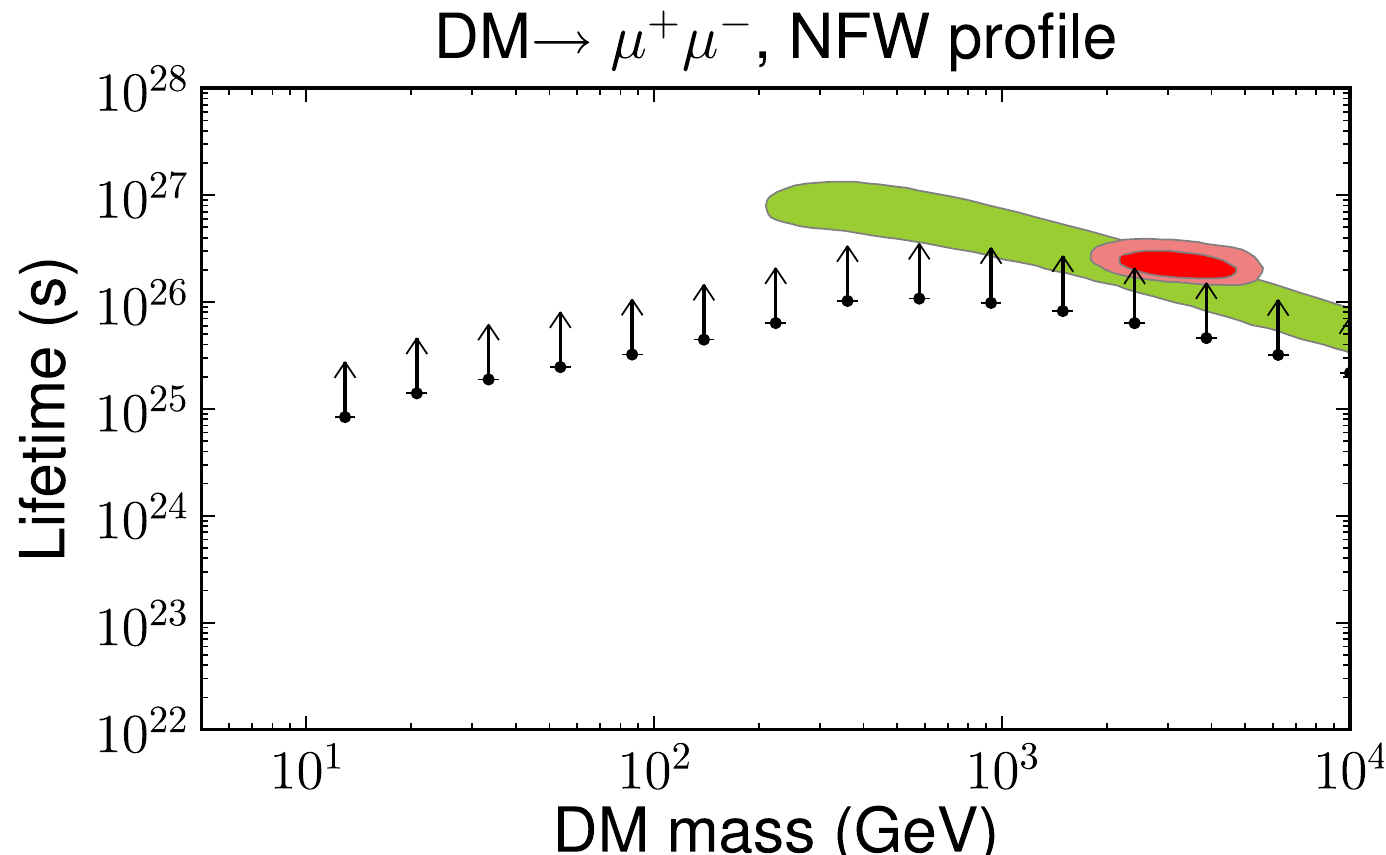}}
\subfigure{\includegraphics[scale=0.58]{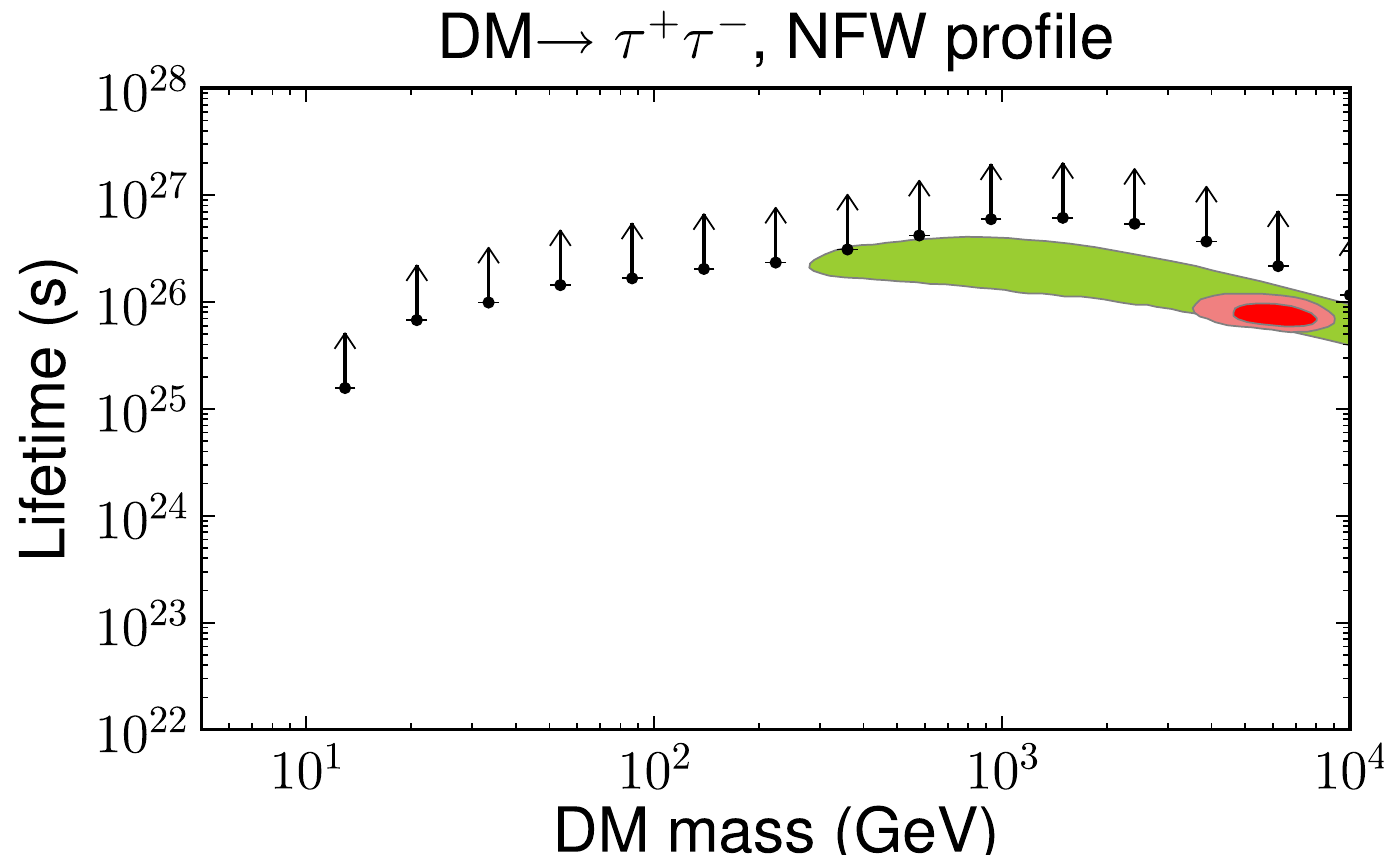}}\\
\vskip 3mm
\subfigure{\includegraphics[scale=0.58]{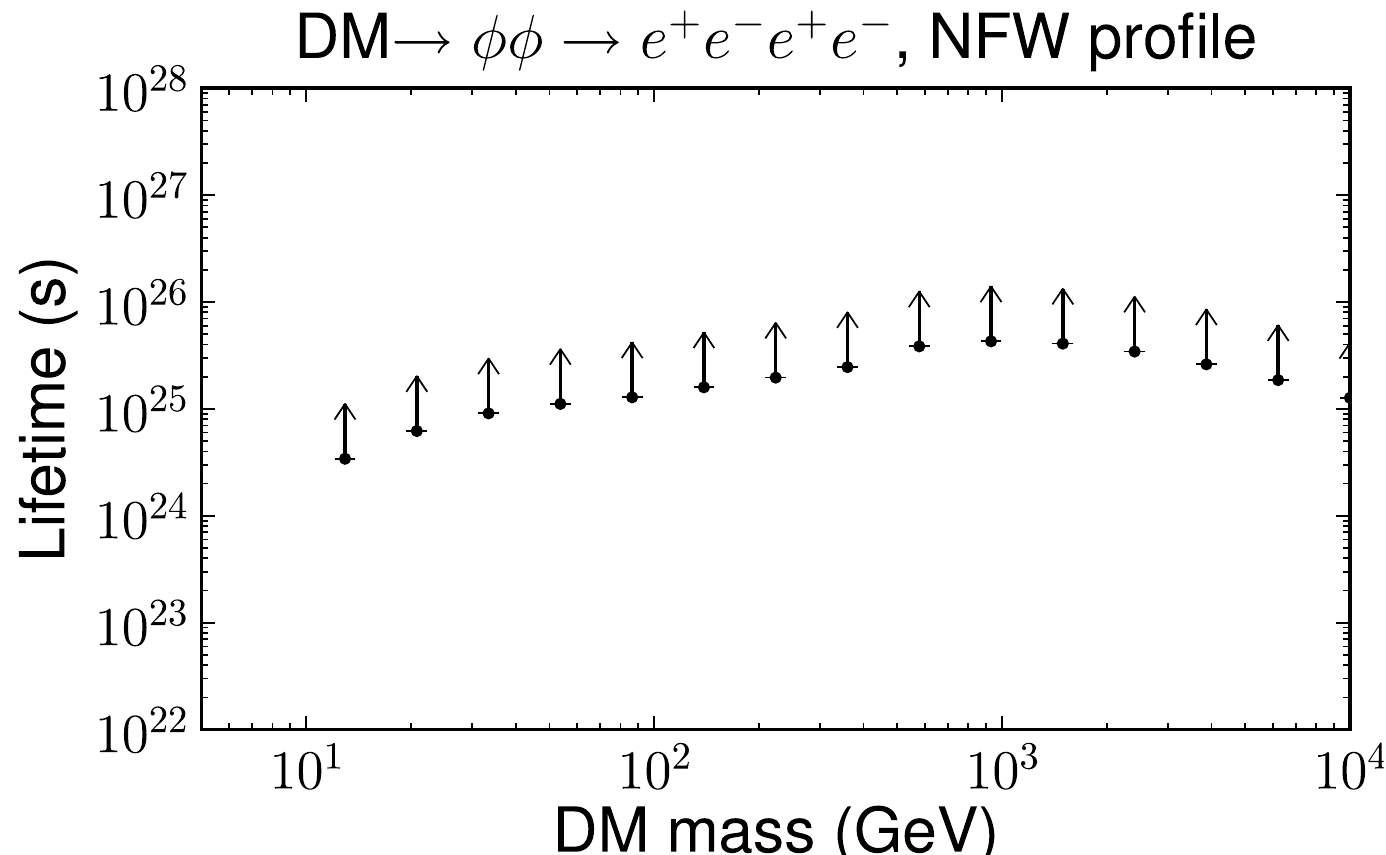}}
\subfigure{\includegraphics[scale=0.58]{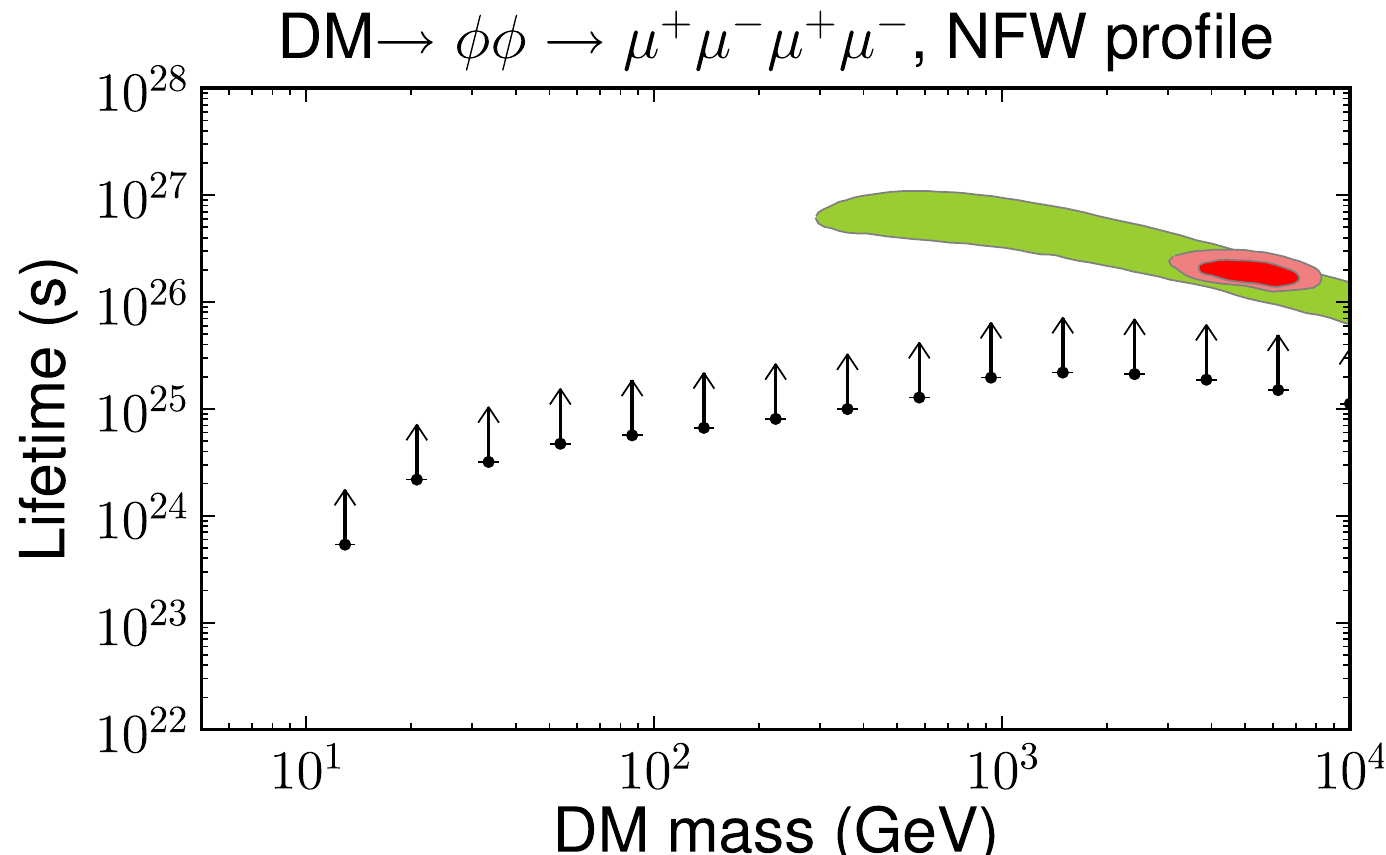}}
\end{center}
\caption{As in Fig.~\ref{fig:flux_lifetime_ll1} but now for dark matter decay to 
$\mu^+\mu^-$, $\tau^+\tau^-$, or $\phi\phi$, where $\phi$ is a new low-mass force mediator with mass 
0.1 GeV (1 GeV), and decays to $e^+e^-$ ($\mu^+\mu^-$) in the figure on the left bottom (right bottom). 
The green and red shaded regions can explain the PAMELA and \emph{Fermi} CR data, respectively. 
They are taken from \citep{Meade2009}, but we have rescaled their regions by $4/3$ to a local 
density of $\rho_{\odot}=0.4$ GeV cm$^{-3}$ from $0.3$ GeV cm$^{-3}$ used in \citep{Meade2009}.
 }
\label{fig:flux_lifetime_ll2}
\end{figure*}

As an example, we show in Fig.~\ref{fig:all_flux_ann_wpwm} the measured intensity and the 
largest allowed flux for WIMP annihilation to $W^+W^-$ for WIMP masses from 86 GeV to 10 TeV.  
Figures \ref{fig:flux_xsection_ul1}--\ref{fig:flux_xsection_ul4} show the annihilation cross 
section upper limits for each WIMP model.  
In the green and red shaded regions (taken from \citep{Meade2009}), dark matter could 
explain the PAMELA and \emph{Fermi} CR data, respectively.  
Since \citep{Meade2009} assume a local density of $\rho_{\odot}=0.3$ GeV cm$^{-3}$, their 
preferred cross section values are multiplied by $(3/4)^2$ to agree with the local density 
used here, $0.4$ GeV cm$^{-3}$.  Figures \ref{fig:flux_lifetime_ll1} and \ref{fig:flux_lifetime_ll2} 
show the lower limits on the decay lifetime.  Here the PAMELA/\emph{Fermi} regions in \citep{Meade2009} 
are scaled by $4/3$.  

We can compare the conservative limits derived here from the inclusive spectrum 
(assuming \emph{no} background photons and \emph{no} ICS contribution) with the 
\emph{Fermi} and Atmospheric Cherenkov Telescope (ACT) constraints from observations 
of Milky Way dwarf spheroidal galaxies 
\cite{Essig:2009jx,Essig:2010em,Abdo:2010ex,Aleksic:2011jx,collaboration:2011wa,GeringerSameth:2011iw} 
and galaxy clusters \cite{Ackermann:2010rg}. 
For a soft photon spectrum obtained from, e.g., annihilation to $b\bar{b}$ (softer than an FSR spectrum), 
the conservative inclusive spectrum constraints are only slightly weaker than the \emph{Fermi} 
stacked dwarf constraints for dark matter masses between $\sim 100 - 1000$ GeV 
(they are significantly weaker below 100 GeV).  
Above 100 GeV, they are stronger than the cluster constraints, assuming no substructure 
enhancement for either.  
The ACT dwarf constraints are weaker for masses up to 
several GeV.  
For a harder FSR spectrum obtained from, e.g., annihilation to $\mu^+\mu^-$, 
the inclusive spectrum constraints are similar to the \emph{Fermi} 
stacked dwarf constraints for masses $\gtrsim 100$ GeV.  The ACT dwarf constraints are only 
available for masses above a few hundred GeV, where they are similar to the 
inclusive spectrum constraints.   The cluster constraints are weaker than the inclusive spectrum 
constraints, again assuming no substructure enhancement for either.  

The limits are strong enough to constrain several interesting dark matter models.  
For example, the wino LSP model \cite{Kane2009} not only produces gamma-ray lines as 
discussed in \S \ref{subsec:DMlines}, but also contributes to the diffuse spectrum through 
annihilation into $W^+W^-$ with a cross section of 
$\sim 2.5 \times10^{-24}$ cm$^3$s$^{-1}$.  This is significantly larger than a thermally 
produced WIMP's cross section of $\sim3\times10^{-26}$ cm$^3$s$^{-1}$.  
Fig.~\ref{fig:flux_xsection_ul2} shows that the limit from the inclusive spectrum 
(assuming no background photons) is of the same order.  

Dark matter models motivated by the PAMELA and \emph{Fermi} CR data produce relatively 
hard photons from final state radiation (FSR).  The figures show that in the annihilation case, 
these do not contribute enough to disfavor the models.  
For $\tau^+\tau^-$, many more prompt photons are produced from $\pi^0$ decays, and the 
inclusive spectrum constraints disfavor this channel as an explanation for the CR anomalies.  
In the decay case, a large signal could be seen at high Galactic latitudes above backgrounds 
(see, e.g., \cite{Arvanitaki:2008hq,Arvanitaki:2009yb}).  The constraints on decaying 
dark matter are already strong enough to probe the relevant regions.  
A much stronger, but more model-dependent, constraint for the electron and muon channels 
is obtained by including the ICS contribution, see also 
\cite{Papucci:2009gd,Cirelli:2009dv}. 

Our results clearly show the sensitivity of the inclusive spectrum to indirect dark matter searches.  
The sensitivity of the inclusive spectrum to various dark matter models may be improved 
further by analyzing the differential photon spectrum as a function of Galactic latitude and longitude 
(as opposed to integrating over most of the sky), modeling the astrophysical contribution to 
the background photon spectrum, and including ICS effects.  This is beyond the scope of this paper. 


\section{Summary and Conclusions}\label{sec:conclusions}

We performed a spectral line search for gamma rays from WIMP annihilation/decay in the Milky Way using data from the \emph{Fermi} LAT. 
We detected no spectral lines from 7 to 200 GeV, and reported 95\% CL upper limits to the flux from spectral lines, which range from $4\times10^{-10}$ to $10^{-8}$ cm$^{-2}$s$^{-1}$.  For the NFW, Einasto, and isothermal Milky Way dark matter halo profiles, upper limits were derived for the $\gamma\gamma$ and $Z\gamma$ annihilation cross sections, and lower limits for the lifetime for decay to $\gamma\nu$. The limits employing the NFW profile are 
$\langle\sigma v\rangle_{\gamma\gamma}\simeq (0.03-4.6) \times 10^{-27}$ 
cm$^3$s$^{-1}$ for $m_\chi = 7-200$ GeV, 
$\langle\sigma v\rangle_{Z\gamma} \simeq (0.02-10.1)\times 10^{-27}$ 
for $m_{\chi} = 63-210$ GeV, and $\tau_{\nu\gamma}=(3.6-26.0)\times 10^{28}$ s for 
$m_{\chi} = 14-400$ GeV. 
The derived limits are strong enough to constrain the parameter space of several dark matter models. 

The inclusive photon flux spectrum can also strongly constrain a wide variety of dark matter models that 
produce a continuous spectrum of photons.  Ignoring any contribution to the inclusive spectrum 
from continuum astrophysical background sources, we presented very conservative upper limits on the annihilation 
cross section and lower limits on the decay lifetime for the channels $e^+e^-$, $\mu^+\mu^-$, 
$\tau^+\tau^-$, $W^+W^-$, $b\overline{b}$, $gg$ (gluons), $\phi\phi\to e^+e^-e^+e^-$, and 
$\phi\phi\to\mu^+\mu^-\mu^+\mu^-$, where $\phi$ is a new force mediator (scalar or vector).  
The models with final state leptons are motivated by a dark matter explanation to the 
PAMELA and \emph{Fermi} CR data, and include models with a new force mediator coupled to 
dark matter. 

\acknowledgments

The \textit{Fermi} LAT Collaboration acknowledges generous ongoing support
from a number of agencies and institutes that have supported both the
development and the operation of the LAT as well as scientific data analysis.
These include the National Aeronautics and Space Administration and the
Department of Energy in the United States, the Commissariat \`a l'Energie Atomique
and the Centre National de la Recherche Scientifique / Institut National de Physique
Nucl\'eaire et de Physique des Particules in France, the Agenzia Spaziale Italiana
and the Istituto Nazionale di Fisica Nucleare in Italy, the Ministry of Education,
Culture, Sports, Science and Technology (MEXT), High Energy Accelerator Research
Organization (KEK) and Japan Aerospace Exploration Agency (JAXA) in Japan, and
the K.~A.~Wallenberg Foundation, the Swedish Research Council and the
Swedish National Space Board in Sweden.

Additional support for science analysis during the operations phase is gratefully
acknowledged from the Istituto Nazionale di Astrofisica in Italy and the Centre National d'\'Etudes Spatiales in France.


\clearpage
\bibliographystyle{apsrev4-1}
\bibliography{refs4}

\end{document}

%% file: PhotonLinePaperAuthorList_011812.tex
\author{M.~Ackermann}
\affiliation{Deutsches Elektronen Synchrotron DESY, D-15738 Zeuthen, Germany}
\author{M.~Ajello}
\affiliation{W. W. Hansen Experimental Physics Laboratory, Kavli Institute for Particle Astrophysics and Cosmology, Department of Physics and SLAC National Accelerator Laboratory, Stanford University, Stanford, CA 94305, USA}
\author{A.~Albert}
\affiliation{Department of Physics, Center for Cosmology and Astro-Particle Physics, The Ohio State University, Columbus, OH 43210, USA}
\author{L.~Baldini}
\affiliation{Istituto Nazionale di Fisica Nucleare, Sezione di Pisa, I-56127 Pisa, Italy}
\author{G.~Barbiellini}
\affiliation{Istituto Nazionale di Fisica Nucleare, Sezione di Trieste, I-34127 Trieste, Italy}
\affiliation{Dipartimento di Fisica, Universit\`a di Trieste, I-34127 Trieste, Italy}
\author{K.~Bechtol}
\affiliation{W. W. Hansen Experimental Physics Laboratory, Kavli Institute for Particle Astrophysics and Cosmology, Department of Physics and SLAC National Accelerator Laboratory, Stanford University, Stanford, CA 94305, USA}
\author{R.~Bellazzini}
\affiliation{Istituto Nazionale di Fisica Nucleare, Sezione di Pisa, I-56127 Pisa, Italy}
\author{B.~Berenji}
\affiliation{W. W. Hansen Experimental Physics Laboratory, Kavli Institute for Particle Astrophysics and Cosmology, Department of Physics and SLAC National Accelerator Laboratory, Stanford University, Stanford, CA 94305, USA}
\author{R.~D.~Blandford}
\affiliation{W. W. Hansen Experimental Physics Laboratory, Kavli Institute for Particle Astrophysics and Cosmology, Department of Physics and SLAC National Accelerator Laboratory, Stanford University, Stanford, CA 94305, USA}
\author{E.~D.~Bloom}
\email{elliott@slac.stanford.edu}
\affiliation{W. W. Hansen Experimental Physics Laboratory, Kavli Institute for Particle Astrophysics and Cosmology, Department of Physics and SLAC National Accelerator Laboratory, Stanford University, Stanford, CA 94305, USA}
% ...
